\newif\ifLong
\Longtrue          %
\newif\ifJournal
\Journaltrue          %
\ifJournal
  \ifLong\else
    \stop
  \fi
\fi

\newcommand{\ifshort}[1]{\ifLong\else#1\fi}
\newcommand{\iflong}[1]{\ifLong#1\else\fi}

\ifLong
 \ifJournal
  \documentclass{fac}
  \usepackage{proof}
 \else
  \documentclass{article}
  \usepackage{msrtr}
  \usepackage{proof}%
  \usepackage{times}%
 \fi
\else
  \documentclass{sig-alternate}
  \usepackage{times}%
\fi

\usepackage{amssymb}                %
\usepackage{latexsym}               %
\usepackage{amsmath} 
\allowdisplaybreaks[2]
\usepackage{url}
\newif{\ifMarginalComments}
\MarginalCommentstrue   %
\newenvironment{prog}{\begin{array}[t]{@{}l@{}}}{\end{array}}

\let\slash=/
\catcode`\/=\active
\def/{\discretionary{\slash}{}{\slash}}

\makeatletter
  \newcommand{\addToLabel}[1]{%
    \protected@edef\@currentlabel{\@currentlabel#1}%
  }
\makeatother

\newtheorem{theorem}{Theorem}

\newtheorem{lemma}{Lemma}

\ifLong
  \newcommand{\qed}{\hspace*{\fill}\nolinebreak[1]\hspace*{\fill}$\Box$}
  \ifJournal\else
    \newenvironment{proof}%
    {\begin{trivlist}\item[]{\normalsize\bf Proof}\hspace*{4mm}}%
    {\qed\end{trivlist}}
  \fi
\fi
\newenvironment{proofx}%
  {\begin{trivlist}\item[]{\normalsize\bf Proof}\hspace*{4mm}}%
  {\end{trivlist}}

\newcounter{rule}

\newcommand{\staterule}[4][]{%
  \refstepcounter{rule}%
  \addToLabel{#2}%
  $\begin{array}[b]{@{}l}%
   \mbox{#2#1}\\%
   \begin{array}{c}
   #3\\
   \hline
   \raisebox{0ex}[2.5ex]{\strut}#4%
   \end{array}
  \end{array}$}

\newcommand{\GAP}{2ex}

\newcommand{\hbra}{
\hbox to \ifshort{.475}\iflong{.995}\textwidth{\vrule width0.3mm height 1.8mm depth-0.3mm
                    \leaders\hrule height1.8mm depth-1.5mm\hfill
                    \vrule width0.3mm height 1.8mm depth-0.3mm}}
\newcommand{\hket}{
\hbox to \ifshort{.475}\iflong{.995}\textwidth{\vrule width0.3mm height1.5mm
                    \leaders\hrule height0.3mm\hfill
                    \vrule width0.3mm height1.5mm}}

\newcommand{\ratio}{.4}
\newenvironment{display}[1]{%
  \ifshort{\vspace{-0.7ex}}%
  \begin{tabbing}
  \hspace{1.5em} \= \hspace{\ratio\linewidth-1.5em} \= \hspace{1.5em} \= \kill
  \textbf{\ifshort{\sc}#1}\\[-.8ex]
  \hbra\\[-.8ex]
  }{\\[-.8ex]\hket\nopagebreak
  \end{tabbing}%
  \ifshort{\vspace{-1.1ex}}%
  }
\newcommand{\entry}[2]{\>$#1$\>\>#2}
\newcommand{\clause}[2]{$#1$\>\>#2}
\newcommand{\Category}[2]{\clause{#1::=}{#2}}

\newcommand{\SUB}[1]{\{#1\}}
\newcommand{\GETS}{\mo{\leftarrow}}

\newcommand{\If}{\mathit{if}\ }
\newcommand{\Then}{\ \mathit{then}\ }
\newcommand{\Else}{\ \mathit{else}\ }
\newcommand{\SpiIf}{\kw{if}\ }
\newcommand{\SpiThen}{\ \kw{then}\ }
\newcommand{\SpiElse}{\ \kw{else}\ }

\newcommand{\Body}{\mathit{Body}}
\newcommand{\CallerId}{\mathit{callerid}}
\newcommand{\Class}{\mathit{Class}}
\newcommand{\Field}{\mathit{Field}}
\newcommand{\Id}{\mathit{Id}}
\newcommand{\In}{\mathit{in}\ } 
\newcommand{\Let}[2]{\mathit{let}\ #1\mo{=}#2\ }
\newcommand{\Meth}{\mathit{Meth}}
\newcommand{\New}{\mathit{new}}
\newcommand{\Prin}{\mathit{Prin}}
\newcommand{\Sig}{\mathit{Sig}}

\newcommand{\Type}{\mathit{Type}}

\newcommand{\fields}{\mathit{fields}}
\newcommand{\finmap}{\stackrel{\mathrm{fin}}{\rightarrow}}

\newcommand{\methods}{\mathit{methods}}
\newcommand{\mo}{\mathord}
\newcommand{\owner}{\mathit{owner}}
\newcommand{\sig}{\mathit{sig}}
\newcommand{\ty}{\mo{:}}
\newcommand{\tyq}{\mathrel{<:>}}  %
\newcommand{\infix}{\mathrel}
\newcommand{\For}[1]{\:{}^{#1}}                 %
\newcommand{\fv}{\mathit{fv}}               %
\newcommand{\WebService}{\mathit{WebService}}
\newcommand{\dom}{\mathit{dom}}

\newcommand{\Value}{\mathit{Value}}
\newcommand{\Null}{\mathit{null}}
\newcommand{\proxy}{\mathit{proxy}}

\newcommand{\class}{\mathit{class}}
\newcommand{\kw}[1]{\mathsf{#1}}

\newcommand{\As}{A\!s}
\newcommand{\xs}{x\!s}
\newcommand{\es}{e\!s}
\newcommand{\fs}{f\!s}

\newcommand{\Snd}[2]{\kw{out}\ #1\ #2}
\newcommand{\Rcv}[2]{\kw{inp}\ #1\ (#2);}
\newcommand{\Res}[1]{\kw{new}\ (#1);}        %
\newcommand{\parpop}{\infix{\mid}}              %
\newcommand{\Repl}{\kw{repeat}\ }
\newcommand{\Stop}{\kw{stop}}
\newcommand{\Split}[3]{\kw{split}\ #1\ \kw{is}\ (#2,#3);}
\newcommand{\Match}[3]{\kw{match}\ #1\ \kw{is}\ (#2,#3);}
\newcommand{\Case}[1]{\kw{case}\ #1\ }
\newcommand{\IsTag}[3]{\kw{is}\ #1(#2);#3}
\newcommand{\Cast}[2]{\kw{cast}\ #1\ \kw{is}\ (#2);}
\newcommand{\Witness}[1]{\kw{witness}\ #1;}
\newcommand{\Trust}[2]{\kw{trust}\ #1\ \kw{is}\ (#2);}

\newcommand{\SplitPoly}[2]{\kw{split}\ #1\ \kw{is}\ #2;}

\newcommand{\bdy}{\mathit{bdy}}
\newcommand{\this}{\mathit{this}}

\newcommand{\sencr}[2]{\{#1\}_{#2}}
\newcommand{\asencr}[2]{\{\!|#1|\!\}_{#2}}

\newcommand{\encrypt}[2]{\{#1\}_{#2}}
\newcommand{\Decrypt}[3]{\kw{decrypt}\ #1\ \kw{is}\
  \{\begin{array}[t]{@{}l@{}}#2\}_{#3};\end{array}%
}
\renewcommand{\check}[1]{\kw{check}\ #1}
\newcommand{\CheckNonce}[2]{\kw{check}\ #1\ \kw{is}\ #2;}

\newcommand{\Begin}[1]{\beginn{#1};}
\newcommand{\End}[1]{\endd{#1};}
\newcommand{\beginn}[1]{\kw{begin}\ #1}
\def\endd#1{\kw{end}\ #1}
\newcommand{\J}{\mathcal{J}}
\newcommand{\ka}{k}

\newcommand{\mset}[1]{[#1]}

\newcommand{\extract}[2]{#1\ (#2)}
\newcommand{\Enc}{\kw{Encrypt}}
\newcommand{\Dec}{\kw{Decrypt}}
\newcommand{\public}{\kw{Public}}
\newcommand{\private}{\kw{Private}}

\newcommand{\pencrypt}[2]{\asencr{#1}{#2}}

\newcommand{\pDecrypt}[3]{\kw{decrypt}\ #1\ \kw{is}\
  \asencr{\begin{array}[t]{@{}l@{}}#2}{#3^{-1}};\end{array}%
}

\newcommand{\CSKey}{\kw{CSKey}}

\newcommand{\Sndp}[2]{\kw{out}\ #1\ #2;}
\renewcommand{\check}[2]{\kw{check}\ #1\ #2}
\newcommand{\trust}[1]{\kw{trust}\:#1}
\newcommand{\Top}{\kw{Top}}

\newcommand{\annotate}[2]{#1\ #2}
\newcommand{\Challenge}[1]{\kw{Challenge}\ #1}
\newcommand{\Response}[1]{\kw{Response}\ #1}

\newcommand{\Key}[1]{\kw{Key}(#1)}
\newcommand{\SharedKey}[1]{\kw{SharedKey}(#1)}
\newcommand{\KeyPair}[1]{\kw{KeyPair}(#1)}

\newcommand{\WS}{\mathit{WS}}

\newcommand{\typerule}[4][]{%
  \refstepcounter{rule}%
  \addToLabel{#2}%
  $\begin{array}[b]{@{}l}%
   \mbox{#2#1}\\%
   \begin{array}{c}
   #3\\
   \hline
   \raisebox{0ex}[2.5ex]{\strut}#4%
   \end{array}
  \end{array}$}

\newcommand{\Judge}[2]{#1 \vdash #2}
\newcommand{\JudgeOK}[1]{\Judge{#1}{\diamond}}

\renewcommand{\emptyset}{\varnothing}

\newcommand{\intension}[1]{[\![#1]\!]}
\newcommand{\qq}[1]{[\![#1]\!]}
\newcommand{\TransV}[1]{\intension{#1}}
\newcommand{\TransB}[3]{\intension{#1}^{#2}_{#3}}

\newcommand{\IndexedPar}[2]{\textstyle\prod_{#1} #2}
\newcommand{\Def}[2]{#1 \triangleq #2}

\newcommand{\ClassMethods}{\mathit{ClMeth}}
\newcommand{\request}{\mathit{req}}
\newcommand{\response}{\mathit{res}}
\newcommand{\getnonce}{\mathit{getnonce}}
\newcommand{\LetCall}[3]{\kw{let}\ #1\mo{=}\kw{call}_{#2}(#3);}
\newcommand{\Un}{\kw{Un}}

\newcommand{\Union}{\kw{Union}}

\newcommand{\hash}{\mathit{Hash}}

\newcommand{\SOAPsize}{\scriptsize}
\newcommand{\unSOAPsize}{\normalsize}

\newcommand{\citem}[1]{\item[-]Case #1:}

\newcommand{\DS}[1]{\mathit{DS#1}}
\newcommand{\VK}[1]{\mathit{VK#1}}
\newcommand{\SK}[1]{\mathit{SK#1}}
\newcommand{\CertVK}[1]{\mathit{CertVK#1}}
\newcommand{\AuthKeys}[1]{\kw{AuthKeys}(#1)}
\newcommand{\AuthMsg}[1]{\kw{AuthMsg}(#1)}
\newcommand{\AuthCert}{\kw{AuthCert}}
\newcommand{\AuthCertKeys}{\kw{AuthCertKeys}}
\newcommand{\CA}{\mathit{CA}}
\newcommand{\PK}[1]{\mathit{PK#1}}
\newcommand{\EK}[1]{\mathit{EK#1}}
\newcommand{\DK}[1]{\mathit{DK#1}}
\newcommand{\CertEK}[1]{\mathit{CertEK#1}}
\newcommand{\AuthEncKeys}[1]{\kw{AuthEncKeys}(#1)}
\newcommand{\AuthEncMsg}[1]{\kw{AuthEncMsg}(#1)}
\newcommand{\AuthEncCert}{\kw{AuthEncCert}}
\newcommand{\AuthEncCertKeys}{\kw{AuthEncCertKeys}}
\newcommand{\msgI}{\mathit{msg_2}}
\newcommand{\msgII}{\mathit{msg_3}}
\newcommand{\SKey}[1]{\kw{SKey}(#1)}

\begin{document}

\ifLong
 \ifJournal
  \title{Validating a Web Service Security Abstraction by Typing}
  \author[A. D. Gordon and R. Pucella]
         {Andrew D. Gordon$^1$ and Riccardo Pucella$^2$\\
          $^1$Microsoft Research\\
          $^2$Cornell University}
  \correspond{Andrew D. Gordon, Microsoft Research, Roger Needham
    Building, 7 J J Thomson Ave, Cambridge CB3 0FB, UK. E-mail: adg@microsoft.com}
  \maketitle
  \makecorrespond
 \else
  \title{Validating a Web Service\\Security Abstraction by Typing}
  \author{Andrew D. Gordon\\
  Microsoft Research
  \and
  Riccardo Pucella\\
  Cornell University}
  \date{December 2002}
  \msrtrno{MSR--TR--2002--108}
  \msrtrmaketitle

  \subsection*{Publication History}

  A portion of this work appears in the proceedings of
  the \emph{ACM Workshop on XML Security 2002}, Washington DC,
  November 22, 2002.

  \subsection*{Affiliation}

  Riccardo Pucella is with Cornell University.  The two authors completed
  this work at Microsoft Research in Cambridge.

  \thispagestyle{empty}
  \null\clearpage
  \pagestyle{empty}
 \fi

\else
  \CopyrightYear{2002}
  \crdata{1-58113-632-3} 
  \conferenceinfo{ACM Workshop on XML Security, Nov.\ 22, 2002,}{Fairfax VA. USA}
  \title{Validating a Web Service\\Security Abstraction by Typing}
  \author{Andrew D. Gordon\\
  Microsoft Research
  \and
  Riccardo Pucella\\
  Cornell University}
  \maketitle
\fi

\begin{abstract}
An XML web service is, to a first approximation, an RPC service in which
requests and responses are encoded in XML as SOAP envelopes, and transported
over HTTP.
We consider the problem of authenticating requests and responses at the
SOAP-level, rather than relying on transport-level security.
We propose a security abstraction, inspired by earlier work on secure RPC, in
which the methods exported by a web service are annotated with one of three
security levels: none, authenticated, or both authenticated and encrypted.
We model our abstraction as an object calculus with primitives
for defining and calling web services.
We describe the semantics of our object calculus by translating to a
lower-level language with primitives for message passing and
cryptography.
To validate our semantics, we embed correspondence assertions that specify the
correct authentication of requests and responses.
By appeal to the type theory for cryptographic protocols of Gordon and
Jeffrey's Cryptyc, we verify the correspondence assertions simply by
typing.
Finally, we describe an implementation of our semantics via custom SOAP
headers.
\end{abstract}

\ifLong
 \ifJournal
  \begin{keywords}
  Web services, remote procedure call, authentication, type systems
  \end{keywords}
 \else
  \thispagestyle{empty}
  \newpage
  \thispagestyle{empty}
  \null\clearpage
  \tableofcontents
  \null\clearpage
  \pagestyle{plain}
  \pagenumbering{arabic}
 \fi
\else
  \terms{Languages, security, theory, verification} 
\fi

\section{Introduction}
\label{sec:intro}

It is common to provide application-level developers with security
abstractions that hide detailed implementations at lower levels of a
protocol stack.  For example, the identity of the sender of a message
may be exposed directly at the application-level, but computed via a
hidden, lower level cryptographic protocol.  The purpose of this paper
is to explore how to build formal models of such security
abstractions, and how to validate their correct implementation in
terms of cryptographic primitives.  Our setting is an experimental
implementation of SOAP security headers for XML web services.

\subsection{Motivation: Web Services and SOAP}

A crisp definition, due to the builders of the TerraService.NET
service, is that ``a web service is a web site intended for use by
computer programs instead of human beings''
\cite{BGSER02:TerraServiceDotNet}.  Each request to or response from a
web service is encoded in XML as a SOAP \emph{envelope}~\cite{soap}.
An envelope consists of a \emph{header}, containing perhaps routing or
security information, and a \emph{body}, containing the actual data of
the request or response.
A promising application for web services is to support direct retrieval of XML
documents from remote databases, without resorting to unreliable ``screen
scraping'' of data from HTML pages.
For example,
Google already offers programmatic access
to its database via a web service~\cite{google}.
Another major application is to support systems interoperability within an
enterprise's intranet.

The interface exported by a web service can be captured as an
XML-encoded service description, in WSDL format~\cite{wsdl}, that
describes the methods---and the types of their arguments and
results---that make up the service.  Tools exist for application-level
developers to generate a WSDL description from the code of a service,
and then to generate proxy code for convenient client access to the
web service.  Like tools for previous RPC mechanisms, these tools
abstract from the details of the underlying messaging infrastructure.
They allow us to regard calling a web service, for many if not all
purposes, as if it were invoking a method on a local object.  Our goal
is to augment this abstraction with security guarantees.

There are many signs of fervour over web services: there is widespread
tool support from both open source and commercial software suppliers,
and frequent news of progress of web service standards at bodies such
as OASIS and the W3C.  Many previous systems support RPC, but one can
argue that what's new about web services is their combination of
vendor-neutral interoperability, internet-scale, and toolsets for
``mere mortals'' \cite{BGSER02:TerraServiceDotNet}.  Still, there are
some reasons for caution.  The XML format was not originally designed
for messaging; it allows for interoperability but is inefficient
compared to binary encodings.
Moreover, it would be useful to use web services for inter-organisational
communication, for example, for e-commerce, but SOAP itself does not define
any security mechanisms.

In fact, there is already wide support for security at the
transport-level, that is, for building secure web services using HTTPS
and SSL.  Still, SSL encrypts all traffic between the client and the
web server, so that it is opaque to intermediaries.  Hence, messages
cannot be monitored by firewalls and cannot be forwarded by
intermediate untrusted SOAP-level routers.  There are proposals to
avoid some of these difficulties by placing security at the
SOAP-level, that is, by partially encrypting SOAP bodies and by
including authenticators, such as signatures, in SOAP headers.  In
particular, the WS-Security~\cite{ws-security} specification describes an
XML syntax for including such information in SOAP envelopes.

Hence, the immediate practical goal of this work is to build and
evaluate an exploratory system for SOAP-level security.

\subsection{Background: Correspondences and Spi}

Cryptographic protocols, for example, protocols for authenticating SOAP
messages, are hard to get right.  Even if we assume perfect
cryptography, exposure to various replay and impersonation attacks may
arise because of flaws in message formats.  A common and prudent
procedure is to invite expert analysis of any protocol, rather than
relying on security through obscurity.  Moreover, it is a useful
discipline to specify and verify protocol goals using formal
notations.  Here, we specify authenticity goals of our protocol using
Woo and Lam's correspondence assertions~\cite{WL93:SemanticModel}, and
verify them, assuming perfect cryptography in the sense of Dolev and
Yao~\cite{DY83:SecurityOfPublicKeyProtocols}, using type theories developed as part of the
Cryptyc
project~\cite{GJ03:AuthenticityByTyping,GJ03:TypingCorrespondenceCommunication,GJ02:TypesAndEffectsForAsymmetricCryptographicProtocols}.

Woo and Lam's correspondence assertions~\cite{WL93:SemanticModel} are a simple and
precise method for specifying authenticity properties.  The idea is to
specify labelled events that mark progress through the protocol.
There are two kinds: begin-events and end-events.  The
assertion is that every end-event should correspond to a distinct,
preceding begin-event with the same label.  For example, Alice
performs a begin-event with label ``Alice sending Bob message $M$'' at
the start of a session when she intends to send $M$ to Bob.  Upon
receiving $M$ and once convinced that it actually comes from Alice,
Bob performs an end-event with the same label.  If the correspondence
assertion can be falsified, Bob can be manipulated into thinking a
message comes from Alice when in fact it has been altered, or came
from someone else, or is a replay.  On the other hand, if the
correspondence assertion holds, such attacks are ruled out.

There are several techniques for formally specifying and verifying
correspondence assertions.  Here, we model SOAP messaging within a
process calculus, and model correspondence assertions by begin- and
end-statements within the calculus.  We use a form of the
spi-calculus~\cite{GJ03:AuthenticityByTyping}, equipped with a type
and effect system able to prove by typechecking that correspondence
assertions hold in spite of an arbitrary attacker.  Spi~\cite{AG99:spi} is
a small concurrent language with primitives for message passing and
cryptography, derived from the $\pi$-calculus~\cite{Milner99:pi}.

\subsection{Contributions of this Paper}

Our approach is as follows:
\begin{itemize}
\item Section~\ref{sec:sec-abs} describes our high-level abstraction
  for secure messaging.
\item Section~\ref{sec:formal-model} models the abstraction as an
  object calculus with primitives for creating and calling web services.
\item Section~\ref{sec:spi-calculus-semantics} defines the semantics
  of our abstraction by translating to the spi-calculus.
  Correspondence assertions specify the authenticity guarantees offered
  to caller and callee, and are verified by typechecking.
\item Section~\ref{sec:soap-level-implementation} describes a SOAP-based
  implementation using Visual Studio .NET.
\ifJournal
  \item Section~\ref{sec:asymmetric} shows how we can accommodate
    public-key infrastructures to implement the abstraction of 
    Section~\ref{sec:sec-abs}. 
\fi
\end{itemize}

Our main innovation is the idea of formalizing the authentication
guarantees offered by a security abstraction by embedding
correspondence assertions in its semantics.  On the other hand, our
high-level abstraction is fairly standard, and is directly inspired by
work on secure network objects~\cite{vanDoorn96}.  Although the rather
detailed description of our model and its semantics may seem complex,
the actual cryptographic protocol is actually quite simple.  Still, we
believe our framework and its implementation are a solid foundation
for developing more sophisticated protocols and their abstractions.

\ifLong
  \ifJournal
    Many
  \else
    Most
  \fi
formal details, as well as the proofs of our formal results, have
been relegated to the appendices. Specifically,
Appendix~\ref{app:soap-messages} gives sample messages exchanged
during web service method calls using our abstractions,
Appendix~\ref{app:semantics} gives a formal description of our object
calculus, Appendix~\ref{app:spi} gives a formal definition of the
spi-calculus used in the paper, 
 \ifJournal
   Appendix~\ref{app:proofs} gives
   the proofs of our formal results, and
   Appendix~\ref{app:first-class-services} describes an extension of
   our object calculus to capture a form of first-class web services.

   A part of this article, in preliminary form,
   appears as a conference paper~\cite{GP02:ValidatingWSSecurityAbstraction-conf}.
 \else
   Appendix~\ref{app:translation} gives
   the formal details of the translation of our object calculus into
   the spi-calculus, and 
 \item Appendix~\ref{app:asymmetric} gives an
   account of our security abstractions using asymmetric cryptography.
  \fi
\else

Some formal details are relegated to the appendices.
Appendix~\ref{app:soap-messages} gives sample messages exchanged during web
service methods calls using our abstractions, and Appendix~\ref{app:spi} gives
a detailed overview of the spi-calculus used in the paper.  A technical
report~\cite{GP02:ValidatingWSSecurityAbstraction-tr} includes details omitted
from this conference paper, such as the formal details of the translation of
our object calculus into the spi-calculus, as well as some extensions, such as
a translation relying on certificates, and all proofs.
\fi

\section{A Security Abstraction}
\label{sec:sec-abs}

We introduce a security abstraction for web services, where the
methods exported by a web service are annotated by one of three
security levels: 
\begin{center}
\begin{tabular}{ll}
\texttt{None} &  unauthenticated call\\
\texttt{Auth} &  authenticated call \\
\texttt{AuthEnc} &  authenticated and encrypted call\\
\end{tabular}
\end{center}
A \emph{call} from a client to a web service is made up of two
messages, the \emph{request} from the client to the web service, and
the \emph{response} from the web service to the client. The
inspiration for the security levels, and the guarantees they provide,
comes from SRC Secure Network Objects \cite{vanDoorn96}. An
authenticated web method call provides a guarantee of \emph{integrity}
(that the request that the service receives is exactly the one sent
by the client and that the response that the client receives is
exactly the one sent by the service as a response to this request) and
\emph{at-most-once semantics} (that the service receives the request
most once, and that the client receives the response at most once). An
authenticated and encrypted web method call provides all the
guarantees of an authenticated call, along with a guarantee of
\emph{secrecy} (that an eavesdropper does not obtain any part of the
method name, the arguments, or the results of the call).

\ifJournal
  We use the language C${}^\#$ to present our security
 abstraction. (There is nothing specific to C${}^\#$ in our approach,
 although the implementation we describe in this section and in
 Section~\ref{sec:soap-level-implementation} takes advantage of some
 features of the language.)
\fi
In C${}^\#$, where users can specify \emph{attributes} on various
entities, our security annotations take the form of an attribute on
web methods, that is, the methods exported by a web service. The attribute
is written
\texttt{[SecurityLevel(}\textit{level}\texttt{)]}, where
\textit{level} is one of \texttt{None}, \texttt{Auth}, or
\texttt{AuthEnc}. For example, consider a simple interface to a
banking service, where \texttt{[WebMethod]} is an attribute used to
indicate a method exported by a web service:

\begin{verbatim}
 class BankingServiceClass {

   string callerid;

   [WebMethod] [SecurityLevel(Auth)]
   public int Balance (int account);

   [WebMethod] [SecurityLevel(AuthEnc)]
   public string Statement (int account);

   [WebMethod] [SecurityLevel(Auth)]
   public void Transfer (int source, 
                         int dest, 
                         int amount);
 }
\end{verbatim}

The annotations get implemented by code to perform the authentication
and encryption, at the level of SOAP envelopes, transparently from the
user. The annotations on the web service side will generate a method
on the web service that can be used to establish a security
context. This method will never be invoked by the user, but
automatically by the code implementing the annotations.  For the
purpose of this paper, we assume a simple setting for authentication and
secrecy, namely that the principals involved possess shared
keys. Specifically, we assume a distinct key $K_{pq}$ shared between every pair
of principals $p$ and $q$. We use the key $K_{pq}$
when $p$ acts as the client and $q$ as the web service. (Notice
that $K_{pq}$ is different from $K_{qp}$.) It is
straightforward to extend our approach to different settings such as
public-key infrastructures or certificate-based authentication
\ifLong 
 \ifJournal
mechanisms (see Section~\ref{sec:asymmetric}).
 \else
mechanisms (see Appendix~\ref{app:asymmetric}).
 \fi
\else
mechanisms (see our technical report~\cite{GP02:ValidatingWSSecurityAbstraction-tr}).
\fi

An authenticated call by $p$ to a web method $\ell$ on a web service $w$ owned
by $q$ with arguments $u_1,\ldots,u_n$ producing a result $r$ uses the
following protocol:
\[\begin{array}{l}
p \rightarrow q: \mbox{request nonce}\\
q \rightarrow p: n_q\\
p \rightarrow q: \begin{prog}
                 p,\request(w,\ell(u_1,\ldots,u_n),s,n_q),n_p,\ifJournal\else \\ \quad \fi
                 \hash(\request(w,\ell(u_1,\ldots,u_n),s,n_q),K_{pq})
                 \end{prog}\\
q \rightarrow p: q,\response(w,\ell(r),s,n_p),\hash(\response(w,\ell(r),s,n_p),K_{pq})
\end{array}\]
Here, $\hash$ is a cryptographic hash function (a one-way message
digest function such as MD5).  We tag the request and the
response messages to be able to differentiate them. We also tag the
response with the name of the method that was originally called. We
include a unique \emph{session tag} $s$ in both the request and
response message to allow the caller $p$ to match the response with
the actual call that was performed. 

An authenticated and encrypted call by $p$ to a web method $\ell$ on a web
service $w$ owned by $q$ with arguments $u_1,\ldots,u_n$ producing a result
$r$ uses a similar protocol, with the difference that the third and fourth
messages are encrypted using the shared key instead of signed:
\[\begin{array}{l}
p \rightarrow q: \mbox{request nonce}\\
q \rightarrow p: n_q\\
p \rightarrow q: p,\sencr{\request(w,\ell(u_1,\ldots,u_n),s,n_q)}{K_{pq}},n_p\\
q \rightarrow p: q,\sencr{\response(w,\ell(r),s,n_p)}{K_{pq}}
\end{array}\]

To convince ourselves that the above protocols do enforce the
guarantees prescribed by the security abstraction, we typically argue
as follows. Let's consider the authenticated and encrypted case, the
authenticated case being similar. When the web service $w$ run by
principal $q$ receives a request $w,\ell(u_1,\ldots,u_n),s,n_q$ encrypted with
$K_{pq}$ ($q$ uses the identity $p$ in the request to determine which
key to use), it knows that only $p$ could have created the message,
assuming that the shared key $K_{pq}$ is kept secret by both $p$ and
$q$. This enforces the integrity of the request. Since the message 
also contains the nonce $n_q$ that the web service can check has never
appeared in a previous message, it knows that the message is not a
replayed
message, hence enforcing at-most-once semantics. Finally, the
secrecy of the shared key $K_{pq}$ implies the secrecy of the
request. A similar argument shows that the protocol satisfies
integrity, at-most-once-semantics, and secrecy for the response.

What do we have at this point? We have an informal description of a security
abstraction, we have an implementation of the abstraction in terms of
protocols, and an informal argument that the guarantees prescribed by the
abstraction are enforced by the implementation. How do we make our security
abstraction precise, and how do we ensure that the protocols do indeed enforce
the required guarantees? In the next section, we give a formal model to make
the abstraction precise. Then, we formalize the implementation by showing how
to translate the abstractions into a lower level calculus that uses the above
protocols. We use types to show that guarantees are formally met by the
implementation, via correspondence assertions.

\section{A Formal Model}
\label{sec:formal-model}

\newcommand{\Alice}{\mathit{Alice}}
\newcommand{\Bob}{\mathit{Bob}}

We model the application-level view of authenticated messaging as an object
calculus.  Object
calculi~\cite{Abadi96,GS01:TypingAMultilanguageIntermediateCode,IPW99:FeatherweightJava}
are object-oriented languages in miniature, small enough to make formal proofs
feasible, yet large enough to study specific features.  As in
FJ~\cite{IPW99:FeatherweightJava}, objects are typed, class-based, immutable,
and deterministic.  As in some of Abadi and Cardelli's object calculi
\cite{Abadi96}, we omit subtyping and inheritance for the sake of simplicity.
In spite of this simplicity, our calculus is Turing complete.  We can define
classes to implement arithmetic, lists, collections, and so on.

To model web services, we assume there are finite sets $\Prin$ and
$\WebService$ of principal identifiers and web service identifiers,
respectively.  We think of each $w \in \WebService$ as a URL referring
to the service; moreover, $\class(w)$ is the name of the class that
implements the service, and $\owner(w) \in \Prin$ is the principal
running the service.

To illustrate this model, we express the banking service interface
introduced in the last section in our calculus. Suppose 
there are
two principals $\Alice,\Bob\in\Prin$, and a web service
$w=\textit{http://bob.com/BankingService}$, where we have $\owner(w)=\Bob$ and
$\class(w)=\mathit{BankingServiceClass}$.  Suppose we wish to implement
the $\mathit{Balance}$ method so that given an account number, it checks that
it has been called by the owner of the account, and if so returns the
balance.  If $\Alice$'s account number is $12345$, we might achieve
this as follows:

\[\begin{prog}
\mathit{class}\ \mathit{BankingServiceClass}\\ \quad
  \begin{prog}
  \Id\:\mathit{CallerId}\\
  \mathit{Num}\:\mathit{Balance}(\mathit{Num}\:\mathit{account})\\ \quad
    \begin{prog}
    \If \mathit{account}=12345 \Then \\ \quad
     \If \mathit{this}.\mathit{CallerId}=\Alice \Then 100 \Else \Null \\
    \Else \ldots
    \end{prog}
  \end{prog}
\end{prog}\]
There are a few points to note about this code. First, as in
BIL~\cite{GS01:TypingAMultilanguageIntermediateCode}, method bodies
conform to a single applicative syntax, rather than there being
separate grammars for statements and expressions. Second, while the
C${}^\#$ code relies on attributes to specify exported methods and security
levels, there are not attributes in our calculus. For simplicity, we
assume that all the methods of a class implementing a web service are
exported as web methods. Furthermore, we assume that all these
exported methods are authenticated and encrypted, as if they had been
annotated \texttt{AuthEnc}. (It is straightforward to extend our
calculus to allow per-method annotations but it complicates the
presentation of the translation in the next section.) 

Every class implementing a web service has exactly one field, named
$\mathit{CallerId}$, which exposes the identity of the caller, and allows
application-level authorisation checks.

We write $w\mo{:}\mathit{Balance}(12345)$ for a client-side call to
method $\mathit{Balance}$ of the service $w$.  The semantics of such a
web service call by $\Alice$ to a service owned by $\Bob$ is that $\Bob$
evaluates the local \ifshort{method} call
$\New\:\mathit{BankingServiceClass}(\Alice).\mathit{Balance}(12345)$ as
$\Bob$.  In other words, $\Bob$ creates a new object \ifshort{of the form}\ifJournal of the form\fi
$\New\:\mathit{BankingServiceClass}(\Alice)$ (that is, an instance of the
class $\mathit{BankingServiceClass}$ with $\mathit{CallerId}$ set to $\Alice$)
and then calls the $\mathit{Balance}$ method.  This would terminate with
$100$, since the value of $\mathit{this}.\mathit{CallerId}$ is $\Alice$.
(For simplicity, we assume every class in the object calculus has a single
constructor whose arguments are the initial values of the object's fields.)
This semantics guarantees to the server $\Bob$ that the field
$\mathit{CallerId}$ contains the identity of his caller, and guarantees to the
client $\Alice$ that only the correct owner of the service receives the
request and returns the result.

In a typical environment for web services, a client will not invoke
web services directly. Rather, a client creates a proxy object
corresponding to the web service, which encapsulates the remote
invocations. Those proxy objects are generally created automatically
by the programming environment. Proxy objects are easily expressible
in our calculus, by associating with every web service $w$ a proxy
class $\proxy(w)$. The class $\proxy(w)$ has a method for every method
of the web service class, the implementation for which simply calls
the corresponding web service method. The proxy class also has a field
$\mathit{Id}$ holding the identity of the owner of the web service. 
Here is the client-side proxy class for our example service:
\[\begin{prog}
\mathit{class}\ \mathit{BankingServiceProxy}\\ \quad
  \begin{prog}
  \Id\:\mathit{Id}()\\ \quad
    \Bob\\
  \mathit{Num}\:\mathit{Balance}(\mathit{Num}\:\mathit{account})\\ \quad
    w\mo{:}\mathit{Balance}(\mathit{account})
  \end{prog}
\end{prog}\]

The remainder of this section details the syntax and informal
semantics of our object calculus.

\subsection{Syntax}
In addition to $\Prin$ and $\WebService$, we assume finite sets
$\Class$, $\Field$, $\Meth$ of class, field, and method names,
respectively.
\begin{renewcommand}{\ratio}{.5}
\begin{display}{Classes, Fields, Methods, Principals, Web Services:}
\clause{c \in \Class}{class name}\\
\clause{f \in \Field}{field name}\\
\clause{\ell \in \Meth}{method name}\\
\clause{p \in \Prin}{principal name}\\
\clause{w \in \WebService}{web service name}
\end{display}
\end{renewcommand}

There are two kinds of data type: $\Id$ is the type of principal
identifiers, and $c \in \Class$ is the type of instances of class $c$.
A method signature specifies the types of its arguments and result.
\begin{display}{Types and Method Signatures:}
\Category{A,B \in \Type}{type}\\
\entry{\Id}{principal identifier}\\
\entry{c}{object}\\
\ifLong
\clause{\sig \in \Sig ::= B(A_1\:x_1,\ldots,A_n\:x_n)}
  {method signature ($x_i$ distinct)}
\else
\clause{\sig \in \Sig}{method signature}\\
\entry{B(A_1\:x_1,\ldots,A_n\:x_n)}{($x_i$ distinct)}
\fi
\end{display}

An execution environment defines the services and code available in
the distributed system.  In addition to $\owner$ and $\class$, described above,
the maps $\fields$ and $\methods$ 
specify
the types of each field and the
signature and body of each method, respectively.
We write $X \to Y$ and $X \finmap Y$ for the sets of total functions and finite
maps, respectively, from $X$ to $Y$.
\begin{display}{Execution Environment: $(\fields,\methods,\owner,\class)$}
\clause{\fields \in \Class \to (\Field \finmap \Type)}
  {fields of a class}\\
\ifLong
  \clause{\methods \in \Class \to (\Meth \finmap \Sig \times \Body)}
    {methods of a class}\\
\else
  \clause{\methods \in \Class \to (\Meth \finmap \Sig \times \Body)}{}\\
  \clause{}{methods of a class}\\
\fi
\clause{\owner \in \WebService \to \Prin}
  {service owner}\\
\clause{\class \in \WebService \to \Class}
  {service implementation}
\end{display}

We complete the syntax by giving the grammars for \emph{method bodies}
and for \emph{values}.
\newcommand{\ifour}{\mathit{i4}}
\begin{renewcommand}{\ratio}{.34}
\begin{display}{Values and Method Bodies:}
\clause{x,y,z}{name: variable, argument}\\
\Category{u,v \in \Value}{value}\\
\entry{x}{variable}\\
\entry{\Null}{null}\\
\entry{\New\:c(v_1,\ldots,v_n)}{object}\\
\entry{p}{principal identifier}\\
\Category{a,b \in \Body}{method body}\\
\entry{v}{value}\\
\entry{\Let{x}{a} \In{b}}{let-expression}\\
\entry{\If u=v \Then a \Else b}{conditional}\\
\entry{v.f}{field lookup}\\
\entry{v.\ell(u_1,\ldots,u_n)}{method call}\\
\entry{w\mo{:}\ell(u_1,\ldots,u_n)}{service call}
\end{display}
\end{renewcommand}

The free variables $\fv(a)$ of a method body are defined in the usual
way, where the only binder is $x$ being bound in $b$ in the expression
$\Let{x}{a}\In{b}$. We write $a\SUB{x\GETS b}$ for the outcome of a
capture-avoiding substitution of $b$ for each free occurence of the
variable $x$ in method body $a$. We view method bodies as being equal
up to renaming of bound variables. Specifically, we take
$\Let{x}{a}\In{b}$ to be equal to $\Let{x'}{a}\In{b\SUB{x\GETS x'}}$,
if $x'\not\in\fv(b)$.

Our syntax for bodies is in a reduced form that simplifies its
semantics; in examples, it is convenient to allow a more liberal
syntax.
For instance, let \ifJournal the term \fi $\If a_1=a_2\Then b_1 \Else b_2$
be short for $\Let{x_1}{a_1}\In\Let{x_2}{a_2}\In \If x_1=x_2\Then b_1
\Else b_2$.
We already used this \ifshort{shorthand} when writing
$\If \mathit{this}.\mathit{CallerId}=\Alice \Then 100 \Else \Null$
in our example. Similarly, we assume a class $\mathit{Num}$ for
numbers, and write integer literals such as $100$ as shorthand for
objects of that class.

Although objects are values, in this calculus, web services are not.  This
reflects the fact that current WSDL does not allow for web services to be
passed as requests or results. 
\ifLong 
  We explore an extension of our model to account for web services as
``first-class values'' in Appendix~\ref{app:first-class-services}.
\fi

We assume all method bodies in our execution environment are well-typed.  If
$\methods(c)(\ell) = (\sig,b)$ and the signature $\sig=B(A_1\:x_1, \ldots,
A_n\:x_n)$ we assume that the body $b$ has type $B$ given a typing environment
$\mathit{this}\ty c, x_1\ty A_1, \ldots, x_n \ty A_n$.  The variable
$\mathit{this}$ refers to the object on which the $\ell$ method was
invoked.
\ifJournal
  The type system is given by a typing judgment $E\vdash a:A$, saying
  that $a$ has type $A$ in an environment $E$ of the form $x_1\ty
  A_1,\ldots,x_n\ty A_n$ that gives a type to the free variables in
  $a$. The domain $\dom(E)$ of $E$ is the set of variables
  $\{x_1,\ldots,x_n\}$ given a type in $E$. 
\fi
\ifLong
The typing rules, which are standard, are given in
Appendix~\ref{app:semantics}. 
\else
The typing rules, which are standard, appear in the
technical report~\cite{GP02:ValidatingWSSecurityAbstraction-tr}.
\fi
We also assume the class $\class(w)$ corresponding to each web service
$w$ has a single field $\CallerId$.  

\subsection{Informal Semantics of our Model}
We explain informally the outcome of evaluating a method body $b$
as principal $p$, that is, on a client or server  machine controlled by $p$.
(Only the semantics of web service calls depend on $p$.) 
\ifLong
A formal account of this semantics, as well as the typing rules of the 
calculus, can be found in Appendix~\ref{app:semantics}.
\fi

To evaluate a value $v$ as $p$, we terminate at once with $v$ itself.

To evaluate a let-expression $\Let{x}{a} \In{b}$ as $p$, we first
evaluate $a$ as $p$.  If $a$ terminates with a value $v$, we proceed
to evaluate $b\SUB{x \GETS v}$, that is, $b$ with each occurrence of
the variable $x$ replaced with $v$.  The outcome of evaluating
$b\SUB{x \GETS v}$ as $p$ is the outcome of evaluating the whole
expression.

To evaluate a conditional $\If u=v \Then a \Else b$ as $p$, we
evaluate $a$ as $p$ if $u$ and $v$ are the same; else we evaluate
$b$ as $p$.

To evaluate a field lookup $v.f$ as $p$, when $v$ is an object value
$\New\:c(v_1,\ldots,v_n)$, we check $f$ is the $j$th field of class
$c$ for some $j \in 1..n$ (that is, that $\fields(c) = f_i \mapsto A_i
\For{i \in 1..n}$ and that $f=f_j$), and then return $v_j$.  If $v$ is
null or if the check fails, evaluation has gone wrong.

To evaluate a method call $v.\ell(u_1,\ldots,u_n)$ as $p$, when $v$ is an
object $\New\:c(v_1,$ \ldots, $v_n)$, we check $\ell$ is a method of class $c$
(that is, that $\methods(c) = \ell_i \mapsto (\sig_i,b_i) \For{i \in 1..m}$
and that $\ell=\ell_j$ for some $j \in 1..m$) and we check the arity of its
signature is $n$ (that is, that $\sig_j=B(A_1\:x_1,\ldots,A_n\:x_n)$) and then
we evaluate the method body as $p$, but with the object $v$ itself in place of
the variable $\this$, and actual parameters $u_1$, \ldots, $u_n$ in place of
the formal parameters $x_1$, \ldots, $x_n$ (that is, we evaluate the
expression $b_i\SUB{\this\GETS v,x_1 \GETS u_1,$ \ldots, $x_n \GETS u_n}$).
If $v$ is null or if either check fails, evaluation has gone wrong.

To evaluate a service call $w\mo{:}\ell(u_1,\ldots,u_n)$ as $p$, we evaluate
the \iflong{local} method call $\New\:c(p).\ell(u_1,\ldots,u_n)$ as $q$, where
$c=\class(w)$ is the class implementing the service, and $q=\owner(w)$ is the
principal owning the service.  (By assumption, $c$'s only field is
$\mathit{CallerId}$ of type $\Id$.)  This corresponds directly to creating a
new object on $q$'s web server to process the incoming request.

\section{A Spi-Calculus Semantics}
\label{sec:spi-calculus-semantics}

We confer a formal semantics on our \ifshort{object} calculus by translation
to the spi-calculus~\cite{AG99:spi,GJ03:AuthenticityByTyping}, a lower-level
language with primitives for message-passing (to model SOAP requests and
responses) and cryptography (to model encryption and decryption of SOAP
headers and bodies).

\subsection{A Typed Spi-Calculus (Informal Review)}

To introduce the spi-calculus, we formalize the situation where Alice
sends a message to Bob using a shared key, together with a
correspondence assertion concerning authenticity of the message, as
outlined in Section~\ref{sec:intro}.  A \emph{name} is an identifier
that is atomic as far as our analysis is concerned.  In this example,
the names $\Alice$ and $\Bob$ identify the two principals, the name
$K$ represents a symmetric key known only to $\Alice$ and $\Bob$, and
the name $n$ represents a public communication channel.  A
\emph{message}, $M$ or $N$, is a data structure such as a name, a
tuple $(M_1,\ldots,M_n)$, a tagged message $t(M)$, or a ciphertext
$\encrypt{M}{N}$ (that is, a message $M$ encrypted with a key $N$,
which is typically a name).  A \emph{process}, $P$ or $Q$, is a
program that may perform local computations such as encryptions and
decryptions, and may communicate with other processes by
message-passing on named channels.  For example, the process $P_\Alice
=
\Begin{\mathit{sending}(\Alice,\Bob,M)} \Snd{n}{\encrypt{M}{K}}$ defines
Alice's behaviour.  First, she performs a begin-event labelled by the tagged
tuple $\mathit{sending}(\Alice,\Bob,M)$, and then she sends the ciphertext
$\encrypt{M}{K}$ on the channel $n$.  The process $P_\Bob =
\Rcv{n}{x}\Decrypt{x}{y}{K}\End{\mathit{sending}(\Alice,\Bob,y)}$ defines
Bob's behaviour.  He blocks till a message $x$ arrives on the channel $n$.
Then he attempts to decrypt the message with the key $K$.  We assume there is
sufficient redundancy, such as a checksum, in the ciphertext that we can tell
whether it was encrypted with $K$.  If so, the plaintext message is
bound to $y$, and he performs an end-event labelled
$\mathit{sending}(\Alice,\Bob,y)$.  The process $\Res{K}(P_\Alice \parpop
P_\Bob)$ defines the complete system.  The composition $P_\Alice \parpop
P_\Bob$ represents Alice and Bob running in parallel, and able to communicate
on shared channels such as $n$.  The binder $\kw{new}(K)$ restricts the scope
of the key $K$ to the process $P_\Alice \parpop P_\Bob$ so that no external
process may use it.  
Appendix~\ref{app:spi} contains the grammar of spi messages and
processes. The grammar includes the type annotations that are required
to appear in spi terms. 
\ifJournal\else In this section, we omit the type annotations
in spi terms for the purpose of illustrating our approach.\fi

We include begin- and end-events in processes simply to specify correspondence
assertions.  We say a process is \emph{safe} to mean that in every run, and
for every $L$, there is a distinct, preceding $\beginn{L}$ event for every
$\endd{L}$ event.  Our example is safe, because Bob's end-event can only
happen after Alice's begin-event.

For correspondence assertions to be interesting, we need to model the
possibility of malicious attacks.  Let an \emph{opponent} be a spi-calculus
process $O$, arbitrary except that $O$ itself cannot perform begin- or
end-events.  We say a process $P$ is \emph{robustly safe} if and only if $P
\parpop O$ is safe for every opponent $O$.  Our example system $\Res{K}(P_\Alice
\parpop P_\Bob)$ is not robustly safe.  The opponent cannot acquire the key
$K$ since its scope is restricted, but it can intercept messages on the public
channel $n$ and mount a replay attack.  The opponent
$\Rcv{n}{x}\Snd{n}{x};\Snd{n}{x}$ duplicates the encrypted message so that Bob
may mistakenly accept $M$ and perform the end-event
$\mathit{sending}(\Alice,\Bob,M)$ twice.  To protect against replays, and to
achieve robust safety, we can add a nonce handshake to the protocol.

In summary, spi lets us precisely represent the behaviour of protocol
participants, and specify authenticity guarantees by process
annotations.  Robust safety is the property that no opponent at the
level of the spi-calculus may violate these guarantees.  We omit the
details here, but a particular type and effect system verifies robust
safety: if a process can be assigned the empty effect, then it is
robustly safe.  The example above is simple, but the general method
works for a wide range of protocol
examples~\cite{GJ03:AuthenticityByTyping,GJ02:TypesAndEffectsForAsymmetricCryptographicProtocols}.

\ifLong
\ifJournal
For the sake of clarity, we defer some of the technical
details to the appendices. Specifically, Appendix~\ref{app:spi}
contains more details on the spi-calculus and the type and effect
sytem, as well as a formal definition of robust safety;
Appendix~\ref{app:proofs} gives a proof of our technical
results.
\else
For the sake of clarity, we defer most of the technical
details to the appendices. Specifically, Appendix~\ref{app:spi}
contains more details on the spi-calculus and the type and effect
sytem, as well as a formal definition of robust safety;
Appendix~\ref{app:translation} gives a complete description of the
translation from our object calculus to spi, including all the type
annotations omitted in this section, and a proof of our technical
results.
\fi
\fi

\subsection{A Semantics for Local Computation}

\newcommand{\Princ}{\kw{Prin}}

\ifJournal
 We translate the types, values, and method bodies of our object
 calculus to types, messages,
\else
 We translate the values and method bodies of our object calculus to
 messages
\fi
and processes, respectively, of the spi calculus.  To begin with, we omit web
services.  Many computational models can be studied by translation to 
process calculi; our translation of local computation follows a 
fairly standard pattern.

We use the notation $\TransV{\,}$ to represent the translation of the
types and terms of our object calculus to appropriate types, messages,
and processes in the spi calculus. In many places, we also define
abbreviations in the spi calculus (for instance, we define
$\LetCall{x}{w}{p,\mathit{args}}P$ as shorthand for a more complex spi
calculus process); these do not use the $\TransV{\,}$ notation.

We assume that $\Prin$ are spi-calculus names, and that $\Field \cup
\Meth \cup \Class \cup \{\Null\}$ are message tags. 
\ifJournal
The translations for types is straightforward. 
Since principal identifiers are presumably known to the opponent, the
type of identifiers corresponds to the spi type $\Un$. A value of
class $c$ is either the value $\Null$, or a tagged tuple
$\New\:c(v_1,\ldots,v_n)$. As we shall see below, we translate $\Null$
to a tagged empty tuple $\Null()$, and an object to a tagged tuple
$c(v_1,\ldots,v_n)$. Thus, a class $c$ translates to a tagged union
type with components $\Null(\Un)$ and $c(\Un)$. (The types $\Un$
indicate that the content of the tuples are presumably known to the
opponent.) 

\begin{display}{Type Translation:}
\clause{\Princ \triangleq \Un}\\
\clause{\Def{\TransV{\Id}}{\Princ}}\\
\clause{\Def{\TransV{c}}
  {\mbox{$\Union(\Null(\Un), c(\Un))$}}}
\end{display}

\begin{display}{Environment Translation:}
\clause{\TransV{x_1\ty A_1, \ldots, x_n\ty A_n} \triangleq
  x_1\ty \TransV{A_1}, \ldots, x_n\ty \TransV{A_n}}
\end{display}

If $\As=A_1,\ldots,A_n$ and $\xs=x_1,\ldots,x_n$
we sometimes write $B(\As\:\xs)$ as shorthand
for the signature $B(A_1\:x_1,\ldots,A_n\:x_n)$.
We define two shorthands for types corresponding to web method
calls. The type $\mathit{Req}(w)$ represents the type of possible calls
to web methods provided by the service $w$; the type of a call is
simply the translated type of the arguments of the web method, tagged
with the name of the method. Similarly, the type $\mathit{Res}(w)$
represents the type of the results of web methods provided by the
service $w$; the type of a result of a call is simply the translated
type of the result of the web method, tagged once again with the name
of the method. 

\begin{display}{Request and Response Types:}
\clause{\qq{A_1,\ldots,A_m} \triangleq \qq{A_1},\ldots,\qq{A_m}}\\
\clause{
  \begin{prog}
  \mathit{Req}(w) \triangleq
   \Union(\ell_i(\qq{\As_i}) \For{i \in 1..n}) \\ \quad
  \mbox{where $\class(w)=c$ and $\methods(c) =
    \ell_i \mapsto (B_i(\As_i \xs_i),b_i) \For{i \in 1..n}$}
  \end{prog}
} \\
\clause{
  \begin{prog}
  \mathit{Res}(w) \triangleq
   \Union(\ell_i(\qq{B_i}) \For{i \in 1..n}) \\ \quad
  \mbox{where $\class(w)=c$ and $\methods(c) =
    \ell_i \mapsto (B_i(\As_i \xs_i),b_i) \For{i \in 1..n}$}
  \end{prog}
}
\end{display}

The translation of expressions really acts on the type derivation of an
expression, not just the expression itself. This means that during the
translation of an expression, we have access to the types of the
subexpressions appearing in the expression. To reduce clutter, we write the
translation as though it is acting on the expression itself, except that when
we need access to the type of a subexpression, we annotate the appropriate
subexpression with its type. For example, the translation of
$\Let{x}{a}\In{b}$ depends on the type of $a$, which is available through the
type derivation of $\Judge{E}{\Let{x}{a}\In{b}:B}$. We write
$\Let{x}{a_A}\In{b}$ to indicate that the type of $a$ is $A$, according to the
type derivation.
\fi
Values translate
easily; in particular, an object translates to a tagged tuple
containing the values of its fields. 

\begin{display}{Translation of a Value $v$ to a Message $\qq{v}$:}
\clause{\Def{\TransV{x}}{x}}\\
\clause{\Def{\TransV{\Null}}{\Null()}}\\
\clause{\Def{\TransV{\New\:c(v_1,\ldots,v_n)}}{c(\TransV{v_1},\ldots,\TransV{v_n})}}\\
\clause{\Def{\TransV{p}}{p}}
\end{display}

We translate a body $b$ to a process $\TransB{b}{p}{k}$ that
represents the evaluation of $b$ as principal $p$.  The name $k$ is a
continuation, a communications channel on which we send $\qq{v}$ to
represent termination with value $v$.  Since our focus is representing
safety rather than liveness properties, we represent an evaluation
that goes wrong simply by the inactive process $\Stop$; it would be
easy---but a complication---to add an exception mechanism.  We use
standard $\kw{split}$ and $\kw{case}$ statements to analyse tuples and
tagged messages, respectively.  To call a method $\ell$ of an object
$v$ of class $c$, with arguments $u_1$, \ldots, $u_n$ we send the
tuple $(p,\TransV{v},\TransV{u_1},\ldots,\TransV{u_n},k)$ on the
channel $c\_\ell$.  The name $p$ is the caller, and channel $k$ is the
continuation for the call.  We translate method $\ell$ of class $c$ to
a process that repeatedly awaits such messages, and triggers
evaluations of its body. 
We defer the translation of web method calls until
Section~\ref{s:semantics-web-services}. 
Our translation depends in part on type
information; we write $v_c$ in the translation of field lookups and
method calls to indicate that $c$ is the type of $v$. 

\ifJournal
\begin{display}{Translation of a Method Body $b$ to a Process
$\TransB{b}{p}{k}$:}
\clause{\Def{\TransB{v}{p}{k}}
            {\Snd{k}{\TransV{v}}}}\\
\clause{\Def{\TransB{\Let{x}{a_A} \In{b}}{p}{k}}
            {\Res{k'\ty \Un}
              (\TransB{a}{p}{k'} \parpop \Rcv{k'}{x\ty\Un}\TransB{b}{p}{k})}}\\
\clause{\Def{\TransB{\If u=v \Then a \Else b}{p}{k}}
            {\SpiIf \TransV{u}=\TransV{v}
             \SpiThen \TransB{a}{p}{k} \SpiElse \TransB{b}{p}{k} }}\\
\clause{\Def{\TransB{v_c.f_j}{p}{k}}
            {\Case{\TransV{v}}
             \begin{prog}
               \IsTag{\Null}{y\ty\Un}{\Stop} \\
               \IsTag{c}{y\ty\Un}{\SplitPoly{y}{(x_1\ty\qq{A_1},\ldots,x_n\ty\qq{A_n})}\Snd{k}{x_j}} \\ \qquad
               \mbox{where $\fields(c)=f_i \mapsto A_i \For{i \in 1..n}$, and $j \in 1..n$}
        \end{prog}}}\\
\clause{\Def{\TransB{v_c.\ell(u_1,\ldots,u_n)}{p}{k}}
            {\Case{\TransV{v}}
             \begin{prog}
               \IsTag{\Null}{y\ty\Un}{\Stop} \\
               \IsTag{c}{y\ty\Un}{\Snd{c\_\ell}{(p,\TransV{v},\TransV{u_1},\ldots,\TransV{u_n},k)}}
        \end{prog}}}
\end{display}
\else
\begin{display}{Translation of a Method Body $b$ to a Process
$\TransB{b}{p}{k}$:}
\clause{\Def{\TransB{v}{p}{k}}
            {\Snd{k}{\TransV{v}}}}\\
\clause{\Def{\TransB{\Let{x}{a} \In{b}}{p}{k}}
            {\Res{k'}
              (\TransB{a}{p}{k'} \parpop \Rcv{k'}{x}\TransB{b}{p}{k})}}\\
\clause{\Def{\TransB{\If u=v \Then a \Else b}{p}{k}}
            {\SpiIf \TransV{u}=\TransV{v}
             \SpiThen \TransB{a}{p}{k} \SpiElse \TransB{b}{p}{k} }}\\
\ifLong
\clause{\Def{\TransB{v_c.f_j}{p}{k}}
            {\Case{\TransV{v}}
             \begin{prog}
               \IsTag{\Null}{y}{\Stop} \\
               \IsTag{c}{y}{\SplitPoly{y}{(x_1,\ldots,x_n)}\Snd{k}{x_j}} \\ \qquad
               \mbox{where $\fields(c)=f_i \mapsto A_i \For{i \in 1..n}$, and $j \in 1..n$}
        \end{prog}}}\\
\clause{\Def{\TransB{v_c.\ell(u_1,\ldots,u_n)}{p}{k}}
            {\Case{\TransV{v}}
             \begin{prog}
               \IsTag{\Null}{y}{\Stop} \\
               \IsTag{c}{y}{\Snd{c\_\ell}{(p,\TransV{v},\TransV{u_1},\ldots,\TransV{u_n},k)}}
        \end{prog}}}
\else
\clause{\begin{prog}\TransB{v_c.f_j}{p}{k} \triangleq\\
        \quad \begin{prog}
             \Case{\TransV{v}}
             \begin{prog}
               \IsTag{\Null}{y}{\Stop} \\
               \IsTag{c}{y}{\SplitPoly{y}{(x_1,\ldots,x_n)}\Snd{k}{x_j}} \\ \qquad
               \mbox{where $\fields(c)=f_i \mapsto A_i \For{i \in 1..n}$, and $j \in 1..n$}
             \end{prog}
              \end{prog}
        \end{prog}}\\
\clause{\begin{prog}\TransB{v_c.\ell(u_1,\ldots,u_n)}{p}{k} \triangleq\\
        \quad\begin{prog}
            \Case{\TransV{v}}
             \begin{prog}
               \IsTag{\Null}{y}{\Stop} \\
               \IsTag{c}{y}{\Snd{c\_\ell}{(p,\TransV{v},\TransV{u_1},\ldots,\TransV{u_n},k)}}
             \end{prog}
             \end{prog}
        \end{prog}}
\fi
\end{display}
\fi

\ifJournal
\begin{display}{Translation of Method $\ell$ of Class $c$:}
\clause{
I_{\mathit{class}}(c,\ell) \triangleq
     \begin{prog}
     \Repl\Rcv{c\_\ell}{z\ty\Un}\\
     \SplitPoly{z}{(p\ty\Princ,\this\ty\Un,x_1\ty\qq{A_1},\ldots,x_n\ty\qq{A_n},k\ty\Un)}
        \TransB{b}{p}{k}\\ \quad
     \mbox{where $\methods(c)(\ell) = (B(A_1\:x_1,\ldots,A_n\:x_n),b)$}
     \end{prog}}
\end{display}
\else
\begin{display}{Translation of Method $\ell$ of Class $c$:}
\ifLong
\clause{I_{\mathit{class}}(c,\ell) \triangleq
     \begin{prog}
     \Repl\Rcv{c\_\ell}{z}\SplitPoly{z}{(p,\this,x_1,\ldots,x_n,k)}
        \TransB{b}{p}{k}\\ \quad
     \mbox{where $\methods(c)(\ell) = (B(A_1\:x_1,\ldots,A_n\:x_n),b)$}
     \end{prog}}
\else
\clause{\begin{prog}
I_{\mathit{class}}(c,\ell) \triangleq\\
\quad
     \begin{prog}
     \Repl\Rcv{c\_\ell}{z}\SplitPoly{z}{(p,\this,x_1,\ldots,x_n,k)}
        \TransB{b}{p}{k}\\ \quad
     \mbox{where $\methods(c)(\ell) = (B(A_1\:x_1,\ldots,A_n\:x_n),b)$}
     \end{prog}
\end{prog}}
\fi
\end{display}
\fi

\subsection{A Semantics for Web Services}\label{s:semantics-web-services}
We complete the semantics for our object calculus by translating our
cryptographic protocol for calling a web service to the spi-calculus.  A new
idea is that we embed begin- and end-events in the translation to
represent the abstract authenticity guarantees offered by the object calculus.

We assume access to all web methods is at the highest security level
\texttt{AuthEnc} from Section~\ref{sec:sec-abs}, providing both
authentication and secrecy.  Here is the protocol, for $p$ making a
web service call $w\mo{:}\ell(u_1,\ldots,u_n)$ to service $w$ owned by
$q$, including the names of continuation channels used at the spi
level.  Recall that the protocol assumes that the client has a way to
query the web service for a nonce. Therefore, we assume that in
addition to the methods of $\class(w)$, each web service also supports
a method $\getnonce$, which we implement specially.
\[\begin{array}{l}
p \rightarrow q ~\mbox{on}~ w: \request(\getnonce()),k_1\\
q \rightarrow p ~\mbox{on}~ k_1: \response(\getnonce(n_q))\\
p \rightarrow q ~\mbox{on}~ w: p, \sencr{\request(w,\ell(u_1,\ldots,u_n),t,n_q)}{K_{pq}},n_p,k_2\\
q \rightarrow p ~\mbox{on}~ k_2: q, \sencr{\response(w,\ell(r),t,n_p)}{K_{pq}}
\end{array}\]

We are assuming there is a shared key $K_{pq}$ for each pair of principals
$p,q \in \Prin$.  For the sake of brevity, we omit the formal description of
the type and effect system
\cite{GJ02:TypesAndEffectsForAsymmetricCryptographicProtocols} we rely on, but
see Appendix~\ref{app:spi} for a detailed overview.  Still, to give a
flavour, we can define the type of a shared key $K_{pq}$ as follows:

\begin{display}{Type of Key Shared Between Client $p$ and Server $q$:}
\clause{
  \begin{prog}
    \CSKey(p,q) \triangleq \\ \quad
    \begin{prog}
    \SharedKey {\Union( \\ \quad
      \begin{prog}
        \request (
        \begin{prog}
           w \ty \Un,
           a \ty \Un,
           t \ty \Un,\\
           n_q \ty \annotate{\public}{\Response{[\endd{\request(p,q,w,a,t)}]}}),
   \end{prog}\\
        \response (
        \begin{prog}
          w \ty \Un,
          r \ty \Un,
          t \ty \Un,\\
          n_p \ty \annotate{\public}{\Response{[\endd{\response(p,q,w,r,t)}]}}))}
        \end{prog}
      \end{prog}
    \end{prog}
   \end{prog}}
\end{display}

The type says we can use the key in two modes.  First, we may encrypt a
plaintext tagged $\request$ containing four components: a public name $w$ of a
service, an argument $a$ suitable for the service, a session tag $t$, and a
nonce $n_q$ proving that a begin-event labelled $\request(p,q,w,a,t)$ has
occurred, and therefore that an end-event with that label would be safe.
Second, we may encrypt a plaintext tagged $\response$ containing four
components: a service $w$, a result $r$ from that service, the session tag
$t$, and a nonce $n_p$ proving that a begin-event labelled
$\response(p,q,w,r,t)$ has occurred.

We translate a service call to the client-side of our cryptographic protocol
as follows.  We start by embedding a begin-event labelled
$\request(p,q,w,\ell(\qq{u_1},\ldots,\qq{u_n}),t)$ to record the details of
client $p$'s call to server $q=\owner(w)$.  We request a nonce $n_q$, and use
it to freshen the encrypted request, which we send with our own nonce $n_p$,
which the server uses to freshen its response.  If the response indeed
contains our nonce, we embed an end-event to record successful authentication.
For the sake of brevity, we rely on some standard shorthands for
pattern-matching.

\ifJournal
\else
\pagebreak
\fi

\ifJournal
\begin{display}{Translation of Web Method Call:}
\clause{\begin{prog}
          \TransB{w\mo{:}\ell(u_1,\ldots,u_n)}{p}{k} \triangleq {} \\ \quad
             \Res{k_1 \ty \Un, k_2 \ty \Un, t\ty\Un,
                  n_p \ty \annotate{\public}{\Challenge{[\,]}}} \\ \quad
             \Begin{\request(p,q,w,\ell(\qq{u_1},\ldots,\qq{u_n}),t)} \\ \quad
              \Snd{w}{(\request(\getnonce()),k_1)}; \\ \quad
              \Rcv{k_1}{\response(\getnonce(n_q \ty \Un))} \\ \quad
              \Cast{n_q}{n'_q \ty
                \annotate{\public}{\Response{[\endd{\request(p,q,w,\ell(\qq{u_1},\ldots,\qq{u_n}),t)}]}}} \\ \quad
              \Snd{w}{(p,\sencr{\request(w,\ell(\qq{u_1},\ldots,\qq{u_n}),t,n'_q)}{K_{pq}},n_p,k_2)}; \\ \quad
              \Rcv{k_2}{q'\ty\Un,\bdy \ty \Un}
              \Decrypt{\bdy}{\response(\mathit{plain})}{K_{pq}}\\ \quad
              \Match{\mathit{plain}}{w}{\mathit{rest} \ty
                (r \ty \mathit{Res}(w), t'\ty\Un,
                 \annotate{\public}{\Response{[\endd{\response(p,q,w,r,t')}]}})}\\ \quad
              \Split{\mathit{rest}}{r\ty\mathit{Res}(w)}{rest'\ty(t'\ty\Un,n_p'\ty\annotate{\public}{\Response{[\endd{\response(p,q,w,r,t')}]}})}\\ \quad
         \Match{\mathit{rest'}}{t}{n_p'\ty\annotate{\public}{\Response{[\endd{\response(p,q,w,r,t)}]}}}\\ \quad
              \CheckNonce{n_p}{n_p'}
              \End{\response(p,q,w,r,t)}
              \Case{r} \IsTag{\ell}{x}{\Snd{k}{x}}\\
             \mbox{where $q=\owner(w)$}
       \end{prog}}
\end{display}
\else
\begin{display}{Semantics of Web Method Call:}
\clause{\begin{prog}
          \TransB{w\mo{:}\ell(u_1,\ldots,u_n)}{p}{k} \triangleq {} \\ \quad
             \Res{k_1, k_2, t, n_p} \\ \quad
             \Begin{\request(p,q,w,\ell(\qq{u_1},\ldots,\qq{u_n}),t)} \\ \quad
             \Snd{w}{(\request(\getnonce()),k_1)}; \\ \quad
              \Rcv{k_1}{\response(\getnonce(n_q))} \\ \quad
              \Snd{w}{(p,\sencr{\request(w,\ell(\qq{u_1},\ldots,\qq{u_n}),t,n'_q)}{K_{pq}},n_p,k_2)}; \\ \quad
              \Rcv{k_2}{q,\bdy}\\ \quad
              \Decrypt{\bdy}{\response(\mathit{plain})}{K_{pq}}\\ \quad
              \Match{\mathit{plain}}{w}{\mathit{rest}} \\ \quad
         \Split{\mathit{rest}}{r}{rest'}\\ \quad
              \Match{\mathit{rest'}}{t}{n_p'}\\ \quad
         \CheckNonce{n_p}{n_p'}\\ \quad
              \End{\response(p,q,w,r,t)}\\ \quad
              \Case{r} \IsTag{\ell}{x}{\Snd{k}{x}} \\
             \mbox{where $q=\owner(w)$}
       \end{prog}}
\end{display}
\fi

Our server semantics relies on a shorthand notation defined below;
\iflong{\ifJournal\else the process\fi}
$\LetCall{x}{w}{p,\ell(u_1,\ldots,u_n)}P$ runs the method $\ell$
of the class $\class(w)$ implementing the service $w$, with arguments $u_1$,
\ldots, $u_n$, and with its $\mathit{CallerId}$ field set to $p$,
binds the result to $x$ and runs $P$.
\ifJournal
\begin{display}{Server-Side Invocation of a Web Method:}
\clause{
\begin{prog}
\LetCall{x}{w}{p, \mathit{args}}P \triangleq {} \\ \quad
   \begin{prog}
   \Res{k}\\
   (\Case{\mathit{args}} (\IsTag{\ell_i}{\xs_i}{}
          \Res{k'}
          (\Snd{c\_\ell_i}{(q,c(p),\xs_i,k')}
          \parpop \Rcv{k'}{r}\Snd{k}{\ell_i(r)})
          ) \For{i \in 1..n}\\
    \quad \parpop \Rcv{k}{x}P)
   \end{prog}\\
    \quad \begin{prog}
          \mbox{where $c=\class(w)$, $q=\owner(w)$,}\\
          \mbox{ and $\methods(c)=\ell_i \mapsto
           (B_i(\As_i,\xs_i),b_i) \For{i \in 1..n}$}
          \end{prog}
\end{prog}
}
\end{display}
\else
\begin{display}{Server-Side Invocation of a Web Method:}
\clause{
\begin{prog}
\LetCall{x}{w}{p, \mathit{args}}P \triangleq {} \\ \quad
   \begin{prog}
   \Res{k}\\
   \quad \Case{\mathit{args}} \\
   \quad (\begin{prog}
          \IsTag{\ell_i}{\xs_i}{} \\ \quad
          \Res{k'}
\ifLong
          (\Snd{c\_\ell_i}{(q,c(p),\xs_i,k')}
          \parpop \Rcv{k'}{r}\Snd{k}{\ell_i(r)})\\
\else
          (\begin{prog}
           \Snd{c\_\ell_i}{(q,c(p),\xs_i,k')} \parpop \\
           \Rcv{k'}{r}\Snd{k}{\ell_i(r)})
           \end{prog}\\
\fi
          ) \For{i \in 1..n}
          \end{prog} \\
    \quad \parpop \Rcv{k}{x}P
   \end{prog}\\
    \quad \begin{prog}
          \mbox{where $c=\class(w)$, $q=\owner(w)$,}\\
          \mbox{ and $\methods(c)=\ell_i \mapsto
           (B_i(\As_i,\xs_i),b_i) \For{i \in 1..n}$}
          \end{prog}
\end{prog}
}
\end{display}
\fi

Finally, we implement each service $w$ by a process
$I_{\mathit{ws}}(w)$.  We repeatedly listen for nonce requests, reply
with one, and then await a web service call freshened by the nonce.
If we find the nonce, it is safe to perform an end-event labelled
$\request(p,q,w,a,t)$, where $p$ is the caller, $q=\owner(w)$ is the
service owner, $a$ is the received method request, and $t$ is the
session tag.  We use the shorthand above to invoke $a$.  If $r$ is the
result, we perform a begin-event labelled $\response(p,q,w,r,t)$ to
record we are returning a result, and then send a response, freshened
with the nonce we received from the client.  In general, the notation
$\textstyle\prod_{i \in 1..n}P_i$ means $P_1 \parpop\cdots\parpop
P_n$.

\ifLong
 \ifJournal
 \pagebreak
 \else
 \pagebreak}
 \fi
\fi

\ifJournal
\begin{display}{Web Service Implementation:}
\clause{I_{\mathit{ws}}(w) \triangleq 
     \begin{prog}
     \Repl\Rcv{w}{\bdy \ty \Un,k_1 \ty \Un}\\
         \Case{\bdy} \IsTag{\request}{\getnonce()}{}\\
         \Res{n_q \ty \annotate{\public}{\Challenge{[\,]}}}\\
         \Snd{k_1}{(\response(\getnonce(n_q)))};\\
         \Rcv{w}{p'\ty\Un,\mathit{cipher}\ty\Un,n_p \ty \Un, k_2 \ty \Un}\\
         \textstyle\prod_{p \in \Prin}\SpiIf p=p' \SpiThen \\
         \Decrypt{\mathit{cipher}}{\request(\mathit{plain})}{K_{pq}}\\
         \Match{\mathit{plain}}{w}{\mathit{rest} \ty \\ \quad
                (a \ty \mathit{Req}(w),t\ty\Un,
                  \annotate{\public}{\Response{[\endd{\request(p,q,w,a,t)}]}})}\\
         \Split{\mathit{rest}}{a\ty\mathit{Req}(w)}{t\ty\Un,n_q'\ty \\ \quad
           \annotate{\public}{\Response{[\endd{\request(p,q,w,a,t)}]}}}\\
         \CheckNonce{n_q}{n_q'}
         \End{\request(p,q,w,a,t)}\\
         \LetCall{r \ty \mathit{Res}(w)}{w}{p,a}\\
         \Begin{\response(p,q,w,r,t)} \\
         \Cast{n_p}{n'_p \ty
            \annotate{\public}{\Response{[\endd{\response(p,q,w,r,t)}]}}}
         \Snd{k_2}{(q,\sencr{\response(w,r,t,n'_p)}{K_{pq}})}\\
     \mbox{where $q=\owner(w)$}
     \end{prog}}
\end{display}
\else
\begin{display}{Web Service Translation:}
\clause{I_{\mathit{ws}}(w) \triangleq 
     \begin{prog}
     \Repl\Rcv{w}{\bdy, k_1}\\ \quad
       \begin{prog}
         \Case{\bdy} \IsTag{\request}{\getnonce()}{}\\
         \Res{n_q}\\
         \Snd{k_1}{(\response(\getnonce(n_q)))};\\
         \Rcv{w}{p',\mathit{cipher},n_p, k_2}\\
         \textstyle\prod_{p \in \Prin}\SpiIf p=p' \SpiThen \\
         \Decrypt{\mathit{cipher}}{\request(\mathit{plain})}{K_{pq}}\\
         \Match{\mathit{plain}}{w}{\mathit{rest}} \\
    \Split{\mathit{rest}}{a}{t,n_q'}\\
         \CheckNonce{n_q}{n_q'}\\
    \End{\request(p,q,w,a,t)}\\
         \LetCall{r}{w}{p,a}\\
         \Begin{\response(p,q,w,r,t)} \\
         \Snd{k_2}{(q,\sencr{\response(w,r,t,n'_p)}{K_{pq}})}
       \end{prog} \\
     \mbox{where $q=\owner(w)$}
     \end{prog}}
\end{display}
\fi

This semantics is subject to more deadlocks than a realistic implementation,
since we do not have a single database of outstanding nonces.  Still, since we
are concerned only with safety properties, not liveness, it is not a problem
that our semantics is rather more nondeterministic than an actual
implementation.

\subsection{Security Properties of a Complete System}

\ifJournal

We define the process $\mathit{Sys}(b,p,k)$ to model a piece of code
$b$ being run by principal $p$ (with continuation $k$) in the context
of implementations of all the classes and web services in $\Class$ and
$\WebService$. The implementation of the classes and web services are
given as follows.
\begin{display}{Implementation of Classes and Web Services:}
\clause{\ClassMethods \triangleq
  \{(c,\ell) ~:~ c\in\Class, \ell\in\dom(\methods(c))\}}\\
\clause{
I_{\mathit{class}} \triangleq
     \IndexedPar{(c,\ell) \in \ClassMethods} I_{\mathit{class}}(c,\ell)}\\
\clause{
  I_{\mathit{ws}} \triangleq
     \IndexedPar{w\in\WebService} I_{\mathit{ws}}(w)}
\end{display}

The process $\mathit{Sys}(b,p,k)$ is defined with respect to an
environment that specifies the type of its free variables, such as the
names of the web services, principals, classes and methods, and keys. 
\begin{display}{Top-Level Environments:}
\clause{
E_{\mathit{class}} \triangleq
  (c\_\ell \ty \Un)\For{(c,\ell) \in \ClassMethods}}\\
\clause{E_{\mathit{keys}}\triangleq(K_{pq}\ty\CSKey(p,q))\For{p,q\in\Prin}}\\
\clause{E_{\mathit{ws}} \triangleq (w\ty\Un) \For{w \in \WebService}}\\
\clause{E_{\mathit{prin}} \triangleq
         p_1 \ty  \Princ, \ldots, p_n \ty  \Princ}
  {where $\Prin=\{p_1, \ldots, p_n\}$}\\
 \clause{E_0 \triangleq E_{\mathit{ws}}, E_{\mathit{prin}},
                      E_{\mathit{class}}, E_{\mathit{keys}}}
\end{display}

The process $\mathit{Sys}(b,p,k)$ is defined as follows:
\[
\mathit{Sys}(b,p,k) \triangleq \Res{E_{\mathit{class}}, E_{\mathit{keys}}}
    (I_{\mathit{class}} \parpop
     I_{\mathit{ws}} \parpop
     \Res{k \ty \Un} \qq{b}^p_k)
\]
We claim that the ways an opponent $O$ can interfere with the behaviour of
$\mathit{Sys}(b,p,k)$ correspond to the ways in which an actual opponent lurking
on a network could interfere with SOAP-level messages being routed between web
servers.  The names $c\_\ell$ of methods are hidden, so $O$ cannot interfere
with calls to local methods.  The keys $K_{pq}$ are also hidden, so $O$ cannot
decrypt or fake SOAP-level encryption.  On the other hand, the names $w$ on
which $\mathit{Sys}(b,p,k)$ sends and receives our model of SOAP envelopes are
public, and so $O$ is free to intercept, replay, or modify such envelopes.

Our main result is that an opponent cannot disrupt the authenticity properties
embedded in our translation.  The proof is by showing the translation
preserves types.
\begin{theorem}\label{t:robust-safety-translation}
If $\Judge{\emptyset}{b : B}$ and $p \in \Prin$ and $k \notin \dom(E_0)$ then the system
$\mathit{Sys}(b,p,k)$ is robustly safe.
\end{theorem}
\begin{proof} See Appendix~\ref{app:proof-one}. \hfill \end{proof}

\else
We define the following process $\mathit{Sys}(b,p)$ to model a piece of code
$b$ being run by principal $p$ in the context of implementations of all the
classes and web services in $\Class$ and $\WebService$.
\begin{eqnarray*}
\lefteqn{\mathit{Sys}(b,p)}\\
&\triangleq&
  \begin{prog}
  \Res{c\_\ell \For{c\in\Class, \ell\in\dom(\methods(c))}}
  \Res{K_{pq} \For{p,q \in \Prin}} \\ \quad
    (\begin{prog}
     \IndexedPar{c\in\Class, \ell\in\dom(\methods(c))} I_{\mathit{class}}(c,\ell) \parpop\\
     \IndexedPar{w\in\WebService} I_{\mathit{ws}}(w) \parpop\\
     \Res{k} \qq{b}^p_k)
     \end{prog}
  \end{prog}
\end{eqnarray*}

We claim that the ways an opponent $O$ can interfere with the behaviour of
$\mathit{Sys}(b,p)$ correspond to the ways in which an actual opponent lurking
on a network could interfere with SOAP-level messages being routed between web
servers.  The names $c\_\ell$ of methods are hidden, so $O$ cannot interfere
with calls to local methods.  The keys $K_{pq}$ are also hidden, so $O$ cannot
decrypt or fake SOAP-level encryption.  On the other hand, the names $w$ on
which $\mathit{Sys}(b,p)$ sends and receives our model of SOAP envelopes are
public, and so $O$ is free to intercept, replay, or modify such envelopes.

Our main result is that an opponent cannot disrupt the authenticity properties
embedded in our translation.  The proof is by showing the translation
preserves types.
\begin{theorem}\label{t:robust-safety-translation}
If body $b$ is well-typed and $p \in \Prin$ then $\mathit{Sys}(b,p)$ is robustly safe.
\end{theorem}

\fi %

\section{A SOAP-Level Implementation}
\label{sec:soap-level-implementation}

We have implemented the security abstraction introduced in
Section~\ref{sec:sec-abs} and formalized in
Sections~\ref{sec:formal-model} and \ref{sec:spi-calculus-semantics}
on top of the Microsoft Visual Studio .NET implementation of web services, as a
library that web service developers and clients can use. A web
service developer adds security attributes to the web methods of the
service. The developer also needs to provide a web method to supply a
nonce to the client. On the client side, the client writer is provided
with a modified proxy class that encapsulates the implementation of
the security abstraction and takes into account the security level of
the corresponding web service methods. Hence, from a client's point of
view, there is no fundamental difference between accessing a web
service with security annotations and one without. 

Consider an implementation of our running example of a banking
service. Here is what (an extract of) the class implementing the web
service looks like:
\ifJournal
\begin{verbatim}
 class BankingServiceClass : System.Web.Services.WebService
 {
   ...
   [WebMethod]
   public int RequestNonce () { ... }

   public DSHeader header;

   [WebMethod]
   [SecurityLevel(Level=SecLevel.Auth)]
   [SoapHeader("header", Direction=Direction.InOut,Required=true)]
   public int Balance (int account) { ... }
 }
\end{verbatim}
\else
\ifLong
\begin{verbatim}
 class BankingServiceClass :
         System.Web.Services.WebService
 {
   ...
   [WebMethod]
   public int RequestNonce () { ... }

   public DSHeader header;

   [WebMethod]
   [SecurityLevel(Level=SecLevel.Auth)]
   [SoapHeader(``header'',
               Direction=Direction.InOut,Required=true)]
   public int Balance (int account) { ... }
 }
\end{verbatim}
\else
\begin{verbatim}
 class BankingServiceClass : 
         System.Web.Services.WebService
 {
   ...
   [WebMethod]
   public int RequestNonce () { ... }

   public DSHeader header;

   [WebMethod]
   [SecurityLevel(Level=SecLevel.Auth)]
   [SoapHeader(``header'',
               Direction=Direction.InOut,
               Required=true)]
   public int Balance (int account) { ... }
 }
\end{verbatim}
\fi
\fi
This is the code we currently have, and it is close to the idealized
interface we gave in Section~\ref{sec:sec-abs}. The differences are
due to implementation restrictions imposed by the development
environment.
The extract
shows that the web service implements the \texttt{RequestNonce} method
required by the authentication protocol. The \texttt{Balance} method is
annotated as an authenticated method, and is also annotated to
indicate that the headers of the SOAP messages used during a call will
be available through the \texttt{header} field of the interface. (The
class \texttt{DSHeader} has fields corresponding to the headers of the
SOAP message.) As we shall see shortly, SOAP headers are used to carry
the authentication information. Specifically, the authenticated
identity of the caller is available in a web method through
\texttt{header.callerid}.

To implement the security abstraction on the web service side, we use a
feature of Visual Studio .NET called SOAP Extensions. Roughly speaking, a SOAP
Extension acts like a programmable ``filter''. It can be installed on either
(or both) of a client or a web service. It gets invoked on every incoming and
outgoing SOAP message, and can be used to examine and modify the content of
the message before forwarding it to its destination. In our case, the
extension will behave differently according to whether the message is incoming
or outgoing, and depending on the security level specified. For an outgoing
message, if the security level is \texttt{None}, the SOAP message is
unchanged. If the security level is \texttt{Auth}, messages are signed as
specified by the protocol: a cryptographic hash of the SOAP body and the
appropriate nonce is stored in a custom header of the messages. If the
security level is \texttt{AuthEnc}, messages are encrypted as specified by the
protocol, before being forwarded. For incoming messages, the messages are
checked and decrypted, if required. If the security level is \texttt{Auth},
the signature of the message checked.  If the security level is
\texttt{AuthEnc}, the message is decrypted before being forwarded. Our
implementation uses the SHA1 hash function for signatures, and the RC2
algorithm for symmetric encryption.

To implement the security abstraction on the client side, we provide the
client with a new proxy class.
The new proxy class provides methods \texttt{None}, \texttt{Auth}, and
\texttt{AuthEnc}, that are called by the proxy methods to initiate the
appropriate protocol. The method \texttt{None} simply sets up the
headers of the SOAP message to include the identity of the caller and
the callee. \texttt{Auth} and \texttt{AuthEnc} do the same, but also
make a call to the web service to get a nonce and add it (along with a
newly created nonce) to the headers. The actual signature and
encryption of the SOAP message is again performed using SOAP
Extensions, just as on the web service side.

Our implementation uses a custom SOAP header \texttt{DSHeader} to carry information such
as nonces, identities, and signatures. It provides the following elements:
\begin{center}
\begin{tabular}{ll}
\texttt{callerid} & identity of the client\\
\texttt{calleeid} & identity of the web service provider\\
\texttt{np} & client nonce\\
\texttt{nq} & web service nonce\\
\texttt{signature} & cryptographic signature of the message
\end{tabular}
\end{center}
Not all of those elements are meaningful for all messages. In addition
to these headers, in the cases where the message is encrypted, the
SOAP body is replaced by the encrypted
body. Appendix~\ref{app:soap-messages} gives actual SOAP messages
exchanged between the client and web service during an authenticated
call to \texttt{Balance}, and an authenticated and encrypted call to
\texttt{Statement}.

\ifJournal
\else
Our implementation is meant as a preliminary design of a C${}^\#$ abstraction for
secure RPC, a starting point to explore abstractions for more general security
policies. There are still issues that need to be addressed, even in a setting
as simple as the one presented in this paper. First, we plan to adopt
recognized formats for encryption and signature of XML data, such as
XML-Encryption and XML-Signature (though our validation does not depend on the
exact XML syntax for cryptography).  Second, it would be valuable to generate
the new proxy class automatically.
\fi

\ifJournal
\section{A Semantics Using Asymmetric Cryptography}
\label{sec:asymmetric}
The security abstraction we describe in Section~\ref{sec:sec-abs}
relies on shared keys between principals. This is hardly a reasonable
setup in modern systems. In this section, we show that our approach
can easily accommodate public-key infrastructures.

\subsection{Authenticated Web Methods} We start by describing the
protocol and implementation for authenticated web methods. Hence, for
now, we assume that all the exported methods of a web service are
annotated with \texttt{Auth}.

Consider a simple public-key infrastructure for digital
signatures. Each principal $p$ has a signing key $\SK{p}$ and a
verification key $\VK{p}$. The signing key is kept private, while the
verification key is public. To bind the name of a principal with their
verification key, we assume a \emph{certification authority} $\CA$
(itself with a signing key $\SK{CA}$ and verification key $\VK{CA}$)
that can sign certificates $\CertVK{p}$ of the form
$\asencr{p,\VK{p}}{\SK{CA}}$. (The notation $\asencr{\cdot}{K}$
is used to represent both asymmetric encryption and signature, 
differentiating it from symmetric encryption. In the case where
$\asencr{M}{K}$ represent a signature, this is simply notation for $M$
along with a token representing the signature of $M$ with asymmetric
key $K$.)

Here is a protocol that uses digital signatures to authenticate
messages, for $p$ making a web service call
$w\mo{:}\ell(u_1,\ldots,u_n)$ to service $w$ owned by $q$, including
the names of continuation channels used at the spi level.  Again, we
assume that in addition to the methods of $\class(w)$, each web
service also supports a method $\getnonce$, which we implement
specially. 
\[\begin{array}{l}
p \rightarrow q ~\mbox{on}~ w: \CertVK{p}, n_p, \request(\getnonce()),k_1\\
q \rightarrow p ~\mbox{on}~ k_1: \CertVK{q}, \response(\getnonce(n_q))\\
p \rightarrow q ~\mbox{on}~ w: p, \asencr{\request(w,\ell(u_1,\ldots,u_n),t,q,n_q)}{\SK{p}},k_2\\
q \rightarrow p ~\mbox{on}~ k_2: q, \asencr{\response(w,\ell(r),t,p,n_p)}{\SK{q}}
\end{array}\]

\begin{display}{Type of Signing Keys:}
\clause{
  \begin{prog}
    \AuthMsg{p} \triangleq \\ \quad
    \begin{prog}
    \Union (\begin{prog}
                  \request (\begin{prog}
                            w \ty \Un,
                            a \ty \Un,
                            t \ty \Un,
                            q \ty \Un,
                            n_q \ty \annotate{\public}{\Response{[\endd{\request(p,q,w,a,t)}]}}),
             \end{prog}\\
                  \response (\begin{prog}
                             w \ty \Un,
                             r \ty \Un,
                             t \ty \Un,
                             q \ty \Un,
                             n_q \ty \annotate{\public}{\Response{[\endd{\response(q,p,w,r,t)}]}}))   
              \end{prog}
             \end{prog}
    \end{prog}
  \end{prog}}    \\
\clause{\AuthKeys{p} \triangleq \KeyPair{\AuthMsg{p}}}\\
\clause{\AuthCert \triangleq (p:\Un, \annotate{\Dec}{\Key{\AuthMsg{p}}})}\\
\clause{\AuthCertKeys \triangleq \KeyPair{\AuthCert}}
\end{display}

We will represent the key pair of a signing key and verification key
for principal $p$ by a pair $\DS{p}$, of type $\AuthKeys{p}$. The key
pair for the certification authority will be represented by a pair
$\DS{CA}$. We use the following abbreviations: 
\begin{display}{Key and Certificates Abbreviations:}
\clause{\SK{p} \triangleq \extract{\Enc}{\DS{p}}}{$p$'s signing key}\\
\clause{\VK{p} \triangleq \extract{\Dec}{\DS{p}}}{$p$'s verification key}\\
\clause{\CertVK{p} \triangleq \asencr{p,\VK{p}}{\SK{CA}}}{$p$'s certificate}
\end{display}

With that in mind, we can amend the translation of
Section~\ref{sec:spi-calculus-semantics} to accommodate the new
protocol. First, we give a new translation for a web method call
$w\mo{:}\ell(u_1,\ldots,u_n)$:
\begin{display}{New Translation of Web Method Call:}
\clause{\begin{prog}
          \TransB{w\mo{:}\ell(u_1,\ldots,u_n)}{p}{k} \triangleq {} \\ \quad
             \Res{k_1 \ty \Un, k_2 \ty \Un, t\ty\Un,
                  n_p \ty \annotate{\public}{\Challenge{[\,]}}} \\ \quad
             \Begin{\request(p,q,w,\ell(\qq{u_1},\ldots,\qq{u_n}),t)} \\ \quad
              \Snd{w}{(\CertVK{p},n_p,\request(\getnonce()),k_1)}; \\ \quad
              \Rcv{k_1}{c\ty\Un,\response(\getnonce(n_q \ty \Un))} \\ \quad
              \pDecrypt{c}{\mathit{cert}\ty(q'\ty\Un,\annotate{\Dec}{\Key{\AuthMsg{q'}}})}{\VK{CA}}\\ \quad
              \Match{\mathit{cert}}{q}{\mathit{vkq}\ty\annotate{\Dec}{\Key{\AuthMsg{q}}}}\\ \quad
              \Cast{n_q}{n'_q \ty
                \annotate{\public}{\Response{[\endd{\request(p,q,w,\ell(\qq{u_1},\ldots,\qq{u_n}),t)}]}}} \\ \quad
              \Snd{w}{(p,\asencr{\request(w,\ell(\qq{u_1},\ldots,\qq{u_n}),t,q,n'_q)}{\SK{p}},k_2)}; \\ \quad
              \Rcv{k_2}{q''\ty\Un,\bdy \ty \Un}\\ \quad
              \begin{prog}
                \pDecrypt{\bdy}{\response(\mathit{plain}\ty(w'\ty\Un,r\ty\Un,t'\ty\Un,p'\ty\Un,\\
                        \qquad\qquad\quad\annotate{\public}{\Response{[\endd{\response(p',q,w',r,t')}]}}))}{\mathit{vkq}}
         \end{prog}\\ \quad
              \Match{\mathit{plain}}{w}{\mathit{rest} \ty
                (\begin{prog}
                 r \ty \mathit{Res}(w), t'\ty\Un, p'\ty\Un,\\
                 \annotate{\public}{\Response{[\endd{\response(p',q,w,r,t')}]}})}
                 \end{prog}\\ \quad
         \Split{\mathit{rest}}{r\ty\mathit{Res}(w)}{\mathit{rest'}\ty
                (\begin{prog}
                 t'\ty\Un,p'\ty\Un,\\
                 \annotate{\public}{\Response{[\endd{\response(p',q,w,r,t')}]}})}
       \end{prog}\\ \quad
              \Match{\mathit{rest'}}{t}{\mathit{rest''}\ty(p'\ty\Un,\annotate{\public}{\Response{[\endd{\response(p',q,w,r,t)}]}})}\\ \quad
              \Match{\mathit{rest''}}{p}{n_p'\ty\annotate{\public}{\Response{[\endd{\response(p,q,w,r,t)}]}}}\\ \quad
         \CheckNonce{n_p}{n_p'}\\ \quad
              \End{\response(p,q,w,r,t)}\\ \quad
              \Case{r} \IsTag{\ell}{x}{\Snd{k}{x}}\\
             \mbox{where $q=\owner(w)$}
       \end{prog}}
\end{display}

We also need to give a new implementation for web services, again to
take into account the different messages being exchanged:
\begin{display}{New Web Service Implementation:}
\clause{\begin{prog}
I_{\mathit{ws}}(w) \triangleq \\
   \quad  \begin{prog}
     \Repl\Rcv{w}{c\ty\Un, n_p\ty\Un,\bdy \ty \Un,k_1 \ty \Un}\\ %

         \Case{\bdy} \IsTag{\request}{\getnonce()}{}\\
         \pDecrypt{c}{p\ty\Un,\mathit{vkp}\ty\annotate{\Dec}{\Key{\AuthMsg{p}}}}{\VK{CA}}\\
         \Res{n_q \ty \annotate{\public}{\Challenge{[\,]}}}\\
         \Snd{k_1}{(\CertVK{q},\response(\getnonce(n_q)))};\\
         \Rcv{w}{p'\ty\Un,\mathit{cipher}\ty\Un,k_2 \ty \Un}\\
         \SpiIf p=p' \SpiThen \\
         \begin{prog}
             \pDecrypt{\mathit{cipher}}{\request(\mathit{plain}\ty(w\ty\Un,a\ty\Un,t\ty\Un,q'\ty\Un,\\
                   \qquad\qquad\quad\annotate{\public}{\Response{[\endd{\request(p,q',w,a,t)}]}}))}{\mathit{vkp}}
         \end{prog}\\
         \Match{\mathit{plain}}{w}{\mathit{rest} \ty
                (\begin{prog}
                 a \ty \mathit{Req}(w),t\ty\Un,q'\ty\Un,\\
                  \annotate{\public}{\Response{[\endd{\request(p,q',w,a,t)}]}})}
                 \end{prog}\\
\end{prog}\end{prog}} \\ \clause{\begin{prog}\quad\begin{prog}
    \Split{\mathit{rest}}{\begin{prog}a\ty\mathit{Req}(w)}{\\t\ty\Un,\mathit{rest'}\ty(q'\ty\Un,\annotate{\public}{\Response{[\endd{\request(p,q',w,a,t)}]}})}\end{prog}\\
         \Match{\mathit{rest'}}{q}{n_q'\ty\annotate{\public}{\Response{[\endd{\request(p,q,w,a,t)}]}}}\\
         \CheckNonce{n_q}{n_q'}\\
    \End{\request(p,q,w,a,t)}
\end{prog}\end{prog}$

\\

$\begin{prog}\quad\begin{prog}
         \LetCall{r \ty \mathit{Res}(w)}{w}{p,a}\\
         \Begin{\response(p,q,w,r,t)} \\
         \Cast{n_p}{n'_p \ty
            \annotate{\public}{\Response{[\endd{\response(p,q,w,r,t)}]}}}\\
         \Snd{k_2}{(q,\asencr{\response(w,r,t,p,n'_p)}{\SK{q}})}

     \end{prog}\\
     \mbox{where $q=\owner(w)$}
   \end{prog}}
\end{display}

Finally, we need to change the top-level environment to account for
the new keys, and to add a channel through which we will publish the public keys. 
\begin{display}{Top-Level Environments:}
\clause{
E_{\mathit{class}} \triangleq
  (c\_\ell \ty \Un)\For{(c,\ell) \in \ClassMethods}}\\
\clause{E_{\mathit{keys}}\triangleq \DS{CA}\ty\AuthCertKeys,(\DS{p}\ty\AuthKeys{p})\For{p\in\Prin}}\\
\clause{E_{\mathit{ws}} \triangleq (w\ty\Un) \For{w \in \WebService}}\\
\clause{E_{\mathit{prin}} \triangleq
         p_1 \ty  \Princ, \ldots, p_n \ty  \Princ}
  {where $\Prin=\{p_1, \ldots, p_n\}$}\\
\clause{E_{\mathit{net}} \triangleq \mathit{net}\ty\Un}\\
\clause{E_0 \triangleq E_{\mathit{ws}}, E_{\mathit{prin}},E_{\mathit{net}},
                       E_{\mathit{class}}, E_{\mathit{keys}}}
\end{display}

Publishing can be achieved by simply sending the public keys on a
public channel, here $\mathit{net}$:
\begin{display}{Public Keys Publishing:}
\clause{
I_{\mathit{net}} \triangleq \Snd{\mathit{net}}{(\VK{CA},(\VK{p})\For{p\in\Prin})}}
\end{display}

We can now establish that the resulting system is robustly safe: 
\begin{theorem}\label{t:robust-safety-asymm-one}
If $\Judge{\emptyset}{a : A}$ and $p \in \Prin$ and $k \notin \dom(E_0)$ then the system
\[\Res{E_{\mathit{class}}, E_{\mathit{keys}}}
    (I_{\mathit{net}} \parpop I_{\mathit{class}} \parpop
     I_{\mathit{ws}} \parpop
     \Res{k \ty \Un} \qq{a}^p_k)\]
is robustly safe.
\end{theorem}
\begin{proof} See Appendix~\ref{app:proof-two}. \hfill \end{proof}

The protocol we give above to provide authentication has some
undesirable properties. Specifically, it requires the server to
remember the certificate $\CertVK{p}$ and nonce $n_p$ at the time
when a nonce is requested. Since anyone can request a nonce, and no
authentication is performed at that stage of the protocol, this makes
the server severely vulnerable to denial-of-service attacks.  The
following variation on the protocol achieves the same guarantees, but 
pushes the exchange of certificates and nonces to later messages,
basically just when they are needed.
\[\begin{array}{l}
p \rightarrow q ~\mbox{on}~ w: \request(\getnonce()),k_1\\
q \rightarrow p ~\mbox{on}~ k_1: \response(\getnonce(n_q))\\
p \rightarrow q ~\mbox{on}~ w: p, \CertVK{p}, n_p, \asencr{\request(w,\ell(u_1,\ldots,u_n),t,q,n_q)}{\SK{p}},k_2\\
q \rightarrow p ~\mbox{on}~ k_2: q, \CertVK{q}, \asencr{\response(w,\ell(r),t,p,n_p)}{\SK{q}}
\end{array}\]

\subsection{Authenticated and Encrypted Web Methods} We now describe a protocol and 
implementation for authenticated and encrypted web methods. Hence, for
now, we assume that all the exported methods of a web service are
annotated with \texttt{AuthEnc}.

The public-key infrastructure we consider for this case is similar
to the one for authenticated web methods, except that now we have
encryption and decryption keys, as opposed to signing and verification
keys. Each principal $p$ has an encryption key $\EK{p}$ and a
decryption key $\DK{p}$. The decryption key is kept private, while the
encryption key is public. To bind the name of a principal with their
encryption key, we again assume a \emph{certification authority} $\CA$
(with a signing key $\SK{CA}$ and verification key $\VK{CA}$)
that can sign certificates $\CertEK{p}$ of the form
$\asencr{p,\EK{p}}{\SK{CA}}$.

Here is a protocol for $p$ making a web service call
$w\mo{:}\ell(u_1,\ldots,u_n)$ to service $w$ owned by $q$, including
the names of continuation channels used at the spi level.  Again, we
assume that in addition to the methods of $\class(w)$, each web
service also supports a method $\getnonce$, which we implement
specially. 
\[\begin{array}{l}
p \rightarrow q ~\mbox{on}~ w: \CertEK{p}, \request(\getnonce()),k_1\\
q \rightarrow p ~\mbox{on}~ k_1: \CertEK{q}, \asencr{\msgI(q,n_K)}{\EK{p}}, \response(\getnonce(n_q))\\
p \rightarrow q ~\mbox{on}~ w: \asencr{\msgII(w,p,K,n_K)}{\EK{q}},n_p, \sencr{\request(\ell(u_1,\ldots,u_n),t,n_q)}{K},k_2\\
q \rightarrow p ~\mbox{on}~ k_2: \sencr{\response(\ell(r),t,n_p)}{K}
\end{array}\]
This protocol is similar to that for authenticated web methods, except
that public key encryption is used to exchange a session-specific
shared key $K$ used to encrypt the actual method call. Specifically,
in the third message, $p$ chooses a session-specific shared key $K$,
and sends it to $q$ encrypted with $q$'s public key $\EK{q}$; this
session key $K$ is used to encrypt the web method call. The result of
the web method call is also encrypted with this shared key. To prevent
replay attacks, the shared key is bound to a nonce $n_K$ sent by $q$
in the second message.

\begin{display}{Type of Keys:}
\clause{
  \begin{prog}
    \SKey{p,q,w} \triangleq \\ \quad
      \SharedKey{\Union(\begin{prog}
        \request(a \ty \Un, 
                 t \ty \Un,
                 n_q \ty \annotate{\public}{\Response{[\endd{\request(p,q,w,a,t)}]}}),\\
        \response(r \ty \Un, 
                  t \ty \Un,
                  n_p \ty \annotate{\public}{\Response{[\endd{\response(p,q,w,r,t)}]}}))}
         \end{prog}
  \end{prog}}\\
\clause{
  \begin{prog}
    \AuthEncMsg{p} \triangleq \\ \quad
    \begin{prog}
    \Union (\begin{prog}
                  \msgI (\begin{prog}
                         q \ty \Un,
                         n_K \ty \annotate{\private}{\Challenge{[\,]}}),
                    \end{prog}\\
                  \msgII (\begin{prog}
                             w \ty \Un,
                             q \ty \Un,
                             K \ty \Top,\\
                             n_K \ty \annotate{\private}{\Response{[\trust{K\ty\SKey{p,q,w}}]}}))
              \end{prog}
             \end{prog}
    \end{prog}
  \end{prog}}    \\
\clause{\AuthEncKeys{p} \triangleq \KeyPair{\AuthEncMsg{p}}}\\
\clause{\AuthEncCert \triangleq (p\ty\Un, \annotate{\Enc}{\Key{\AuthEncMsg{p}}})}\\
\clause{\AuthEncCertKeys \triangleq \KeyPair{\AuthEncCert}}
\end{display}

We will represent the key pair of an encryption key and decryption key
for principal $p$ by a pair $\PK{p}$, of type $\AuthEncKeys{p}$. The
signing key pair for the certification authority will be represented
by a pair $\DS{CA}$. We use the following abbreviations:
\begin{display}{Key and Certificates Abbreviations:}
\clause{\EK{p} \triangleq \extract{\Enc}{\PK{p}}}{$p$'s encryption key}\\
\clause{\DK{p} \triangleq \extract{\Dec}{\PK{p}}}{$p$'s decryption key}\\
\clause{\CertEK{p} \triangleq \asencr{p,\EK{p}}{\SK{CA}}}{$p$'s certificate}
\end{display}

Again, we can amend the translation of
Section~\ref{sec:spi-calculus-semantics} to accommodate the new
protocol. First, we give a new translation for a web method call
$w\mo{:}\ell(u_1,\ldots,u_n)$: 
\begin{display}{New Translation of Web Method Call:}
\clause{\begin{prog}
          \TransB{w\mo{:}\ell(u_1,\ldots,u_n)}{p}{k} \triangleq {} \\ \quad
             \Res{k_1 \ty \Un, k_2 \ty \Un, t\ty\Un,
                  n_p \ty \annotate{\public}{\Challenge{[\,]}}} \\ \quad
             \Begin{\request(p,q,w,\ell(\qq{u_1},\ldots,\qq{u_n}),t)} \\ \quad
              \Snd{w}{(\CertEK{p},\request(\getnonce()),k_1)}; \\ \quad
              \Rcv{k_1}{c\ty\Un,\mathit{cipher}\ty\Un,\response(\getnonce(n_q \ty \Un))}

         \end{prog}}\\
\clause{\begin{prog}\quad

              \pDecrypt{c}{\mathit{cert}\ty(q'\ty\Un,\annotate{\Enc}{\Key{\AuthEncMsg{q'}}})}{\VK{CA}}\\ \quad
              \Match{\mathit{cert}}{q}{\mathit{ekq}\ty\annotate{\Enc}{\Key{\AuthEncMsg{q}}}}\\ \quad
              \pDecrypt{\mathit{cipher}}{\msgI(q'\ty\Un,n_K\ty\Un)}{\DK{p}}\\ \quad
              \SpiIf q=q' \SpiThen\\ \quad
              \Cast{n_q}{n'_q \ty
                \annotate{\public}{\Response{[\endd{\request(p,q,w,\ell(\qq{u_1},\ldots,\qq{u_n}),t)}]}}} \\ \quad
              \Res{K \ty \SKey{p,q,w}}\\ \quad
              \Witness{K\ty\SKey{p,q,w}}\\ \quad
              \Cast{n_K}{n'_K \ty
                \annotate{\private}{\Response{[\trust{K\ty\SKey{p,q,w}}]}}}\\ \quad
              \Snd{w}{(\asencr{\msgII(w,p,t,K,n_K')}{\mathit{ekq}},n_p,
                       \sencr{\request(w,\ell(\qq{u_1},\ldots,\qq{u_n}),t,n'_q)}{K},k_2)}; \\ \quad
              \Rcv{k_2}{\bdy \ty \Un}\\ \quad
              \begin{prog}
                 \Decrypt{\bdy}{\response(\mathit{plain}\ty(r\ty\mathit{Res}(w),t'\ty\Un,\\
                   \qquad\qquad\quad\annotate{\public}{\Response{[\endd{\response(p,q,w,r,t')}]}}))}{K}
         \end{prog}\\ \quad
              \Match{\mathit{plain}}{r\ty\mathit{Res}(w)}
                    {\mathit{rest}\ty(t'\ty\Un,\annotate{\public}{\Response{[\endd{\response(p,q,w,r,t')}]}})}\\ \quad
              \Match{\mathit{rest}}{t}{n_p'\ty\annotate{\public}{\Response{[\endd{\response(p,q,w,r,t)}]}}}\\ \quad
         \CheckNonce{n_p}{n_p'}\\ \quad
              \End{\response(p,q,w,r,t)}\\ \quad
              \Case{r} \IsTag{\ell}{x}{\Snd{k}{x}}\\
             \mbox{where $q=\owner(w)$}
       \end{prog}}
\end{display}

We also need to give a new implementation for web services, again to
take into account the different messages being exchanged:
\begin{display}{New Web Service Implementation:}
\clause{\begin{prog}
I_{\mathit{ws}}(w) \triangleq \\ \quad
     \begin{prog}
     \Repl\Rcv{w}{c\ty\Un,\bdy \ty \Un,k_1 \ty \Un}\\
         \Case{\bdy} \IsTag{\request}{\getnonce()}{}\\
         \pDecrypt{c}{p\ty\Un,\mathit{ekp}\ty\annotate{\Enc}{\Key{\AuthEncMsg{p}}}}{\VK{CA}}\\
         \Res{n_q \ty \annotate{\public}{\Challenge{[\,]}}}\\
         \Res{n_K \ty \annotate{\private}{\Challenge{[\,]}}}\\
         \Snd{k_1}{(\CertEK{q},\asencr{\msgI(q,n_K)}{\mathit{ekp}},\response(\getnonce(n_q)))};\\
         \Rcv{w}{\mathit{cipher}_1\ty\Un,n_p \ty \Un, \mathit{cipher}_2\ty\Un,k_2 \ty \Un}\\
         \begin{prog}
           \pDecrypt{\mathit{cipher}_1\\\qquad}{\msgII(\mathit{plain}_1\ty(w\ty\Un,p'\ty\Un,K\ty\Top,\\
              \qquad\qquad\quad\annotate{\private}{\Response{[\trust{K\ty\SKey{p',q,w}}]}}))}{\DK{q}}
         \end{prog}\\
         \Match{\mathit{plain}_1}{w}{\mathit{rest}\ty
           (\begin{prog}
             p'\ty\Un,K\ty\Top,\\
             \annotate{\private}{\Response{[\trust{K\ty\SKey{p',q,w}}]}})}
            \end{prog}\\
         \Match{\mathit{rest}}{p}{\mathit{rest'}\ty(K\ty\Top,
                   \annotate{\private}{\Response{[\trust{K\ty\SKey{p,q,w}}]}})}\\
         \Split{\mathit{rest'}}{K\ty\Top}{n'_K\ty\annotate{\private}{\Response{[\trust{K\ty\SKey{p,q,w}}]}}}\\
         \CheckNonce{n_K}{n_K'}\\
         \Trust{K}{K'\ty\SKey{p,q,w}}\\
         \begin{prog}
            \Decrypt{\mathit{cipher}_2}{\request(\mathit{plain}_2\ty(a\ty\mathit{Req}(w),t\ty\Un,\\
            \qquad\qquad\quad\annotate{\public}{\Response{[\endd{\request(p,q,w,a,t)}]}}))}{K'}
         \end{prog}\\
    \Split{\mathit{plain}_2}{a\ty\mathit{Req}(w)}{t\ty\Un,n_q'\ty\annotate{\public}{\Response{[\endd{\request(p,q,w,a,t)}]}}}\\
         \CheckNonce{n_q}{n_q'}\\
    \End{\request(p,q,w,a,t)}\\

\end{prog}\end{prog}}\\
\clause{\begin{prog}\quad\begin{prog}

         \LetCall{r \ty \mathit{Res}(w)}{w}{p,a}\\
         \Begin{\response(p,q,w,r,t)} \\
         \Cast{n_p}{n'_p \ty
            \annotate{\public}{\Response{[\endd{\response(p,q,w,r,t)}]}}}\\
         \Snd{k_2}{\sencr{\response(r,t,n'_p)}{K'}}
     \end{prog}\\
     \mbox{where $q=\owner(w)$}
     \end{prog}}
\end{display}

Finally, we need to change the top-level environment to account for
the new keys, and to add a channel through which we will publish the public keys. 
\ifJournal
\clearpage
\fi
\begin{display}{Top-Level Environments:}
\clause{
E_{\mathit{class}} \triangleq
  (c\_\ell \ty \Un)\For{(c,\ell) \in \ClassMethods}}\\
\clause{E_{\mathit{keys}}\triangleq \DS{CA}\ty\AuthEncCertKeys,(\PK{p}\ty\AuthEncKeys{p})\For{p\in\Prin}}\\
\clause{E_{\mathit{ws}} \triangleq (w\ty\Un) \For{w \in \WebService}}\\
\clause{E_{\mathit{prin}} \triangleq
         p_1 \ty  \Princ, \ldots, p_n \ty  \Princ}
  {where $\Prin=\{p_1, \ldots, p_n\}$}\\
\clause{E_{\mathit{net}} \triangleq \mathit{net}\ty\Un}\\
\clause{E_0 \triangleq E_{\mathit{ws}}, E_{\mathit{prin}},E_{\mathit{net}},
                       E_{\mathit{class}}, E_{\mathit{keys}}}
\end{display}

Publishing can be achieved by simply sending the public keys on a
public channel, here $\mathit{net}$:
\begin{display}{Public Keys Publishing:}
\clause{
I_{\mathit{net}} \triangleq \Snd{\mathit{net}}{(\VK{CA},(\EK{p})\For{p\in\Prin})}}
\end{display}

We can now establish that the resulting system is robustly safe: 
\begin{theorem}\label{t:robust-safety-asymm-two}
If $\Judge{\emptyset}{a : A}$ and $p \in \Prin$ and $k \notin \dom(E_0)$ then the system
\[\Res{E_{\mathit{class}}, E_{\mathit{keys}}}
    (I_{\mathit{net}} \parpop I_{\mathit{class}} \parpop
     I_{\mathit{ws}} \parpop
     \Res{k \ty \Un} \qq{a}^p_k)\]
is robustly safe.
\end{theorem}
\begin{proof} See Appendix~\ref{app:proof-three}. \hfill \end{proof}

We can note some further possibilities, with respect to the protocols
implemented in this section:
\begin{itemize}
\item
The protocol implementing authenticated and encrypted invocation uses
certificates to essentially negotiate a symmetric key with which to
actually perform the encryption. It is straightforward to apply the
same idea to the authenticated-only case, negotiating a symmetric key
with which to hash the content of the method call (instead of relying
on public-key signatures).
\item
In the above protocol, a new symmetric key is negotiated at every
method invocation. A more efficient variation would be to re-use a
negotiated symmetric key over multiple web method calls. Once a
symmetric key has been negotiated, it can effectively act as a shared
key between the two principals, which is the case we investigated in
the body of this paper. We can therefore use the above protocol for
the first web method call between a principal and a particular
service, and the shared-key protocol for subsequent web method calls. 
\end{itemize}
\fi

\section{Related Work}
\label{sec:related-work}

There has been work for almost twenty years on secure RPC
mechanisms, going back to Birrell \cite{Birrell85}. More recently,
secure RPC has been studied in the context of distributed object
systems. As we mentioned, our work was inspired by the work of van
Doorn \emph{et al.}\ \cite{vanDoorn96}, itself inspired by
\cite{LABW92:AuthenticationDistributedSystemsTheoryPractice,WABL94:AuthenticationTaOS}. These
techniques (or similar ones) have been applied to CORBA \cite{Lang02},
DCOM \cite{Box97}, and Java \cite{Balfanz00,Eronen01}.

In contrast, little work seems to have been done on formalizing secure
RPC. Of note is the work of Abadi, Fournet, and Gonthier
\cite{AFG98:SecureImplementationOfChannelAbstractions,AFG99:SecureCommunicationsProcessingForDistributedLanguages},
who show how to compile the standard join-calculus into the
sjoin-calculus, and show that the compilation is fully abstract. In a
subsequent paper \cite{AFG00:AuthPrimsCompilation}, they treat
similarly and more simply a join-calculus with authentication
primitives: each message contains its source address, there is a way
to extract the principal owning a channel from the channel, and any
piece of code runs as a particular principal. Their fully abstract
translation gives very strong guarantees: it shows that for all
intents and purposes, we can reason at the highest level (at the level
of the authentication calculus).  
Although our guarantees are weaker, they are easier to establish. 

Duggan \cite{Duggan02:CryptographicTypes} formalizes an application-level
security abstraction by introducing types for signed and encrypted messages;
he presents a fully abstract semantics for the abstraction by translation to a
spi-calculus.

Much of the literature on security in distributed systems studies the question
of \emph{access control}. Intuitively, access control is the process of
determining if the principal calling a particular method has permission to
access the objects that the method refers to, according to a particular access
control policy. There is a distinction to be made between authentication and
access control. Authentication determines whether the principal calling a
method is indeed the principal claiming to be calling the method, while access
control can use this authenticated identity to determine whether that
principal is allowed access. This distinction is made clear in the work of
Balfanz \emph{et al.}~\cite{Balfanz00}, where they provide authenticated and
encrypted communication over Java RMI (using SSL) and use that infrastructure
as a basis for a logic-based access control mechanism. The access control
decisions are based on the authenticated caller identity obtained from the
layer in charge of authentication. This approach is also possible in our
framework, which provides access to an authenticated identity as well. We plan
to study access control abstractions in our framework.  Various forms
of access control mechanisms have been formalized via $\pi$-calculi,
\cite{riely98hlcl,sewell98a,HSY01:ATypedProcessCalculusForFineGrainedResourceACinDC},
and other process calculi \cite{CG00:MobileAmbients,DeNicola99}. An access
control language based on temporal logic has been defined by Sirer and Wang
\cite{Sirer02} specifically for web services.
Damiani \emph{et al.}~\cite{ddps.ijis} describe an implementation of an access
control model for SOAP; unlike our work, and the WS-Security
proposal\ifJournal\else~\cite{ws-security}\fi, it relies on an underlying secure channel, such
as an SSL connection.

\ifJournal
\else
The GRID is a proposed distributed infrastructure with scientific computing as
an important application; consequently, the need arises for a distributed
security architecture~\cite{FKTT88:SecArchForCompGrids} including
authentication and access control.
\fi

\ifJournal
Since this work was completed, a series of specifications for web
services security has been published, as laid out in a whitepaper from IBM and
Microsoft~\cite{security-white-paper}.
In particular, WS-Security~\cite{ws-security} defines how to add signatures, to apply encryption,
and to add principal identities, such as usernames or certificates, to a SOAP
envelope.
It would be straightforward, for example, to adapt our implementation to
produce WS-Security compliant SOAP envelopes.
A recent paper shows how to formalize the authentication goals of protocols
based on WS-Security using the applied
$\pi$-calculus~\cite{BFG04:SemanticsWebServicesAuthentication}.
\else
An intense area of activity in the world of web services is the
definition of standards for web service security. WS-Security is a
standard that describes how to attach signature and encryption headers
to SOAP envelopes. Envisioned standards, described in
\cite{security-white-paper}, will build on the specifications of
WS-Security, for example, to manage and authenticate message exchanges
between participants. Our work has an immediate application in this
context. It is straightforward, for example, to adapt our
implementation to produce WS-Security compliant SOAP envelopes. More
importantly, we can use the techniques in this paper to model security
abstractions provided by emerging standards and study them formally.
\fi

Despite its enjoyable properties, the formal model we use to study
the implementation of our security abstraction suffers from some
limitations. For instance, it makes the usual Dolev-Yao assumptions
that the adversary can compose messages, replay them, or decipher them
if it knows the right key, but cannot otherwise ``crack'' encrypted
messages. A more severe restriction is that we cannot yet model
insider attacks: principals with shared keys are assumed
well-behaved. Work is in progress to extend the Cryptyc type theory to
account for malicious insiders. We have not verified the hash-based
protocol of Section~\ref{sec:sec-abs}.

\section{Conclusions}
\label{sec:conc}

Authenticated method calls offer a convenient abstraction for
developers of both client and server code.  Various authorisation
mechanisms may be layered on top of this abstraction.  This paper
proposes such an abstraction for web services, presents a theoretical
model, and describes an implementation using SOAP-level security.  By
typing our formal semantics, we show no vulnerability exists to
attacks representable within the spi-calculus, given certain
assumptions.  Vulnerabilities may exist outside our model---there are
no methods, formal or otherwise, to guarantee security absolutely.

While our approach is restricted to proving properties of protocols
that can be established using the Cryptyc type and effect system, it
is worth pointing out that it is compatible with alternative methods for
protocol verification. For instance, it is possible to analyze the
protocols we use to implement secure web method calls for security
flaws beyond those that can be uncovered using Cryptyc (for instance,
flaws involving malicious insiders).

Our work shows that by exploiting recent advances in authenticity
types, we can develop a theoretical model of a security abstraction,
and then almost immediately obtain precise guarantees.
(As with many formal analyses, these guarantees concern the design of our
abstraction, and do not rule out code defects in its actual implementation.)
\ifJournal
\else
We intend to exploit these ideas further by exploring enriched programming
models for authentication and authorisation, while simultaneously building
theoretical models and SOAP-level implementations.
\fi

This study furthermore validates the adequacy of the spi-calculus, and
Cryptyc in particular, to formally reason about security properties in
a distributed communication setting. 

\paragraph*{Acknowledgments}
Cryptyc is an ongoing collaboration between Alan Jeffrey and the first author.
Ernie Cohen, C{\'e}dric Fournet, and Alan Jeffrey made useful suggestions
during the writing of this paper.

\ifLong
 \ifJournal\else
 \clearpage
 \fi
\fi

\appendix

\section{Sample SOAP Messages}
\label{app:soap-messages}

We give some sample SOAP messages exchanged during web service method
calls of the web service described in
Section~\ref{sec:soap-level-implementation}. One thing that is
immediately clear is that we are not using standard XML formats for
signing and encrypting messages, such as XML-Encryption and
XML-Signature. There is no intrinsic difficulty in adapting our
infrastructure to use standard formats. The point is that the
validation of the security abstraction does not rely on the exact
syntax of the SOAP envelopes.

\subsection{An Authenticated Call} 

We describe an authenticated call to the \texttt{Balance} method.
The messages exchanged to obtained the nonce are standard SOAP
messages. 
\ifLong\else
To simplify the presentation of these messages, we have removed some 
of the namespace information. More specifially, the \verb+<soap:Envelope>+ 
element carries the following namespaces:
\begin{quote}
\SOAPsize
\begin{verbatim}
xmlns:soap="http://schemas.xmlsoap.org/soap/envelope" 
xmlns:xsi="http://www.w3.org/2001/XMLSchema-instance" 
xmlns:xsd="http://www.w3.org/2001/XMLSchema"
\end{verbatim}
\unSOAPsize
\end{quote}
\fi
The following message is the request from Alice to the web
service to execute the \texttt{Balance} method on argument
12345. Notice the \texttt{DSHeader} element holding the identity of
the principals involved, as well as the nonces and the cryptographic
signature.
\ifJournal\else
 \begin{quote}
 \SOAPsize
\fi
\ifLong
\begin{verbatim}
<?xml version="1.0" encoding="utf-8"?>
<soap:Envelope xmlns:soap="http://schemas.xmlsoap.org/soap/envelope/" 
               xmlns:xsi="http://www.w3.org/2001/XMLSchema-instance" 
               xmlns:xsd="http://www.w3.org/2001/XMLSchema">
  <soap:Header>
    <DSHeader xmlns="http://tempuri.org/">
      <callerid>Alice</callerid>
      <calleeid>Bob</calleeid>
      <np>13</np>
      <nq>42</nq>
      <signature>
        3E:67:75:28:3B:AD:DF:32:E7:6C:D3:66:2A:CF:E7:8A:3F:0A:A6:0D
      </signature>
    </DSHeader>
  </soap:Header>
  <soap:Body>
    <Balance xmlns="http://tempuri.org/">
      <account>12345</account>
    </Balance>
  </soap:Body>
</soap:Envelope>
\end{verbatim}
\else
\begin{verbatim}
<?xml version="1.0" encoding="utf-8"?>
<soap:Envelope>
  <soap:Header>
    <DSHeader xmlns="http://tempuri.org/">
      <callerid>Alice</callerid>
      <calleeid>Bob</calleeid>
      <np>13</np>
      <nq>42</nq>
      <signature>
        3E:67:75:28:3B:AD:DF:32:E7:6C:D3:66:2A:CF:
        E7:8A:3F:0A:A6:0D
      </signature>
    </DSHeader>
  </soap:Header>
  <soap:Body>
    <Balance xmlns="http://tempuri.org/">
      <account>12345</account>
    </Balance>
  </soap:Body>
</soap:Envelope>
\end{verbatim}
\fi
\ifJournal\else
  \unSOAPsize
 \end{quote}
\fi
The response from the web service has a similar form:
\ifJournal\else
  \begin{quote}
  \SOAPsize
\fi
\ifLong
\begin{verbatim}
<?xml version="1.0" encoding="utf-8"?>
<soap:Envelope xmlns:soap="http://schemas.xmlsoap.org/soap/envelope/" 
               xmlns:xsi="http://www.w3.org/2001/XMLSchema-instance" 
               xmlns:xsd="http://www.w3.org/2001/XMLSchema">
  <soap:Header>
    <DSHeader xmlns="http://tempuri.org/">
      <callerid>Alice</callerid>
      <calleeid>Bob</calleeid>
      <np>13</np>
      <nq>42</nq>
      <signature>
        8D:31:52:6E:08:F0:89:7B:1E:12:3F:5E:63:EE:B0:D2:63:89:CA:73
      </signature>
    </DSHeader>
  </soap:Header>
  <soap:Body>
    <BalanceResponse xmlns="http://tempuri.org/">
      <BalanceResult>100</BalanceResult>
    </BalanceResponse>
  </soap:Body>
</soap:Envelope>
\end{verbatim}
\else
\begin{verbatim}
<?xml version="1.0" encoding="utf-8"?>
<soap:Envelope>
  <soap:Header>
    <DSHeader xmlns="http://tempuri.org/">
      <callerid>Alice</callerid>
      <calleeid>Bob</calleeid>
      <np>13</np>
      <nq>42</nq>
      <signature>
        8D:31:52:6E:08:F0:89:7B:1E:12:3F:5E:63:EE:
        B0:D2:63:89:CA:73
      </signature>
    </DSHeader>
  </soap:Header>
  <soap:Body>
    <BalanceResponse xmlns="http://tempuri.org/">
      <BalanceResult>100</BalanceResult>
    </BalanceResponse>
  </soap:Body>
</soap:Envelope>
\end{verbatim}
\fi
\ifJournal\else
  \unSOAPsize
  \end{quote}
\fi

\subsection{Authenticated and Encrypted Call}

We describe an authenticated and encrypted call, this time to the
\texttt{Statement} method.  Again, the messages exchanged to obtained the
nonce are standard SOAP messages. The following message is the request from
Alice to the web service to execute the \texttt{Statement} method on argument
12345. As in the authenticated call above, the \texttt{DSHeader} element holds
identity information. The body of the message itself is encrypted. Note that
the nonce \texttt{nq} must be encrypted according to the protocol, so its
encrypted value is included in the encrypted data, and its element is reset to
a dummy value (here, -1). Similarly, the signature is unused and set to a
dummy value.
\ifJournal\else
  \begin{quote}
  \SOAPsize
\fi
\ifLong
\begin{verbatim}
<?xml version="1.0" encoding="utf-8"?>
<soap:Envelope xmlns:soap="http://schemas.xmlsoap.org/soap/envelope/" 
               xmlns:xsi="http://www.w3.org/2001/XMLSchema-instance" 
               xmlns:xsd="http://www.w3.org/2001/XMLSchema">
  <soap:Header>
    <DSHeader xmlns="http://tempuri.org/">
      <callerid>Alice</callerid>
      <calleeid>Bob</calleeid>
      <np>13</np>
      <nq>-1</nq>
      <signature>4E:00:6F:00</signature>
    </DSHeader>
  </soap:Header>
  <soap:Body>
    9D:8F:95:2B:BC:60:B1:73:A7:C4:82:F5:39:20:97:F7:69:71:66:
    D3:A3:A0:90:B9:9B:FE:71:0A:65:C1:EF:EE:99:CB:4D:8A:40:37:
    CA:1E:D0:03:50:34:76:8C:E3:F3:30:DD:C9:34:19:D4:04:CB:39:
    7D:1A:84:2F:CA:30:DA:68:7E:E1:CB:07:9C:EB:79:F9:E9:4B:47:
    5B:94:56:D7:22:0E:02:CD:AA:F5:D3:40:C1:EC:13:FB:B9:E6:4F:
    13:CD:70:FD:BA:18:80:FC:50:F3:75:F2:2F:95:50:5D:41:7E:C8:
    8B:BB:AB:76:C9:59:BA:E2:3B:E5:4D:79:71:E4:AD:18:5A:4B:EA:
    29:17:30:90:66:08:27:ED:B4:BD:2E:89:06:6D:0B:56:40:43:35:
    A1:77:AE:12:7E:4B:19:26:B5:24:1A:D9:67:3D:A0:91
  </soap:Body>
</soap:Envelope>
\end{verbatim}
\else
\begin{verbatim}
<?xml version="1.0" encoding="utf-8"?>
<soap:Envelope>
  <soap:Header>
    <DSHeader xmlns="http://tempuri.org/">
      <callerid>Alice</callerid>
      <calleeid>Bob</calleeid>
      <np>13</np>
      <nq>-1</nq>
      <signature>4E:00:6F:00</signature>
    </DSHeader>
  </soap:Header>
  <soap:Body>
    9D:8F:95:2B:BC:60:B1:73:A7:C4:82:F5:39:20:97:
    F7:69:71:66:D3:A3:A0:90:B9:9B:FE:71:0A:65:C1:
    EF:EE:99:CB:4D:8A:40:37:CA:1E:D0:03:50:34:76:
    8C:E3:F3:30:DD:C9:34:19:D4:04:CB:39:7D:1A:84:
    2F:CA:30:DA:68:7E:E1:CB:07:9C:EB:79:F9:E9:4B:
    47:5B:94:56:D7:22:0E:02:CD:AA:F5:D3:40:C1:EC:
    13:FB:B9:E6:4F:13:CD:70:FD:BA:18:80:FC:50:F3:
    75:F2:2F:95:50:5D:41:7E:C8:8B:BB:AB:76:C9:59:
    BA:E2:3B:E5:4D:79:71:E4:AD:18:5A:4B:EA:29:17:
    30:90:66:08:27:ED:B4:BD:2E:89:06:6D:0B:56:40:
    43:35:A1:77:AE:12:7E:4B:19:26:B5:24:1A:D9:67:
    3D:A0:91
  </soap:Body>
</soap:Envelope>
\end{verbatim}
\fi
\ifJournal\else
  \unSOAPsize
  \end{quote}
\fi
The response is similarly encoded. Notice that this time the nonce
\texttt{np} must be encrypted, so its value is again included in the
encrypted data, and its element is reset to a dummy value.
\ifJournal\else
  \begin{quote}
  \SOAPsize
\fi
\ifLong
\begin{verbatim}
<?xml version="1.0" encoding="utf-8"?>
<soap:Envelope xmlns:soap="http://schemas.xmlsoap.org/soap/envelope/" 
               xmlns:xsi="http://www.w3.org/2001/XMLSchema-instance" 
               xmlns:xsd="http://www.w3.org/2001/XMLSchema">
  <soap:Header>
    <DSHeader xmlns="http://tempuri.org/">
      <callerid>Alice</callerid>
      <calleeid>Bob</calleeid>
      <np>-1</np>
      <nq>-1</nq>
      <signature>4E:00:6F:00</signature>
    </DSHeader>
  </soap:Header>
  <soap:Body>
    98:FD:6A:5B:38:0A:82:95:3F:01:EC:D3:55:F9:AA:35:4D:18:DB:
    1B:7D:9D:FE:3F:78:52:29:99:C9:41:84:EE:B1:42:12:B2:02:AC:
    63:F5:0C:92:9B:DB:75:FB:6C:8B:65:EB:3C:42:6B:79:70:AF:61:
    2A:C2:7B:ED:96:E1:D6:7A:F6:D2:0C:DF:BC:2A:4C:93:B3:D0:7B:
    7D:2D:83:18:60:D2:D8:05:EB:73:74:2D:75:A2:B2:57:C9:04:B4:
    C1:E6:66:54:BA:42:86:AF:22:72:3D:B7:90:CF:03:22:E5:C4:47:
    03:F0:77:A0:30:01:C9:FE:78:A1:AB:FA:B1:CB:EE:E2:0B:F2:79:
    17:1B:8E:82:E2:13:F4:66:52:76:6D:BA:1B:E9:8E:75:15:90:37:
    0A:64:ED:F3:9C:18:94:EC:4F:CF:61:92:38:EF:A9:46:E8:4E:E9:
    4A:E6:8A:C9:5E:ED:A7:34:72:3E:72:A2:BE:0D:DC:07:22:45:B0:
    E6:79:33:8F:CD:90:B8:97:DB:BA:3B:B2:8B:38:38:B6:5B:F1:11:
    FB:DD:88:CE:9A:3E:B4:E6:31:13:CB:1C:F3:B5:17:D8:9B:CF:2E:
    65:23:4D:BA:ED:72:6D:F4:53:97:B8:7A:D2:9C:2C:10:58:A3:0E:
    FE:48:A2:2A:2A:57:AE:6D:69:4D:97:90:EF:9F:C6:7E:9B
  </soap:Body>
</soap:Envelope>
\end{verbatim}
\else
\begin{verbatim}
<?xml version="1.0" encoding="utf-8"?>
<soap:Envelope>
  <soap:Header>
    <DSHeader xmlns="http://tempuri.org/">
      <callerid>Alice</callerid>
      <calleeid>Bob</calleeid>
      <np>-1</np>
      <nq>-1</nq>
      <signature>4E:00:6F:00</signature>
    </DSHeader>
  </soap:Header>
  <soap:Body>
    98:FD:6A:5B:38:0A:82:95:3F:01:EC:D3:55:F9:AA:
    35:4D:18:DB:1B:7D:9D:FE:3F:78:52:29:99:C9:41:
    84:EE:B1:42:12:B2:02:AC:63:F5:0C:92:9B:DB:75:
    FB:6C:8B:65:EB:3C:42:6B:79:70:AF:61:2A:C2:7B:
    ED:96:E1:D6:7A:F6:D2:0C:DF:BC:2A:4C:93:B3:D0:
    7B:7D:2D:83:18:60:D2:D8:05:EB:73:74:2D:75:A2:
    B2:57:C9:04:B4:C1:E6:66:54:BA:42:86:AF:22:72:
    3D:B7:90:CF:03:22:E5:C4:47:03:F0:77:A0:30:01:
    C9:FE:78:A1:AB:FA:B1:CB:EE:E2:0B:F2:79:17:1B:
    8E:82:E2:13:F4:66:52:76:6D:BA:1B:E9:8E:75:15:
    90:37:0A:64:ED:F3:9C:18:94:EC:4F:CF:61:92:38:
    EF:A9:46:E8:4E:E9:4A:E6:8A:C9:5E:ED:A7:34:72:
    3E:72:A2:BE:0D:DC:07:22:45:B0:E6:79:33:8F:CD:
    90:B8:97:DB:BA:3B:B2:8B:38:38:B6:5B:F1:11:FB:
    DD:88:CE:9A:3E:B4:E6:31:13:CB:1C:F3:B5:17:D8:
    9B:CF:2E:65:23:4D:BA:ED:72:6D:F4:53:97:B8:7A:
    D2:9C:2C:10:58:A3:0E:FE:48:A2:2A:2A:57:AE:6D:
    69:4D:97:90:EF:9F:C6:7E:9B
  </soap:Body>
</soap:Envelope>
\end{verbatim}
\fi
\ifJournal\else
  \unSOAPsize
  \end{quote}
\fi

\ifLong
\section{Semantics of the Object Calculus}\label{app:semantics}

In this appendix, we give a formal description of the operational
semantics and typing rules of the object calculus. We first describe
some encodings showing the expressiveness of the calculus.

\subsection{Encoding Arithmetic}

The calculus is simple enough that questions about whether or not it
is sufficiently expressive to be of interest arise. This is especially
likely since there are no recursive functions in the calculus, and it
is not clear that it is even Turing complete. That the calculus indeed
is Turing complete is a consequence of the fact that we can write
recursive classes and methods, and that we have a null object. The
following example shows an encoding of natural numbers as a class
$\mathit{Num}$, with the typical recursive definition of addition:
\[\begin{prog}
\mathit{class}\ \mathit{Num}\\ \quad
  \begin{prog}
  \mathit{Num}\:\mathit{pred}\\
  \mathit{Num}\:\mathit{succ}()\\ \quad
    \New\:\mathit{Num}(\this)\\
  \mathit{Num}\:\mathit{add}(\mathit{Num} x)\\ \quad
    \begin{prog}
    \If x.\mathit{pred}=\Null \Then \\ \quad
     \this\\
    \Else \this.\mathit{add}(x.\mathit{pred}).\mathit{succ}()
    \end{prog}
  \end{prog}
\end{prog}\]
We define $\mathit{zero}$ as $\New\:\mathit{Num}(\Null)$,
$\mathit{one}$ as $\mathit{zero}.\mathit{succ}()$, and so on. 

\subsection{Formalization of proxy objects}

We mentioned in the text that we can easily express proxy objects
within the calculus. For completeness, here is a detailed
formalization of such proxy objects. First, we assume a map $\proxy\in
\WebService \to \Class$, assigning to every web service $w \in
\WebService$ a proxy class $\proxy(w)$. We further assume that for
each $w \in \WebService$,
\begin{itemize}
\item $\dom(\methods(\class(w))) \cup \{\Id\} =
\dom(\methods(\proxy(w)))$, 
\item $\fields(\proxy(w))=\emptyset$,
\item $\methods(\proxy(w)(\Id)) = (\Id (), \owner(w))$, and
\item for all $\ell \in \dom(\methods(\class(w)))$,
\[\methods(\proxy(w))(\ell) = (B(A_1\:x_1,\ldots,A_n\:x_n),
w\mo{:}\ell(x_1,\ldots,x_n)),\] where $\methods(\class(w))(\ell) =
(B(A_1\:x_1,\ldots,A_n\:x_n),b)$.
\end{itemize}

\subsection{Operational Semantics} 

The operational semantics is defined by a transition relation, written
$a \to^p a'$, where $a$ and $a'$ are method bodies, and $p$ is the
principal evaluating the body $a$.

To specify the semantics, we need to keep track of which principal is
currently running a method body. We add a new method body form to our
object calculus, $p[a]$, meaning $p$ running body $a$. This form does
not appear in code written by the user, but only arises through the
transitions of the semantics.

\begin{display}{Extended Method Bodies:}
\Category{a,b \in \Body}{method body}\\
\entry{\cdots}{as in Section~\ref{sec:formal-model}}\\
\entry{p[a]}{body $a$ running as $p$}
\end{display}

\begin{display}{Transitions:}
\staterule
  {(Red Let 1)}
  {a \to^p a'}
  {\Let{x}{a} \In{b} \to^p \Let{x}{a'} \In{b}}
  \label{Rule:red-let-1}
\staterule
  {(Red Let 2)}
  {}
  {\Let{x}{v} \In{b} \to^p b\SUB{x \GETS v}}
  \label{Rule:red-let-2}
  \\[\GAP]
\staterule
  {(Red If)}
  {}
  {\If u=v \Then a_{\mathit{true}} \Else a_{\mathit{false}} \to^p a_{u=v}}
  \label{Rule:red-if}
  \\[\GAP]
\staterule
  {(Red Field)}
  {\fields(c) = f_i \mapsto A_i \For{i \in 1..n} \quad j \in 1..n}
  {(\New\:c(v_1,\ldots,v_n)).f_j \to^p v_j}
  \\[\GAP]
\staterule
  [(where $v=\New\:c(v_1,\ldots,v_n)$)]
  {(Red Invoke)}
  {\methods(c) = \ell_i \mapsto (\sig_i,b_i) \For{i \in 1..n} \quad
   j \in 1..n \quad \sig_j=B(A_1\:x_1,\ldots,A_m\:x_m)}
  {v.\ell_j(u_1,\ldots,u_m) \to^p
     b_j\SUB{\mathit{this} \GETS v,x_k \GETS u_k \For{k \in 1..m}}}
  \\[\GAP]
\staterule
  {(Red Remote)}
  {\owner(w)=q \quad \class(w)=c}
  {w\mo{:}\ell(u_1,\ldots,u_n) \to^p q[\New\:c(p).\ell(u_1,\ldots,u_n)]}\\
  \\[\GAP]
\staterule
  {(Red Prin 1)}
  {a \to^q a'}
  {q[a] \to^p q[a']}
\staterule
  {(Red Prin 2)}
  {}
  {q[v] \to^p v}
\end{display}

\subsection{Type System} 

The judgments of our type system all depend on an \emph{environment}
$E$, that defines the types of all variables in scope. An environment
takes the form $x_1\ty A_1,\ldots,x_n\ty A_n$ and defines the type
$A_i$ for each variable $x_i$. The domain $\dom(E)$ of an environment
$E$ is the set of variables whose types it defines. 

\begin{display}{Environments:}
\Category{D,E}{environment}\\
\entry{\emptyset}{empty}\\
\entry{E,x\ty A}{entry}\\
\clause{\dom(x_1\ty A_1,\ldots,x_n\ty A_n) \triangleq \{x_1,\ldots,x_n\}}{domain of an environment}
\end{display}

The following are the two judgments of our type system. They are
inductively defined by rules presented in the following tables. 

\begin{display}{Judgments $\Judge{E}{\mathcal{J}}$:}
\clause{\JudgeOK{E}}{good environment}\\
\clause{\Judge{E}{a:A}}{good expression $a$ of type $A$}
\end{display}%
We write $\Judge{E}{\mathcal{J}}$ when we want to talk about both
kinds of judgments, where $\mathcal{J}$ stands for either $\diamond$
or $a:A$.

The following rules define an environment
$x_1\ty A_1,\ldots,x_n\ty A_n$ to be well-formed if each of the
names $x_1,\ldots,x_n$ are distinct. 

\begin{display}{Rules for Environments:}
\typerule
  {(Env $\emptyset$)}
  {}
  {\JudgeOK{\emptyset}}
\typerule
  [(where $x\not\in\dom(E)$)]
  {(Env $x$)}
  {\JudgeOK{E}}
  {\JudgeOK{E,x\ty A}}
\end{display}

We present the rules for deriving the judgment $\Judge{E}{a:A}$ that
assigns a type $A$ to a value or method body $a$. These rules are
split into two tables, one for values, and one for method bodies. 

\begin{display}{Rules for Typing Values:}
\typerule
  {(Val $x$)}
  {E=E_1,x\ty A,E_2 \quad \JudgeOK{E}}
  {\Judge{E}{x : A}}
\typerule
  {(Val $\Null$)}
  {\JudgeOK{E}}
  {\Judge{E}{\Null:c}}
  \\[\GAP]
\typerule
  {(Val Object)}
  {\fields(c)=f_i \mapsto A_i \For{i \in 1..n} \quad
  \Judge{E}{v_i:A_i} \quad \forall i\in 1..n}
  {\Judge{E}{\New\:c(v_1,\ldots,v_n):c}}
\typerule
  {(Val Princ)}
  {\JudgeOK{E}}
  {\Judge{E}{p:\Id}}
\end{display}

\begin{display}{Rules for Typing Method Bodies:}
\typerule
  {(Body Let)}
  {\Judge{E}{a:A} \quad \Judge{E,x\ty A}{b:B}}
  {\Judge{E}{\Let{x}{a} \In{b}:B}}
  \\[\GAP]
\typerule
  {(Body If)}
  {\Judge{E}{u:A} \quad \Judge{E}{v:A} \quad \Judge{E}{a:B} \quad \Judge{E}{b:B}}
  {\Judge{E}{\If u=v \Then a \Else b:B}}
  \\[\GAP]
\typerule
  {(Body Field)}
  {\Judge{E}{v:c} \quad \fields(c)=f_i \mapsto A_i \For{i \in 1..n} \quad j \in 1..n}
  {\Judge{E}{v.f_j : A_j}}
  \\[\GAP]
\typerule
  {(Body Invoke)}
  {\Judge{E}{v:c} \quad \methods(c)=\ell_i \mapsto (\sig_i,b_i) \For{i \in 1..n} \quad j \in 1..n \\
   \sig_j = B (A_1\:x_1,\ldots,A_m\:x_m) \quad
   \Judge{E}{u_k:A_k} \quad \forall k \in 1..m}
  {\Judge{E}{v.\ell_j(u_1,\ldots,u_m) : B}}
  \\[\GAP]
\typerule
  {(Body Remote)}
  {\class(w)= c \quad \methods(c)=\ell_i \mapsto (\sig_i,b_i) \For{i \in 1..n}
  \quad j \in 1..n \\
   \sig_j = B (A_1\:x_1,\ldots,A_m\:x_m) \quad
   \Judge{E}{u_i:A_i} \quad \forall i \in 1..m}
  {\Judge{E}{w\mo{:}\ell_j(u_1,\ldots,u_m) : B}}
\typerule
   {(Body Princ)}
   {\Judge{E}{a : A}}
   {\Judge{E}{p[a] : A}}
\end{display}

We make the following assumption on the execution environment.
\begin{display}{Assumptions on the Execution Environment:}
(1) \mbox{For each $w \in \WebService$, $\fields(\class(w)) =
    \mathit{CallerId}:\Id$.}\\
(2) \begin{prog}
    \mbox{No tagged expression $p[a]$ occurs within the body of any
method;}\\
    \mbox{such expressions occur only at runtime, to track the call
    stack of principals.}
    \end{prog}\\
(3) \begin{prog}
    \mbox{for each $c \in \Class$ and each $\ell \in
       \dom(\methods(c))$,}\\
    \mbox{if $\methods(c)(\ell) = (B
       (A_1\:x_1,\ldots,A_n\:x_n),b)$,}\\
    \mbox{then $\Judge{\this\ty c,x_1\ty A_1,\ldots,x_n\ty A_n}{b:B}$.}
    \end{prog}
\end{display}

It is straightforward to show that our type system is sound, that is,
that the type system ensures that methods that typecheck do not get
stuck when evaluating. Some care is needed to make this precise, since
evaluation can block if one attempts to access a field of a null
object, or to invoke a method on a null object. (We could introduce an
\emph{error token} in the semantics and propagate that error token
when such a case is encountered, but this would needlessly complicate
the semantics, at least for our purposes.) Soundness can be derived as
usual via Preservation and Progress theorems. To establish these, we
first need the following lemmas:
\begin{lemma} The following properties of judgments hold:
\begin{description}
\item[(Exchange)] if $\Judge{E,x\ty A,y\ty B,E'}{\mathcal{J}}$, then
$\Judge{E,y\ty B,x\ty A,E'}{\mathcal{J}}$;
\item[(Weakening)] if $\Judge{E}{\mathcal{J}}$ and $x\not\in\dom(E)$,
then $\Judge{E,x\ty A}{\mathcal{J}}$;
\item[(Strengthening)] if $\Judge{E,x\ty B}{a:A}$ and $x\not\in\fv(a)$,
then $\Judge{E}{a:A}$. 
\end{description}
\end{lemma}
\begin{proof}
Straightforward.\hfill
\end{proof}
We often use the above properties silently in the course of proofs. 
\begin{lemma}[Substitution]
If $\Judge{E,x\ty B}{a:A}$ and $\Judge{E}{v:B}$, then
$\Judge{E}{a\SUB{x\GETS v}:A}$
\end{lemma}
\begin{proofx} This is a straightforward proof by induction on the
height of the typing derivation for $\Judge{E}{a:A}$. We proceed by case analysis on the
form of $a$.
\begin{itemize}

\citem{$a=x$} Since $\Judge{E,x\ty B}{x:A}$, we must have $A=B$. Since
$a\SUB{x\GETS v} = v$ and $\Judge{E}{v:B}$, we have $\Judge{E}{v:A}$,
as required.

\citem{$a=y$, where $y\ne x$} Since $x$ is not free in $y$,
$\Judge{E,x\ty B}{y:A}$ implies $\Judge{E}{y:A}$, by the Strengthening
Lemma, as required. 

\citem{$a=\Null$} Since $x$ is not free in $\Null$,
$\Judge{E,x\ty B}{\Null:A}$ implies $\Judge{E}{\Null:A}$, as required. 

\citem{$a=p$} Since $x$ is not free in $p$,
$\Judge{E,x\ty B}{p:A}$ implies $\Judge{E}{p:A}$, as required. 

\citem{$a=\New\:c(v_1,\ldots,v_n)$} We have the equation
$\New\:c(v_1,\ldots,v_n)\SUB{x\GETS v} = \New\:c(v_1\SUB{x\GETS
v},\ldots,v_n\SUB{x\GETS v})$. We have $\Judge{E,x\ty
B}{\New\:c(v_1,\ldots,v_n):A}$, hence $\Judge{E,x\ty B}{v_i:A_i}$ if
$\fields(c)=f_i\mapsto A_i\For{i\in 1..n}$. By the induction
hypothesis, we know $\Judge{E}{v_i\SUB{x\GETS v}:A_i}$ for all $i\in
1..n$. Hence, we can derive $\Judge{E}{\New\:c(v_1\SUB{x\GETS
v},\ldots,v_n\SUB{x\GETS v}):A}$, as required.

\citem{$a=\Let{y}{a_0}\In{b}$} Without loss of generality, we can take
$y\ne x$, since $y$ is bound in $b$. Note that
$(\Let{y}{a_0}\In{b})\SUB{x\GETS v}=\Let{y}{a_0\SUB{x\GETS
v}}\In{b\SUB{x\GETS v}}$. We have $\Judge{E,x\ty
B}{\Let{y}{a_0}\In{b}:A}$, hence $\Judge{E,x\ty B}{a_0:A_0}$ for some
$A_0$, and $\Judge{E,y\ty A_0,x\ty B}{b:A}$. By the induction hypothesis,
$\Judge{E}{a_0\SUB{x\GETS v}:A_0}$ and $\Judge{E,y\ty A_0}{b\SUB{x\GETS
v}:A}$, and hence $\Judge{E}{\Let{x}{a_0\SUB{x\GETS
v}}\In{b\SUB{x\GETS v}}:A}$, as required.

\citem{$a=\If u_0=u_1 \Then a_0 \Else a_1$} We have $(\If
 u_0=u_1 \Then a_0$ $\Else a_1)\SUB{x\GETS v}=\If u_0\SUB{x\GETS
 v}=u_1\SUB{x\GETS v}\Then a_0\SUB{x\GETS v} \Else a_1\SUB{x\GETS v}$. We have
 $\Judge{E,x\ty B}{\If u_0=u_1\Then a_0 \Else a_1}$, hence $\Judge{E,x\ty
 B}{u_0:A'}$, $\Judge{E,x\ty B}{u_1:A'}$, $\Judge{E,x\ty B}{a_0:A}$ and
 $\Judge{E,x\ty B}{a_1:A}$. Applying the induction hypothesis to these
 judgments, we can derive \[\Judge{E}{\If u_0\SUB{x\GETS v}=u_1\SUB{x\GETS
 v}\Then a_0\SUB{x\GETS v} \Else a_1\SUB{x\GETS v}:A}\] as required.

\end{itemize}
The remaining cases are similar, upon noting that:
\begin{itemize}
\item[-] $(u.f_j)\SUB{x\GETS v}=u\SUB{x\GETS v}.f_j$, 
\item[-] $(u.\ell(u_1,\ldots,u_m))\SUB{x\GETS v}=u\SUB{x\GETS
     v}(u_1\SUB{x\GETS v},\ldots,u_m\SUB{x\GETS v})$,
\item[-] $(w\mo{:}\ell(u_1,\ldots,u_n))\SUB{x\GETS v}=w\mo{:}(u_1\SUB{x\GETS
   v},\ldots,u_n\SUB{x\GETS v})$, and
\item[-]$(p[a])\SUB{x\GETS v}=p[a\SUB{x\GETS v}]$.\qed
\end{itemize}
   \end{proofx}

\begin{theorem}[Preservation]
If $\Judge{E}{a : A}$ and $a \to^p a'$ then $\Judge{E}{ a' : A}$.
\end{theorem}
\begin{proofx}
We proceed by induction on the height of the typing derivation for
$\Judge{E}{a:A}$. Since $a\to^p a'$, $a$ cannot be a value $v$. 
\begin{itemize}

\citem{$a=\Let{x}{v}\In{b}$} Since $\Judge{E}{a:A}$, we have
$\Judge{E}{v:B}$ and $\Judge{E,x\ty B}{b:A}$. We must have
$a'=b\SUB{x\GETS v}$. Applying the Substitution Lemma, we have
$\Judge{E}{b\SUB{x\GETS v}:A}$, as required.

\citem{$a=\Let{x}{a_0}\In{b}$, where $a_0$ is not a value} We have $\Judge{E}{a_0:B}$, and
$\Judge{E,x\ty B}{b:A}$. Since $a\to^p a'$, we must have have
$a_0\to^p a_0'$. By induction hypothesis, $\Judge{E}{a_0':B}$, and
hence $\Judge{E}{\Let{x}{a_0'}\In{b}:A}$, as required.

\citem{$a=\If u=v\Then a_0 \Else a_1$} Note that either $a\to^p
a_0$ or $a\to^p a_1$. In both cases, since $\Judge{E}{\If u=v\Then
a_0\Else a_1:A}$, we have $\Judge{E}{a_0:A}$ and $\Judge{E}{a_1:A}$,
as required. 

\citem{$a=(\New\:c(v_1,\ldots,v_n)).f_j$} We have
$\fields(c)=f_i\mapsto A_i\For{i\in 1..n}$. The type derivation for
$a$ is as follows:
\[ 
\infer{\Judge{E}{(\New\:c(v_1,\ldots,v_n)).f_j:A_j}}
      {\infer{\Judge{E}{\New\:c(v_1,\ldots,v_n):c}}
             {\Judge{E}{v_i:A_i}\For{i\in 1..n}}}
\]
Since $a'=v_j$, we have $\Judge{E}{v_j:A_j}$, as required. 

\citem{$a=(\New\:c(v_1,\ldots,v_n)).\ell_j(u_1,\ldots,u_m)$} Let
$v=\New\:c(v_1,\ldots,v_n)$. We have
$\methods(c)=\ell_i\mapsto(\sig_i,b_i)\For{i\in 1..n}$, where
$\sig_j=B(A_1\:x_1,\ldots,A_m\:x_m)$. By the typing derivation for
$\Judge{E}{a:B}$, we have $\Judge{E}{u_k:A_k}$ for all $k\in 1..m$, and
$\Judge{E}{v:c}$. By assumption 
on the execution environment, we know $\Judge{\this\ty c,x_1\ty
A_1,\ldots,x_m\ty A_m}{b:B}$. Applying the Substitution and the
Weakening Lemmas, we get
$\Judge{E}{b\SUB{\this\GETS v,x_k\GETS u_k\For{k\in 1..m}}:B}$, as
required. 

\citem{$a=w\mo{:}\ell_j(u_1,\ldots,u_n)$} We have $\class(w)=c$,
$\methods(c)=\ell_i\mapsto(\sig_i,b_i)\For{i\in 1..n}$ where
$\sig_j=B(A_1\:x_1,\ldots,A_m\:x_m)$. By the typing derivation for
$\Judge{E}{a:B}$, we have $\Judge{E}{u_i:A_i}$ for all $i\in 1..m$. We can
therefore derive the required type for
$a'=q[\New\:c(p).\ell(u_1,\ldots,u_m)]$: 
\[
\infer{\Judge{E}{q[\New\:c(p).\ell(u_1,\ldots,u_m)]:B}}
      {\infer{\Judge{E}{\New\:c(p).\ell(u_1,\ldots,u_m):B}}
             {\Judge{E}{\New\:c(p):c} & 
              \Judge{E}{u_i:A_i} \quad \forall i\in 1..m}}
\]

\citem{$a=q[v]$} Since $\Judge{E}{q[v]:A}$, we have $\Judge{E}{v:A}$,
and $q[v]\to^p v$, as required.

\citem{$a=q[a_0]$, where $a_0$ is not a value} Since $\Judge{E}{q[a_0]:A}$, we have
$\Judge{E}{a_0:A}$, and since $a\to^p a'$, we must have $a_0\to^q
a_0'$. By induction hypothesis, $\Judge{E}{a_0':A}$, and hence
$\Judge{E}{q[a_0']:A}$, as required.\qed
\end{itemize}
\end{proofx}

To state the Progress Theorem, we need to recognize programs that are
blocked because of a \emph{null} in object position. We say a method
body $a$ is \emph{null-blocked} if, essentially, it is stuck trying to
access a field of a null object, or invoke a method on a null
object. Formally, $a$ is null-blocked if it is of the form
$\Null.f_j$, $\Null.\ell(u_1,\ldots,u_n)$, $\Let{x}{a}\In{b}$ (where
$a$ is null-blocked), or $q[a]$ (where $a$ is null-blocked).

\begin{theorem}[Progress]
If $\Judge{\emptyset}{a : A}$ and $a$ is not a value and is not
null-blocked, and $p \in \Prin$, then $a \to^p a'$ for some $a'$.
\end{theorem}
\begin{proofx}
Again, we proceed by induction on the height of the typing derivation
for $\Judge{\emptyset}{a:A}$. We assume $a$ is not a value, and $a$ is
not null-blocked.
\begin{itemize}

\citem{$a=\Let{x}{a_0}\In{b}$} We consider two subcases, depending on
whether $a_0$ is a value or not. 
\begin{itemize}
\citem{$a_0$ is a value $v$} We have $a\to^p b\SUB{x\GETS v}$.

\citem{$a_0$ is not a value} Since $\Judge{\emptyset}{a:A}$, we have
$\Judge{\emptyset}{a_0:B}$ for some $B$, $a_0$ not a value. Since $a$
is not null-blocked, $a_0$ is not null-blocked. Hence, by induction
hypothesis, we have $a_0\to^p a_0'$. Hence, we have
$a\to^p\Let{x}{a_0'}\In{b}$. 
\end{itemize}

\citem{$a=\If u=v\Then a_0 \Else a_1$} We have $a\to^p a_0$ or
$a\to^p a_1$ depending on the result of $u=v$. 

\citem{$a=v.f_j$} Since $\Judge{\emptyset}{a:A}$ and $a$ is not null-blocked,
we must have $v=\New\:c(u_1,\ldots,u_n)$, and $\fields(c)=f_i\mapsto
A_i\For{i\in 1..n}$. Therefore, we have $v.f_j\to^p u_j$.

\citem{$a=v.\ell_j(u_1,\ldots,u_m)$} Since $\Judge{\emptyset}{a:A}$ and $a$ is
not null-blocked, we must have $v=\New\:c(u_1,\ldots,u_n)$,
$\methods(c)=\ell_i\mapsto (\sig_i,b_i)$, and
$\sig_j=B(A_1\:x_1,\ldots,A_m\:x_m)$. Therefore, we have
$v.\ell_j(u_1,\ldots,u_m)\to^p b_j\SUB{\this\GETS v,x_k\GETS
u_k\For{k\in 1..m}}$. 

\citem{$a=w\mo{:}\ell(u_1,\ldots,u_m)$} The following transition rule
$w\mo{:}\ell(u_1,\ldots,u_m)\to^p$ $q[\New\:c(p).\ell(u_1,\ldots,u_m)]$
applies, with $\owner(w)=q$ and $\class(w)=c$.

\citem{$a=q[a_0]$} We consider two subcases, depending on whether
$a_0$ is a value or not. 
\begin{itemize}
\citem{$a_0$ is a value $v$} We have $q[v]\to^p v$.

\citem{$a_0$ is not a value} Since $\Judge{\emptyset}{q[a_0]:A}$, we have
$\Judge{\emptyset}{a_0:A}$, $a_0$ not a value. Since $a$ is not null-blocked,
$a_0$ is not null-blocked. Hence, by induction hypothesis, we have
$a_0\to^q a_0'$, and $q[a_0]\to^p q[a_0']$. \qed
\end{itemize}

\end{itemize}
\end{proofx}

We can now state soundness formally. We say a method body $a$ is stuck
if $a$ is not a value, $a$ is not null-blocked, and there is no $a'$
and $p$ such that $a\to^p a'$. We write $a\to^{*}a'$ to mean that
there exists a sequence $a_1,\ldots,a_n$ and principals
$p_1,\ldots,p_{n+1}$ such that
$a\to^{p_1}a_1\to^{p_2}\cdots\to^{p_{n}}a_n\to^{p_{n+1}}a'$.
(Hence, $\to^{*}$ is a kind of transitive closure of
$\to^{p}$.) 
\begin{theorem}[Soundness]\label{thm:soundness}
If $\Judge{\emptyset}{a:A}$, and $a\to^{*}a'$, then $a'$ is not
stuck. 
\end{theorem}
\begin{proof}
A straightforward induction on the number of transitions in
$a\to^{*}a'$.  \hfill
\end{proof}
\fi%

\section{The Spi-Calculus in More Detail}
\label{app:spi}

We give an overview of the language and type system on which our analysis of
web services depends.  We give the syntax in detail, but for the sake of
brevity give only an informal account of the operational semantics and type
system.  Full details are in a technical
report~\cite{GJ02:TypesAndEffectsForAsymmetricCryptographicProtocols}, from
which some of the following explanations are drawn.
\ifLong
Some constructs primitive here are actually derived forms in the original calculus.
\fi

\begin{renewcommand}{\ratio}{.3}
\begin{display}{Names, Messages:}
\clause{\ka ::= \Enc \mid \Dec}{key attribute}\\
\clause{m,n,x,y,z}{name: nonce, key, key-pair}\\
\Category{L,M,N}{message}\\
\entry{x}{name}\\
\entry{(M_1,\ldots,M_n)}{record, $n\ge 0$}\\
\entry{t_i(M)}{tagged union}\\
\entry{\encrypt{M}{N}}{symmetric encryption}\\
\entry{\pencrypt{M}{N}}{asymmetric encryption}\\
\entry{\extract{\ka}{M}}{key-pair component}
\end{display}
\end{renewcommand}

The message $x$ is a name, representing a channel, nonce, symmetric key,
or asymmetric key-pair.
We do not differentiate in the syntax or operational semantics between
key-pairs used for public key cryptography and those used for digital
signatures.

The message $(M_1,\ldots,M_n)$ is a record with $n$ fields, $M_1$, \ldots,
$M_n$.

The message $t_i(M)$ is message $M$ tagged with tag $t_i$.
The message $\encrypt{M}{N}$ is the ciphertext obtained by
encrypting the plaintext $M$ with the symmetric key $N$.

The message $\pencrypt{M}{N}$ is the ciphertext obtained by
encrypting the plaintext $M$ with the asymmetric encryption key $N$.

The message $\extract{\Dec}{M}$ is the decryption key (or signing key) component
of the key-pair $M$, and $\extract{\Enc}{M}$ is the encryption key (or
verification key) component of the key-pair $M$.

\begin{renewcommand}{\ratio}{.3}
\begin{display}{Types and Effects:}
\clause{\ell ::= \public \mid \private}{nonce attribute}\\
\Category{S,T,U}{type}\\
\entry{\Un}{data known to the opponent}\\
\entry{(x_1\ty T_1,\ldots,x_n\ty T_n)}{dependent record, $n\ge 0$}\\
\entry{\Union(t_1(T_1),\ldots,t_n(T_n))}{tagged union}\\
\entry{\Top}{top}\\
\entry{\SharedKey{T}}{shared-key type}\\
\entry{\KeyPair{T}}{asymmetric key-pair}\\
\entry{\annotate{\ka}{\Key{T}}}{encryption or decryption part}\\
\entry{\annotate{\ell}{\Challenge{\es}}}{challenge type}\\
\entry{\annotate{\ell}{\Response{\fs}}}{response type}\\
\Category{e,f}{atomic effect}\\
\entry{\endd{L}}{end-event labelled $L$}\\
\entry{\check{\ell}{N}}{name-check for a nonce $N$}\\
\entry{\trust{M\ty T}}{trust that $M \ty T$}\\
\Category{\es,\fs}{effect}\\
\entry{\mset{e_1,\ldots,e_n}}{multiset of atomic effects}
\end{display}
\end{renewcommand}

The type $\Un$ describes messages that may flow to or from the opponent,
which we model as an arbitrary process of the calculus.
We say that a type is \emph{public} if messages of the type may flow to the
opponent.
Dually, we say a type is \emph{tainted} if messages from the opponent may flow
into the type.  The type $\Un$ is both public and tainted.

The type $(x_1\ty T_1,\ldots,x_n\ty T_n)$ describes a record $(M_1,\ldots,M_n)$
where each $M_i : T_i$.  The scope of each variable $x_i$ consists of the
types $T_{i+1}, \ldots, T_n$.  Type $(x_1\ty T_1,$ $\ldots,$ $x_n\ty T_n)$
is public just if all of the types $T_i$ are public,
and tainted just if all of the types $T_i$ are tainted.

The type $\Union(t_1(T_1),\ldots,t_n(T_n))$ describes a tagged message $t_i(M)$
where $i \in 1..n$ and $M : T_i$.
Type $\Union(t_1(T_1),$ $\ldots,$ $t_n(T_n))$
is public just if all of the types $T_i$ are public,
and tainted just if all of the types $T_i$ are tainted.

The type $\Top$ describes all well-typed messages; it is tainted but not public.

The type $\SharedKey{T}$ describes symmetric keys for encrypting messages
of type $T$; it is public or tainted just if $T$ is both public and tainted.

The type $\KeyPair{T}$ describes asymmetric key-pairs for encrypting or
signing messages of type $T$; it is public or tainted just if $T$ is both
public and tainted.
The key-pair can be used for public-key cryptography just if $T$ is tainted,
and for digital signatures just if $T$ is public.

The type $\annotate{\Enc}{\Key{T}}$ describes an encryption or signing key
for messages of type $T$; it is public just if $T$ is tainted, and it is
tainted just if $T$ is public.

The type $\annotate{\Dec}{\Key{T}}$ describes a decryption or verification
key for messages of type $T$; it is public just if $T$ is public, and it is
tainted just if $T$ it tainted.

The types $\annotate{\ell}{\Challenge{\es}}$ and
$\annotate{\ell}{\Response{\fs}}$ describe nonce challenges and responses,
respectively.  The effects $\es$ and $\fs$ embedded in these types represent
certain events.  An outgoing challenge of some type
$\annotate{\ell}{\Challenge{\es}}$ can be cast into a response of type
$\annotate{\ell}{\Response{\fs}}$ and then returned, provided the events in
the effect $\es+\fs$ have been justified, as explained below.
Therefore, if we have created a fresh challenge at type
$\annotate{\ell}{\Challenge{\es}}$, and check that it equals an incoming
response of type $\annotate{\ell}{\Response{\fs}}$, we can conclude that the
events in $\es+\fs$ may safely be performed.  The attribute $\ell$ is either
$\public$ or $\private$; the former means the nonce may eventually be public,
while the latter means the nonce is never made public.
Type $\annotate{\public}{\Challenge{\es}}$ is public, or tainted, just if
$\es=[\,]$.
Type $\annotate{\public}{\Response{\fs}}$ is always public,
but tainted just if $\es=[\,]$.
Neither $\annotate{\private}{\Challenge{\es}}$
nor $\annotate{\private}{\Response{\fs}}$
is public; both are tainted.

An effect $\es$ is a multiset, that is, an unordered list of atomic effects,
$e$ or $f$.  Effects embedded in challenge or response types signify that
certain actions are justified, that is, may safely be performed.  An atomic
effect $\endd{L}$ justifies a single subsequent end-event labelled $L$, and is
justified by a distinct, preceding begin-event labelled $L$.  An atomic effect
$\check{\ell}{N}$ justifies a single subsequent check that an $\ell$ response
equals an $\ell$ challenge named $N$, where $\ell$ is $\public$ or $\private$,
and is justified by freshly creating the challenge $N$.
An atomic effect $\trust{M\ty T}$ justifies casting message $M$ to type $T$,
and is justified by showing that $M$ indeed has type $T$.

\begin{display}{Processes:}
\Category{O,P,Q,R}                {process}\\
\entry{\Snd{M}{N}}                  {output}\\
\entry{\Rcv{M}{x \ty T}P}             {input \ifLong($x$ bound in $P$)\fi}\\
\entry{\Repl{\Rcv{M}{x \ty T} P}}       {replicated input \ifLong($x$ bound in $P$)\fi}\\
\entry{\SplitPoly{M}{(x_1\ty T_1,\ldots,x_n\ty T_n)}P}    {record splitting}\\
\entry{\Match{M}{N}{y\ty T}P}           {pair matching \ifLong($y$ bound in $P$)\fi}\\
\entry{\Case{M} \IsTag{t_i}{x_i \ty T_i}{P_i}\For{i \in 1..n}}
               {tagged union case \ifLong($t_i$ distinct)\fi}\\
\entry{\SpiIf M=N \SpiThen P \SpiElse Q}{conditional (new)}\\
\entry{\Res{x \ty T}P}                  {name generation \ifLong($x$ bound in $P$)\fi}\\
\entry{P \parpop Q}                     {composition}\\
\entry{\Stop}                           {inactivity}\\
\entry{\Decrypt{M}{x \ty T}{N}P} {symmetric decrypt \ifLong($x$ bound in $P$)\fi}\\
\entry{\pDecrypt{M}{x \ty T}{N}P}   {asymmetric decrypt \ifLong($x$ bound in $P$)\fi}\\
\entry{\CheckNonce{M}{N}P}    {nonce-checking}\\
\entry{\Begin{L}P}                      {begin-assertion}\\
\entry{\End{L}P}                       {end-assertion}\\
\entry{\Cast{M}{x \ty T}P}{cast to nonce type}\\
\entry{\Witness{M\ty T}P}{witness testimony}\\
\entry{\Trust{M}{x\ty T}P}{trusted cast}
\end{display}

The processes $\Snd{M}{N}$ and $\Rcv{M}{x \ty T}P$ are output and input,
respectively, along an asynchronous, unordered channel $M$.  If an output
$\Snd{x}{N}$ runs in parallel with an input $\Rcv{x}{y}P$, the two can
interact to leave the residual process $P\SUB{y \GETS N}$, the outcome
of substituting $N$ for each free occurrence of $y$ in $P$.  We write
$\Sndp{x}{(M)}P$ as a simple shorthand for $\Snd{x}{M} \parpop P$.

The process $\Repl{\Rcv{M}{x \ty T}P}$ is replicated input,
which behaves like input, except that each time an input of $N$ is
performed, the residual process $P\SUB{y \GETS N}$ is spawned off
to run concurrently with the original process $\Repl{\Rcv{M}{x \ty T}P}$.

The process $\SplitPoly{M}{(x_1\ty T_1,\ldots,x_n\ty T_n)}P$ splits the record
$M$ into its $n$ components.  If $M$ is $(M_1,\ldots,M_n)$, the process
behaves as $P\SUB{x_1 \GETS M_1}\cdots\SUB{x_n \GETS M_n}$.  Otherwise, it
deadlocks, that is, does nothing.

The process $\Match{M}{N}{y \ty U}P$ splits the pair (binary record) $M$ into
its two components, and checks that the first one is $N$.  If $M$ is $(N,L)$,
the process behaves as $P\SUB{y \GETS L}$.  Otherwise, it deadlocks.

The process $\Case{M} \IsTag{t_i}{x_i \ty T_i}{P_i}\For{i \in 1..n}$ checks the
tagged union $M$.  If $M$ is $t_j(L)$ for some $j \in 1..n$, the process
behaves as $P\SUB{x_i \GETS L}$.  Otherwise, it deadlocks.

The process $\SpiIf M=N \SpiThen P \SpiElse Q$ behaves as $P$ if $M$ and $N$
are the same message, and otherwise as $Q$.  (This process is not present in
the original
calculus~\cite{GJ02:TypesAndEffectsForAsymmetricCryptographicProtocols} but
is a trivial and useful addition.)

The process $\Res{x \ty T}P$ generates a new name $x$, whose scope is $P$, and
then runs $P$.  This abstractly represents nonce or key generation.

The process $P \parpop Q$ runs processes $P$ and $Q$ in parallel.

The process $\Stop$ is deadlocked.

The process $\Decrypt{M}{x \ty T}{N}P$ decrypts $M$ using symmetric key $N$.  If
$M$ is $\encrypt{L}{N}$, the process behaves as $P\SUB{x \GETS L}$.
Otherwise, it deadlocks.  We assume there is enough redundancy in the
representation of ciphertexts to detect decryption failures.

The process $\pDecrypt{M}{x \ty T}{N}P$ decrypts $M$ using asymmetric key $N$.
If $M$ is $\pencrypt{L}{\extract{\Enc}{K}}$ and $N$ is $\extract{\Dec}{K}$,
then the process behaves as $P\SUB{x \GETS L}$.  Otherwise, it deadlocks.

The process $\CheckNonce{M}{N}P$ checks the messages $M$ and $N$ are the same
name before executing $P$.  If the equality test fails, the process deadlocks.

The process $\Begin{L}P$ autonomously performs a begin-event labelled $L$, and
then behaves as $P$.

The process $\End{L}P$ autonomously performs an end-event labelled $L$, and
then behaves as $P$.

The process $\Cast{M}{x \ty T}P$ binds the message $M$ to the variable $x$ of
type $T$, and then runs $P$.  In well-typed programs, $M$ is a challenge
of type $\annotate{\ell}{\Challenge{\es}}$, and $T$ is a response type
$\annotate{\ell}{\Challenge{\fs}}$.  This is the only way to populate a
response type.

The process $\Witness{M\ty T}P$ simply runs $P$, but is well-typed only if $M$
has the type $T$.  This is the only way to justify a $\trust{M\ty T}$ effect.

The process $\Trust{M}{x\ty T}P$ binds the message $M$ to the variable $x$ of
type $T$, and then runs $P$.  In well-typed programs, this cast is justified
by a previous run of a $\Witness{M\ty T}Q$ process.

Next, we recall the notions of process safety, opponents, and robust safety
introduced in Section~\ref{sec:spi-calculus-semantics}.  The notion of a run
of a process can be formalized by an operational semantics.

\begin{display}{Safety:}
  A process $P$ is \emph{safe} if and only if \\\quad
  for every run of the process and for every $L$,\\\qquad
  there is a distinct $\beginn{L}$ event for every
  $\endd{L}$ event.
\end{display}

\begin{display}{Opponents and Robust Safety:}
A process $P$ is \emph{assertion-free} if and only if\\\quad
   it contains no begin- or end-assertions.\\
A process $P$ is \emph{untyped} if and only if \\\quad
  the only type occurring in $P$ is $\Un$.\\
An \emph{opponent} $O$ is an assertion-free untyped process.\\
A process $P$ is \emph{robustly safe} if and only if \\\quad
  $P \parpop O$ is safe for every opponent $O$.
\end{display}

Our problem, then, is to show that processes representing protocols are
robustly safe.  We appeal to a type and effect system to establish robust
safety (but not to define it).  The system involves the following type
judgments.
\begin{renewcommand}{\ratio}{.3}
\begin{display}{Judgments $E \vdash \J$:}
\clause{E \vdash \diamond}{good environment}\\
\clause{E \vdash \es}{good effect $\es$}\\
\clause{E \vdash T}{good type $T$}\\
\clause{E \vdash M : T}{good message $M$ of type $T$}\\
\clause{E \vdash P : \es}{good process $P$ with effect $\es$}
\end{display}
\end{renewcommand}

We omit the rules defining these judgments, which can be found in
\cite{GJ02:TypesAndEffectsForAsymmetricCryptographicProtocols}; our previous
informal explanation of types should give some intuitions.

We made two additions to the language as defined in
\cite{GJ02:TypesAndEffectsForAsymmetricCryptographicProtocols}, namely
the empty record type $()$ (and corresponding empty record message
$()$), and the conditional form $\SpiIf M=N \SpiThen P \SpiElse
Q$. The empty record type can be handled by simply extending the
typing rules for records to the case where there are no elements. The
main consequence of this is that the type $()$ will be isomorphic to
the type $\Un$, by the extended subtyping rules. The extension of spi
to handle the conditional is similarly straightforward, except that we
need to actually add a transition rule to the operational semantics,
and a new typing rule to propagate the effects. For completeness, we
describe the additions here, with the understanding that they rely on
terminology defined and explained in
\cite{GJ02:TypesAndEffectsForAsymmetricCryptographicProtocols}:
\ifLong
\begin{renewcommand}{\ratio}{.65}
\begin{display}{Extensions to Spi for the Conditional:}
\clause{[\SpiIf M=N \SpiThen P_{\mathit{true}} \SpiElse
P_{\mathit{false}}]+\As \to [P_{M=N}]+\As}{transition rule}\\[\GAP]
\clause{
\typerule
  {(Proc If)}
  {\Judge{E}{M:\Top} \quad \Judge{E}{N:\Top} \quad
   \Judge{E}{P:\es} \quad \Judge{E}{Q:\fs}}
  {\Judge{E}{\SpiIf M=N \SpiThen P \SpiElse Q:\es\vee\fs}}}{typing rule}
\end{display}
\end{renewcommand}
\else
\begin{renewcommand}{\ratio}{.6}
\begin{display}{Extensions to Spi for the Conditional:}
\clause{\begin{prog}
        [\SpiIf M=N \SpiThen P_{\mathit{true}} \SpiElse
          P_{\mathit{false}}]+\As \to \\
        \quad [P_{M=N}]+\As
        \end{prog}}{transition rule}\\[\GAP]
\clause{
\typerule
  {(Proc If)}
  {\begin{prog}
   \Judge{E}{M:\Top} \quad \Judge{E}{N:\Top} \\
   \Judge{E}{P:\es} \quad \Judge{E}{Q:\fs}
   \end{prog}}
  {\Judge{E}{\SpiIf M=N \SpiThen P \SpiElse Q:\es\vee\fs}}}{typing rule}
\end{display}
\end{renewcommand}
\fi

The type and effect system can guarantee the robust safety of a process,
according to the following theorem
\cite{GJ02:TypesAndEffectsForAsymmetricCryptographicProtocols}:
\begin{theorem}[Robust Safety]
\label{Thm:robust-safety}
  If $x_1\ty\Un,\ldots,x_n\ty\Un \vdash P : [\,]$
  then $P$ is robustly safe.
\end{theorem}

\ifLong

 \ifJournal

\section{Proofs}\label{app:proofs}

\subsection{Proof of Theorem~\ref{t:robust-safety-translation}}\label{app:proof-one}

A consequence of the types translation for our calculus is that
$\qq{A}$ is isomorphic to $\Un$ for all types $A$. Formally,
\begin{lemma}\label{lemma:iso-un}
$\Judge{\qq{A}\tyq\Un}$ for all types $A$.
\end{lemma}
In practice, this means that we can replace $\qq{A}$ by $\Un$ in type
derivations, and vice versa. 

Some general remarks on typing are in order. A consequence of
Lemma~\ref{lemma:iso-un}, as well as our general use of types, reveals
that we rely on typing exclusively to show security properties, not to
establish standard safety results. For instance, we do not use types
to ensure that the type of the arguments supplied at method invocation
match the type of the parameters to the method. Indeed, the only
channel type in our translation has itself type $\Un$.

In order to prove Theorem~\ref{t:robust-safety-translation}, we first
establish some lemmas.

\begin{lemma}~\label{lemma:type-preservation-2}
\begin{enumerate}
\item
If $E \vdash v : A$ then $E_{\mathit{prin}},\qq{E} \vdash \qq{v} : \qq{A}$.
\item
If $E \vdash a : A$ and $\Judge{E_0,\qq{E}}{p:\Princ}$ and $k \notin \dom(E_0,\qq{E})$
then: \[E_0, \qq{E}, k \ty \Un \vdash \qq{a}^p_k : [\,]\]
\item
If $c\in\Class$ and $\ell\in\dom(\methods(c))$
then $E_0 \vdash I_{\mathit{class}}(c,\ell) : [\,]$.
\item
If $w\in\WebService$ then $E_0 \vdash I_{\mathit{ws}}(w) : [\,]$.
\end{enumerate}
\end{lemma}
\begin{proof}
\begin{enumerate}

\item We prove this by induction on the height of the type derivation
for $\Judge{E}{v:A}$:
\begin{itemize}

\citem{$v=x$} Since $\Judge{E}{x:A}$, we must have $x\ty A\in E$. By
definition of the translation for environment, $x\ty\qq{A}\in \qq{E}$,
hence $\Judge{E_{\mathit{prin}},\qq{E}}{x:\qq{A}}$,
as required.

\citem{$v=\Null$} We have $\Judge{E}{\Null:c}$. Since
$\qq{c}=\Union(\Null(),c(\Un))$ and $\qq{\Null}=\Null()$, we have
$\Judge{E_{\mathit{prin}},\qq{E}}{\Null():\Union(\Null(\Un),$ $c(\Un))}$,
as required.

\citem{$v=\New\:c(v_1,\ldots,v_n)$} Since $\Judge{E}{v:A}$, where
$A=c$, we have $\fields(c)=f_i\mapsto A_i\For{i\in 1..n}$, and
$\Judge{E}{v_i:A_i}$ for all $i\in 1..n$. Let
$E'=E_{\mathit{prin}},\qq{E}$.  By induction
hypothesis, $\Judge{E'}{\qq{v_i}:\qq{A_i}}$ for all $i\in 1..n$. We
can now derive: 
\[
\infer{\Judge{E'}{c(\qq{v_1},\ldots,\qq{v_n}):\Union(\Null(\Un),c(\Un))}}
      {\infer{\Judge{E'}{\qq{v_1},\ldots,\qq{v_n}:\Un}}
             {\infer{\Judge{E'}{\qq{v_1},\ldots,\qq{v_n}:(\Un,\ldots,\Un)}}
                    {\infer{\Judge{E'}{(\qq{v_1},\ldots,\qq{v_n}):(\qq{A_1},\ldots,\qq{A_n})}}
                           {\Judge{E'}{\qq{v_i}:\qq{A_i}} \quad
                            \forall i\in 1..n}}}}
\]
 
\citem{$v=p$} Since $\Judge{E}{p:A}$ (with $A=\Id$), we have
$p\in\Prin$, hence $p\ty\Princ\in E_{\mathit{prin}}$. Since
$\qq{\Id}=\Princ$, we have
$\Judge{E_{\mathit{prin}},\qq{E}}{p:\Princ}$, as
required. 

\end{itemize}

\item Again, we proceed by induction on the height of the type
derivation for $\Judge{E}{a:A}$. 
\begin{itemize}

\citem{$a=v$} We can apply the result of part (1). Since
$\Judge{E}{v:A}$, then
$\Judge{E_{\mathit{prin}},\qq{E}}{\qq{v}:\qq{A}}$.
We can derive: 
\[
\infer{\Judge{E_0,\qq{E},k\ty\Un}{\Snd{k}{\qq{v}}:[\,]}}
      {\Judge{E_0,\qq{E},k\ty\Un}{k:\Un} &
       \infer{\Judge{E_0,\qq{E}}{\qq{v}:\Un}}
             {\Judge{E_0,\qq{E}}{\qq{v}:\qq{A}}}}
\]

\citem{$a=\Let{x}{a_0}\In{b}$} We have $\Judge{E}{a_0:B}$ for some
$B$, and $\Judge{E,x\ty B}{b:A}$. Applying the induction hypothesis,
we derive $\Judge{E_0,\qq{E},k'\ty\Un}{\TransB{a_0}{p}{k'}:[\,]}$ and
$\Judge{E_0,\qq{E},x\ty\qq{B},k\ty\Un}{\TransB{b}{p}{k}:[\,]}$. Let
$E'=E_0,\qq{E},k\ty\Un$. We can now derive:
\[
\infer{\Judge{E'}
             {\Res{k'\ty\Un}(\TransB{a}{p}{k'}\parpop\Rcv{k'}{x\ty\Un}\TransB{b}{p}{k}):[\,]}}
      {\infer{\Judge{E',k'\ty\Un}{\TransB{a}{p}{k'}\parpop\Rcv{k'}{x\ty\Un}\TransB{b}{p}{k}:[\,]}}
             {\Judge{E',k'\ty\Un}{\TransB{a}{p}{k'}:[\,]} &
              \infer{\Judge{E',k'\ty\Un}{\Rcv{k'}{x\ty\Un}\TransB{b}{p}{k}:[\,]}}
                    {\Judge{E',k'\ty\Un}{k':\Un} & 
                     \infer{\Judge{E',k'\ty\Un,x\ty\Un}{\TransB{b}{p}{k}:[\,]}}
                           {\Judge{E',k'\ty\Un,x\ty\qq{B}}{\TransB{b}{p}{k}:[\,]}}}}}
\]

\citem{$a=\If u=v\Then a_0 \Else a_1$} We have $\Judge{E}{u:B}$,
$\Judge{E}{v:B}$, $\Judge{E}{a_0:A}$, and $\Judge{E}{a_1:A}$. Applying
the induction hypothesis, we derive
$\Judge{E_0,\qq{E},k\ty\Un}{\TransB{a_0}{p}{k}:[\,]}$ and
$\Judge{E_0,\qq{E},k\ty\Un}{\TransB{a_0}{p}{k}:[\,]}$. By (1), we also
have $\Judge{E_0,\qq{E}}{\qq{u}:\qq{B}}$ and
$\Judge{E_0,\qq{E}}{\qq{v}:\qq{B}}$. This gives us
$\Judge{E_0,\qq{E},k\ty\Un}{\SpiIf \qq{u}=\qq{v}\SpiThen
\TransB{a_0}{p}{k} \SpiElse \TransB{a_1}{p}{k}:[\,]}$, as required. 

\citem{$a=v.f_j$} We have $\Judge{E}{v.f_j:A_j}$, where
$\Judge{E}{v:c}$ and $\fields(c)=f_i\mapsto A_i\For{i\in 1..n}$. By
(1), $\Judge{E_0,\qq{E}}{\qq{v}:\qq{c}}$. Let
$E'=E_0,\qq{E},k\ty\Un$. First, let us derive that 
$E',y:\Un \vdash \SplitPoly{y}{(x_1\ty\qq{A_1},\ldots,x_n\ty\qq{A_n})}$
$\Snd{k}{x_j} : [\,]$.
Let $E'' = x_1\ty\qq{A_1},\ldots,x_n\ty\qq{A_n}$.
(We trim environments where possible to reduce clutter.)
\[
\infer{\Judge{E',y:\Un}{\SplitPoly{y}{(x_1\ty\qq{A_1},\ldots,x_n\ty\qq{A_n})}\Snd{k}{x_j}:[\,]}}
      {\Judge{E',y:\Un}{y:\Un} & 
       \infer{\Judge{E',y:\Un,E''}{\Snd{k}{x_j}:[\,]}}
             {\Judge{E'}{k:\Un} & 
              \infer{\Judge{E',y:\Un,E''}{x_j:\Un}}
                    {\Judge{E',y:\Un,E''}{x_j:\qq{A_j}}}}}
\]

We can now derive:
\[
\infer{\Judge{E'}{\Case{\qq{v}}
         \begin{prog}
         \IsTag{\Null}{y\ty\Un}{\Stop}\\
         \IsTag{c}{y}{\SplitPoly{y}{(x_1\ty\qq{A_1},\ldots,x_n\ty\qq{A_n})}
                      \Snd{k}{x_j}}:[\,]
    \end{prog}}}
      {\begin{prog}
         \Judge{E'}{\qq{v}:\Union(\Null(\Un),c(\Un))} \\
         \Judge{E',y:\Un}{\Stop:[\,]}\\
         \Judge{E',y:\Un}{\SplitPoly{y}{(x_1\ty\qq{A_1},\ldots,x_n\ty\qq{A_n})}\Snd{k}{x_j}:[\,]}
       \end{prog}}
\]

\citem{$a=v.\ell_j(u_1,\ldots,u_m)$} We have
$\Judge{E}{v.\ell_j(u_1,\ldots,u_m):B}$, where $\Judge{E}{v:c}$,
$\methods(c)=\ell_i\mapsto(\sig_i,b_i)\For{i\in 1..n}$,
$\sig_j=B(A_1\:x_1,$ $\ldots,$ $A_m\:x_m)$, and $\Judge{E}{u_k:A_k}$ for all
$k\in 1..m$. By (1), $\Judge{E_0,\qq{E}}{\qq{u_k}:\qq{A_k}}$ for all
$k\in 1..m$. Let $E'=E_0,\qq{E},k\ty\Un$. First, let us derive that 
$\Judge{E',y\ty\Un}{\Snd{c\_\ell}{(p,\TransV{v},\TransV{u_1},\ldots,\TransV{u_n},k)}:[\,]}$.
\[
\infer{\Judge{E',y\ty\Un}{\Snd{c\_\ell}{(p,\TransV{v},\TransV{u_1},\ldots,\TransV{u_n},k)}:[\,]}}
      {\Judge{E',y\ty\Un}{c\_\ell:\Un} & 
       \Judge{E',y\ty\Un}{(p,\TransV{v},\TransV{u_1},\ldots,\TransV{u_n},k):\Un}}
\]

We can derive:
\[
\infer{\Judge{E'}{\Case{\TransV{v}}
                  \begin{prog}
                  \IsTag{\Null}{y\ty\Un}{\Stop} \\
                  \IsTag{c}{y}{\Snd{c\_\ell}{(p,\TransV{v},\TransV{u_1},\ldots,\TransV{u_n},k)}}:[\,]
        \end{prog}}}
      {\begin{prog}
         \Judge{E'}{\qq{v}:\Union(\Null(\Un),c(\Un))} \\
         \Judge{E',y\ty\Un}{\Stop:[\,]}\\
         \Judge{E',y\ty\Un}{\Snd{c\_\ell}{(p,\TransV{v},\TransV{u_1},\ldots,\TransV{u_n},k)}:[\,]}
       \end{prog}}
\]

\citem{$a=w\mo{:}\ell_j(u_1,\ldots,u_m)$} We have
$\Judge{E}{w\mo{:}\ell_j(u_1,\ldots,u_m):B}$, where $\class(w)=c$,
$\owner(w)=q$, 
$\methods(c)=\ell_i\mapsto(\sig_i,b_i)\For{i\in 1..n}$, 
$\sig_j=B(A_1\:x_1,\ldots,A_m\:x_m)$, and $\Judge{E}{u_k:A_k}$ for all
$k\in 1..m$. By (1), $\Judge{E_0,\qq{E}}{\qq{u_k}:\qq{A_k}}$ for all
$k\in 1..m$. Rather than giving the full type derivation for the
translation of a web service call, we outline the derivation of
effects:

$\begin{prog}\quad
             \Res{k_1 \ty \Un, k_2 \ty \Un, t\ty\Un,
                  n_p \ty \annotate{\public}{\Challenge{[\,]}}} \\ \quad
             \mbox{// Effect: $[\check{\public}{n_p}]$}\\ \quad
             \Begin{\request(p,q,w,\ell(\qq{u_1},\ldots,\qq{u_n}),t)} \\ \quad
             \mbox{// Effect: $[\check{\public}{n_p},\endd{\request(p,q,w,\ell(\qq{u_1},\ldots,\qq{u_n}),t)}]$}\\ \quad
              \Snd{w}{(\request(\getnonce()),k_1)}; \\ \quad
             \mbox{// Effect: $[\check{\public}{n_p},\endd{\request(p,q,w,\ell(\qq{u_1},\ldots,\qq{u_n}),t)}]$}\\ \quad
              \Rcv{k_1}{\response(\getnonce(n_q \ty \Un))} \\ \quad
              \mbox{// Effect: $[\check{\public}{n_p},\endd{\request(p,q,w,\ell(\qq{u_1},\ldots,\qq{u_n}),t)}]$}\\ \quad
              \Cast{n_q}{n'_q \ty
                \annotate{\public}{\Response{[\endd{\request(p,q,w,\ell(\qq{u_1},\ldots,\qq{u_n}),t)}]}}} \\ \quad
              \mbox{// Effect: $[\check{\public}{n_p}]$}\\ \quad
              \Snd{w}{(p,\sencr{\request(w,\ell(\qq{u_1},\ldots,\qq{u_n}),t,n'_q)}{K_{pq}},n_p,k_2)}; \\ \quad
              \mbox{// Effect: $[\check{\public}{n_p}]$}\\ \quad
              \Rcv{k_2}{q'\ty\Un,\bdy \ty \Un}\\ \quad
              \mbox{// Effect: $[\check{\public}{n_p}]$}\\ \quad
              \Decrypt{\bdy}{\response(\mathit{plain})}{K_{pq}}\\ \quad
              \mbox{// Effect: $[\check{\public}{n_p}]$}\\ \quad
              \Match{\mathit{plain}}{w}{\mathit{rest} \ty \\ \qquad
                (r \ty \mathit{Res}(w), t'\ty\Un,
                 \annotate{\public}{\Response{[\endd{\response(p,q,w,r,t')}]}})}\\ \quad
              \mbox{// Effect: $[\check{\public}{n_p}]$}\\ \quad
              \Split{\mathit{rest}}{r\ty\mathit{Res}(w)}{rest'\ty \\ \qquad
                (t'\ty\Un,n_p'\ty\annotate{\public}{\Response{[\endd{\response(p,q,w,r,t')}]}})}\\ \quad
              \mbox{// Effect: $[\check{\public}{n_p}]$}\\ \quad
         \Match{\mathit{rest'}}{t}{n_p'\ty\annotate{\public}{\Response{[\endd{\response(p,q,w,r,t)}]}}}\\ \quad
              \mbox{// Effect: $[\check{\public}{n_p}]$}\\ \quad
         \CheckNonce{n_p}{n_p'}\\ \quad
              \mbox{// Effect: $[\endd{\response(p,q,w,r,t)}]$}\\ \quad
              \End{\response(p,q,w,r,t)}\\ \quad
              \mbox{// Effect: $[\,]$}\\ \quad
              \Case{r} \IsTag{\ell}{x}{\Snd{k}{x}}\\ \quad
              \mbox{// Effect: $[\,]$}
  \end{prog}$

\end{itemize}

\item Recall that we assume that method bodies are well-typed, that
is, we assume for $c,\ell_j$ with
$\methods(c)=\ell_i\mapsto(\sig_i,b_i)$ and
$\sig_j=B(A_1\:a_1,\ldots,A_m\:x_m)$, that $\Judge{\this\ty c,x_1\ty
A_1,\ldots,x_m\ty A_m}{b_j:B}$. By clause (2) above, this means that
$\Judge{E_0,\this\ty\qq{c},x_1\ty \qq{A_1},\ldots,x_m\ty
\qq{A_m},k\ty\Un}{\TransB{b_j}{p}{k}:[\,]}$. Applying
Lemma~\ref{lemma:iso-un}, we derive $\Judge{E_0,\this\ty\Un,x_1\ty
\Un,\ldots,x_m\ty\Un,k\ty\Un}{\TransB{b_j}{p}{k}:[\,]}$. We can now easily derive the following:
\[
\infer{\Judge{E_0}{I_{\mathit{class}}(c,\ell) : [\,]}}
      {\infer{\Judge{E_0}{\Repl{\Rcv{c\_\ell}{z}
                          \SplitPoly{z}{(p\ty\Princ,\this\ty\Un,x_1\ty\Un,\ldots,x_n\ty\Un,k\ty\Un)}
                          \TransB{b_j}{p}{k}:[\,]}}}
             {\Judge{E_0}{c\_\ell:\Un} & 
              \infer{\Judge{E_0,z\ty\Un}{\SplitPoly{z}{(p\ty\Princ,\this\ty\Un,x_1\ty\Un,\ldots,x_n\ty\Un,k\ty\Un)}
                                         \TransB{b_j}{p}{k}:[\,]}}
                    {\begin{prog}
                     \Judge{E_0,z\ty\Un}{z:(\Princ,\Un,\ldots,\Un)} \\
                     \Judge{E_0,z\ty\Un,p\ty\Princ,\this\ty\Un,x_1\ty\Un,\ldots,x_n\ty\Un,k\ty\Un}
                           {\TransB{b_j}{p}{k}:[\,]}
           \end{prog}}}}
\]

\item Let $w\in\WebService$, with $\owner(w)=q$. First, note that the following
derivation is admissible:
\[\infer{\Judge{E_0,E}{\LetCall{r \ty \mathit{Res}(w)}{w}{p,a}P:\es}}
       {\Judge{E_0,E}{p:\Prin} &
        \Judge{E_0,E}{a:\mathit{Req}(w)} &
        \Judge{E_0,E,r\ty\mathit{Res}(w)}{P:\es}}\]
(The proof is a straightforward, if longish, type derivation.)
Rather than giving the full type derivation for the
implementation of web service $w$, we outline the derivation of
effects:

$\begin{prog}\quad
     \Repl\Rcv{w}{\bdy \ty \Un,k_1 \ty \Un}\\ \quad
     \mbox{// Effect: $[\,]$}\\ \quad

         \Case{\bdy} \IsTag{\request}{\getnonce()}{}\\ \quad
         \mbox{// Effect: $[\,]$}\\  \quad
         \Res{n_q \ty \annotate{\public}{\Challenge{[\,]}}}\\ \quad
         \mbox{// Effect: $[\check{\public}{n_q}]$}\\  \quad
         \Snd{k_1}{(\response(\getnonce(n_q)))};\\ \quad
         \mbox{// Effect: $[\check{\public}{n_q}]$}\\  \quad
         \Rcv{w}{p'\ty\Un,\mathit{cipher}\ty\Un,n_p \ty \Un, k_2 \ty \Un}\\ \quad
         \mbox{// Effect: $[\check{\public}{n_q}]$}\\  \quad
         \textstyle\prod_{p \in \Prin}\SpiIf p=p' \SpiThen \\ \quad
         \mbox{// Effect: $[\check{\public}{n_q}]$}\\  \quad
         \Decrypt{\mathit{cipher}}{\request(\mathit{plain})}{K_{pq}}\\ \quad
         \mbox{// Effect: $[\check{\public}{n_q}]$}\\  \quad
         \Match{\mathit{plain}}{w}{\mathit{rest} \ty
                (\begin{prog}
                 a \ty \mathit{Req}(w),t\ty\Un,\\
                  \annotate{\public}{\Response{[\endd{\request(p,q,w,a,t)}]}})}
                 \end{prog}\\ \quad
         \mbox{// Effect: $[\check{\public}{n_q}]$}
  \end{prog}$

$\begin{prog}\quad
    \Split{\mathit{rest}}{a\ty\mathit{Req}(w)}{t\ty\Un,n_q'\ty\annotate{\public}{\Response{[\endd{\request(p,q,w,a,t)}]}}}\\ \quad
         \mbox{// Effect: $[\check{\public}{n_q}]$}\\  \quad
         \CheckNonce{n_q}{n_q'}\\ \quad
         \mbox{// Effect: $[\endd{\request(p,q,w,a,t)}]$}\\ \quad
    \End{\request(p,q,w,a,t)}\\ \quad
         \mbox{// Effect: $[\,]$}\\ \quad
         \LetCall{r \ty \mathit{Res}(w)}{w}{p,a}\\ \quad
         \mbox{// Effect: $[\,]$}\\ \quad
         \Begin{\response(p,q,w,r,t)} \\ \quad
         \mbox{// Effect: $[\endd{\response(p,q,w,r,t)}]$}\\ \quad
         \Cast{n_p}{n'_p \ty
            \annotate{\public}{\Response{[\endd{\response(p,q,w,r,t)}]}}}\\ \quad
         \mbox{// Effect: $[\,]$}\\ \quad
         \Snd{k_2}{(q,\sencr{\response(w,r,t,n'_p)}{K_{pq}})}\\ \quad
         \mbox{// Effect: $[\,]$}
  \end{prog}$

\end{enumerate}
\ifJournal \vspace{1ex} \mbox{} \hfill \fi
\end{proof}

\begin{lemma}\label{lemma:type-preservation-3}
If $\emptyset \vdash a : A$ and $p \in \Prin$ and $k \notin \dom(E_0)$
then: \[E_{\mathit{ws}}, E_{\mathit{prin}} \vdash
  \Res{E_{\mathit{class}}, E_{\mathit{keys}}}
    (I_{\mathit{class}} \parpop
     I_{\mathit{ws}} \parpop
     \Res{k \ty \Un} \qq{a}^p_k) : [\,]\]
\end{lemma}
\begin{proof}
This is a corollary of Lemma~\ref{lemma:type-preservation-2}.
Specifically, we can derive:  
\[
\infer{\Judge{E_{\mathit{ws}}, E_{\mathit{prin}}}
             {\Res{E_{\mathit{class}}, E_{\mathit{keys}}}
              (I_{\mathit{class}} \parpop
               I_{\mathit{ws}} \parpop
               \Res{k \ty \Un} \qq{a}^p_k) : [\,]}}
      {\infer{\Judge{E_0}{I_{\mathit{class}} \parpop
                          I_{\mathit{ws}} \parpop
                          \Res{k \ty \Un} \qq{a}^p_k : [\,]}}
             {\infer{\Judge{E_0}{I_{\mathit{class}}:[\,]}}
                    {\Judge{E_0}{I_{\mathit{class}}(c,\ell):[\,]}\For{(c,\ell)\in\ClassMethods}} &
              \infer{\Judge{E_0}{I_{\mathit{ws}}:[\,]}}
                    {\Judge{E_0}{I_{\mathit{ws}}(w):[\,]\For{w\in\WebService}}} &
              \infer{\Judge{E_0}{\Res{k \ty \Un} \qq{a}^p_k : [\,]}}
                    {\Judge{E_0,k\ty\Un}{\qq{a}^p_k:[\,]}}}}
\]
\ifJournal \vspace{1ex} \mbox{} \hfill \fi
\end{proof}

We can now prove Theorem~\ref{t:robust-safety-translation}.
\begin{theorem*}
If $\Judge{\emptyset}{a : A}$ and $p \in \Prin$ and $k \notin \dom(E_0)$ then the system
\[\Res{E_{\mathit{class}}, E_{\mathit{keys}}}
    (I_{\mathit{class}} \parpop
     I_{\mathit{ws}} \parpop
     \Res{k \ty \Un} \qq{a}^p_k)\]
is robustly safe.
\end{theorem*}
\begin{proof}
By Lemma~\ref{lemma:type-preservation-3},
    \[\Judge{E_{\mathit{ws}},E_{\mathit{prin}}}
    {\Res{E_{\mathit{class}}, E_{\mathit{keys}}}  
                 (I_{\mathit{class}} \parpop
                 I_{\mathit{ws}} \parpop
                 \Res{k \ty \Un} \qq{a}^p_k):[\,]}.\]
 Robust safety of the system follows by Theorem~\ref{Thm:robust-safety}.   
\hfill
\end{proof}

\subsection{Proof of Theorem~\ref{t:robust-safety-asymm-one}}\label{app:proof-two}

\begin{theorem*}
If $\Judge{\emptyset}{a : A}$ and $p \in \Prin$ and $k \notin \dom(E_0)$ then the system
\[\Res{E_{\mathit{class}}, E_{\mathit{keys}}}
    (I_{\mathit{net}} \parpop I_{\mathit{class}} \parpop
     I_{\mathit{ws}} \parpop
     \Res{k \ty \Un} \qq{a}^p_k)\]
is robustly safe.
\end{theorem*}

\begin{proof}
Rather than giving a full proof, we point out the parts of the proof
of Theorem~\ref{t:robust-safety-translation} that need to be
updated. Essentially, we need to show that the new semantics for web
method invocations is effect-free, and similarly for the new
implementation of web services. These occur in the proof of
Lemma~\ref{lemma:type-preservation-2}, part (2) and (4). 

As we did in Lemma~\ref{lemma:type-preservation-2}, rather than giving
the full type derivation for the translation of a web service call, we
outline the derivation of effects:

\vspace{1ex}
$\begin{prog}
             \Res{k_1 \ty \Un, k_2 \ty \Un, t\ty\Un,
                  n_p \ty \annotate{\public}{\Challenge{[\,]}}} \\ 
             \mbox{// Effect: $[\check{\public}{n_p}]$}\\ 
             \Begin{\request(p,q,w,\ell(\qq{u_1},\ldots,\qq{u_n}),t)} \\ 
             \mbox{// Effect: $[\check{\public}{n_p},\endd{\request(p,q,w,\ell(\qq{u_1},\ldots,\qq{u_n}),t)}]$}\\ 
              \Snd{w}{(\CertVK{p},n_p,\request(\getnonce()),k_1)}; \\ 
             \mbox{// Effect: $[\check{\public}{n_p},\endd{\request(p,q,w,\ell(\qq{u_1},\ldots,\qq{u_n}),t)}]$}\\ 
              \Rcv{k_1}{c\ty\Un,\response(\getnonce(n_q \ty \Un))} \\ 
             \mbox{// Effect: $[\check{\public}{n_p},\endd{\request(p,q,w,\ell(\qq{u_1},\ldots,\qq{u_n}),t)}]$}\\ 
              \pDecrypt{c}{\mathit{cert}\ty(q'\ty\Un,\annotate{\Dec}{\Key{\AuthMsg{q'}}})}{\VK{CA}}\\ 
             \mbox{// Effect: $[\check{\public}{n_p},\endd{\request(p,q,w,\ell(\qq{u_1},\ldots,\qq{u_n}),t)}]$}\\ 
              \Match{\mathit{cert}}{q}{\mathit{vkq}\ty\annotate{\Dec}{\Key{\AuthMsg{q}}}}\\ 
             \mbox{// Effect: $[\check{\public}{n_p},\endd{\request(p,q,w,\ell(\qq{u_1},\ldots,\qq{u_n}),t)}]$}\\ 
              \Cast{n_q}{n'_q \ty
                \annotate{\public}{\Response{[\endd{\request(p,q,w,\ell(\qq{u_1},\ldots,\qq{u_n}),t)}]}}} \\ 
             \mbox{// Effect: $[\check{\public}{n_p}]$}\\ 
              \Snd{w}{(p,\asencr{\request(w,\ell(\qq{u_1},\ldots,\qq{u_n}),t,q,n'_q)}{\SK{p}},k_2)}; \\ 
             \mbox{// Effect: $[\check{\public}{n_p}]$}\\ 
              \Rcv{k_2}{q''\ty\Un,\bdy \ty \Un}\\ 
             \mbox{// Effect: $[\check{\public}{n_p}]$}\\ 
              \begin{prog}
                \pDecrypt{\bdy}{\response(\mathit{plain}\ty(w'\ty\Un,r\ty\Un,t'\ty\Un,p'\ty\Un,\\
                        \qquad\qquad\quad\annotate{\public}{\Response{[\endd{\response(p',q,w',r,t')}]}}))}{\mathit{vkq}}
         \end{prog}\\ 
             \mbox{// Effect: $[\check{\public}{n_p}]$}\\ 
              \Match{\mathit{plain}}{w}{\mathit{rest} \ty
                (\begin{prog}
                 r \ty \mathit{Res}(w), t'\ty\Un, p'\ty\Un,\\
                 \annotate{\public}{\Response{[\endd{\response(p',q,w,r,t')}]}})}
                 \end{prog}\\ 
             \mbox{// Effect: $[\check{\public}{n_p}]$}\\ 
         \Split{\mathit{rest}}{r\ty\mathit{Res}(w)}{\mathit{rest'}\ty
                (\begin{prog}
                 t'\ty\Un,p'\ty\Un,\\
                 \annotate{\public}{\Response{[\endd{\response(p',q,w,r,t')}]}})}
       \end{prog}\\ 
             \mbox{// Effect: $[\check{\public}{n_p}]$}\\ 
              \Match{\mathit{rest'}}{t}{\mathit{rest''}\ty(p'\ty\Un,\annotate{\public}{\Response{[\endd{\response(p',q,w,r,t)}]}})}\\ 
             \mbox{// Effect: $[\check{\public}{n_p}]$}\\ 
              \Match{\mathit{rest''}}{p}{n_p'\ty\annotate{\public}{\Response{[\endd{\response(p,q,w,r,t)}]}}}\\ 
             \mbox{// Effect: $[\check{\public}{n_p}]$}\\ 
         \CheckNonce{n_p}{n_p'}\\ 
             \mbox{// Effect: $[\endd{\response(p,q,w,r,t)}]$}\\ 
              \End{\response(p,q,w,r,t)}\\ 
             \mbox{// Effect: $[\,]$}\\ 
              \Case{r} \IsTag{\ell}{x}{\Snd{k}{x}} \\ 
             \mbox{// Effect: $[\,]$}
       \end{prog}$

\vspace{1ex}
For the new implementation of web service $w$, rather than giving the
full type derivation, we outline the derivation of effects:
\vspace{1ex}

$\begin{prog}
     \Repl\Rcv{w}{c\ty\Un, n_p\ty\Un,\bdy \ty \Un,k_1 \ty \Un}\\
     \mbox{// Effect: $[\,]$}\\
         \Case{\bdy} \IsTag{\request}{\getnonce()}{}\\
         \mbox{// Effect: $[\,]$}\\
         \pDecrypt{c}{p\ty\Un,\mathit{vkp}\ty\annotate{\Dec}{\Key{\AuthMsg{p}}}}{\VK{CA}}\\ 
         \mbox{// Effect: $[\,]$}\\
         \Res{n_q \ty \annotate{\public}{\Challenge{[\,]}}}\\
         \mbox{// Effect: $[\check{\public}{n_q}]$}\\
         \Snd{k_1}{(\CertVK{q},\response(\getnonce(n_q)))};\\
         \mbox{// Effect: $[\check{\public}{n_q}]$}\\
         \Rcv{w}{p'\ty\Un,\mathit{cipher}\ty\Un,k_2 \ty \Un}\\
         \mbox{// Effect: $[\check{\public}{n_q}]$}\\
         \SpiIf p=p' \SpiThen \\
         \mbox{// Effect: $[\check{\public}{n_q}]$}\\
         \begin{prog}
             \pDecrypt{\mathit{cipher}}{\request(\mathit{plain}\ty(w\ty\Un,a\ty\Un,t\ty\Un,q'\ty\Un,\\
                   \qquad\qquad\quad\annotate{\public}{\Response{[\endd{\request(p,q',w,a,t)}]}}))}{\mathit{vkp}}
    \end{prog}\\
         \mbox{// Effect: $[\check{\public}{n_q}]$}\\
         \Match{\mathit{plain}}{w}{\mathit{rest} \ty
                (\begin{prog}
                 a \ty \mathit{Req}(w),t\ty\Un,q'\ty\Un,\\
                  \annotate{\public}{\Response{[\endd{\request(p,q',w,a,t)}]}})}
                 \end{prog}\\
         \mbox{// Effect: $[\check{\public}{n_q}]$}\\
    \Split{\mathit{rest}}{\begin{prog}a\ty\mathit{Req}(w)}{\\t\ty\Un,\mathit{rest'}\ty(q'\ty\Un,\annotate{\public}{\Response{[\endd{\request(p,q',w,a,t)}]}})}\end{prog}\\
         \mbox{// Effect: $[\check{\public}{n_q}]$}\\
         \Match{\mathit{rest'}}{q}{n_q'\ty\annotate{\public}{\Response{[\endd{\request(p,q,w,a,t)}]}}}\\
         \mbox{// Effect: $[\check{\public}{n_q}]$}\\
         \CheckNonce{n_q}{n_q'}\\
         \mbox{// Effect: $[\endd{\request(p,q,w,a,t)}]$}\\
    \End{\request(p,q,w,a,t)}\\
         \mbox{// Effect: $[\,]$}\\ 
         \LetCall{r \ty \mathit{Res}(w)}{w}{p,a}\\
         \mbox{// Effect: $[\,]$}\\ 
         \Begin{\response(p,q,w,r,t)} \\
         \mbox{// Effect: $[\endd{\response(p,q,w,r,t)}]$}\\ 
         \Cast{n_p}{n'_p \ty
            \annotate{\public}{\Response{[\endd{\response(p,q,w,r,t)}]}}}\\
         \mbox{// Effect: $[\,]$}\\ 
         \Snd{k_2}{(q,\asencr{\response(w,r,t,p,n'_p)}{\SK{q}})}\\
         \mbox{// Effect: $[\,]$}
     \end{prog}$

\ifJournal \vspace{1ex} \mbox{} \hfill \fi
\end{proof}

\subsection{Proof of Theorem~\ref{t:robust-safety-asymm-two}}\label{app:proof-three}
\begin{theorem*}
If $\Judge{\emptyset}{a : A}$ and $p \in \Prin$ and $k \notin \dom(E_0)$ then the system
\[\Res{E_{\mathit{class}}, E_{\mathit{keys}}}
    (I_{\mathit{net}} \parpop I_{\mathit{class}} \parpop
     I_{\mathit{ws}} \parpop
     \Res{k \ty \Un} \qq{a}^p_k)\]
is robustly safe.
\end{theorem*}
\begin{proof}
Rather than giving a full proof, we point out the parts of the proof
of Theorem~\ref{t:robust-safety-translation} that need to be
updated. Essentially, we need to show that the new semantics for web
method invocations is effect-free, and similarly for the new
implementation of web services. These occur in the proof of
Lemma~\ref{lemma:type-preservation-2}, part (2) and (4). 

As we did in Lemma~\ref{lemma:type-preservation-2}, rather than giving
the full type derivation for the translation of a web service call, we
outline the derivation of effects:
\vspace{1ex}

$\begin{prog}
     \Res{k_1 \ty \Un, k_2 \ty \Un, t\ty\Un,
          n_p \ty \annotate{\public}{\Challenge{[\,]}}} \\
     \mbox{// Effect: $[\check{\public}{n_p}]$}\\
     \Begin{\request(p,q,w,\ell(\qq{u_1},\ldots,\qq{u_n}),t)} \\ 
     \mbox{// Effect: $[\check{\public}{n_p},\endd{\request(p,q,w,\ell(\qq{u_1},\ldots,\qq{u_n}),t)}]$}\\
     \Snd{w}{(\CertEK{p},\request(\getnonce()),k_1)}; \\
     \mbox{// Effect: $[\check{\public}{n_p},\endd{\request(p,q,w,\ell(\qq{u_1},\ldots,\qq{u_n}),t)}]$}
     \Rcv{k_1}{c\ty\Un,\mathit{cipher}\ty\Un,\response(\getnonce(n_q \ty \Un))} \\ 
     \mbox{// Effect: $[\check{\public}{n_p},\endd{\request(p,q,w,\ell(\qq{u_1},\ldots,\qq{u_n}),t)}]$}\\ 
     \pDecrypt{c}{\mathit{cert}\ty(q'\ty\Un,\annotate{\Enc}{\Key{\AuthEncMsg{q'}}})}{\VK{CA}}\\ 
     \mbox{// Effect: $[\check{\public}{n_p},\endd{\request(p,q,w,\ell(\qq{u_1},\ldots,\qq{u_n}),t)}]$}\\ 
     \Match{\mathit{cert}}{q}{\mathit{ekq}\ty\annotate{\Enc}{\Key{\AuthEncMsg{q}}}}\\ 
     \mbox{// Effect: $[\check{\public}{n_p},\endd{\request(p,q,w,\ell(\qq{u_1},\ldots,\qq{u_n}),t)}]$}\\ 
     \pDecrypt{\mathit{cipher}}{\msgI(q'\ty\Un,n_K\ty\Un)}{\DK{p}}\\ 
     \mbox{// Effect: $[\check{\public}{n_p},\endd{\request(p,q,w,\ell(\qq{u_1},\ldots,\qq{u_n}),t)}]$}\\ 
     \SpiIf q=q' \SpiThen\\ 
     \mbox{// Effect: $[\check{\public}{n_p},\endd{\request(p,q,w,\ell(\qq{u_1},\ldots,\qq{u_n}),t)}]$}\\ 
     \Cast{n_q}{n'_q \ty
         \annotate{\public}{\Response{[\endd{\request(p,q,w,\ell(\qq{u_1},\ldots,\qq{u_n}),t)}]}}} \\ 
     \mbox{// Effect: $[\check{\public}{n_p}]$}\\ 
     \Res{K \ty \SKey{p,q,w}}\\ 
     \mbox{// Effect: $[\check{\public}{n_p}]$}\\ 
     \Witness{K\ty\SKey{p,q,w}}\\ 
     \mbox{// Effect: $[\check{\public}{n_p},\trust{K\ty\SKey{p,q,w}}]$}\\ 
     \Cast{n_K}{n'_K \ty
           \annotate{\private}{\Response{[\trust{K\ty\SKey{p,q,w}}]}}}\\ 
     \mbox{// Effect: $[\check{\public}{n_p}]$}\\ 
     \Snd{w}{(\asencr{\msgII(w,p,t,K,n_K')}{\mathit{ekq}},n_p,
              \sencr{\request(w,\ell(\qq{u_1},\ldots,\qq{u_n}),t,n'_q)}{K},k_2)}; \\ 
     \mbox{// Effect: $[\check{\public}{n_p}]$}\\ 
     \Rcv{k_2}{\bdy \ty \Un}\\ 
     \mbox{// Effect: $[\check{\public}{n_p}]$}\\ 
              \begin{prog}
                 \Decrypt{\bdy}{\response(\mathit{plain}\ty(r\ty\mathit{Res}(w),t'\ty\Un,\\
                   \qquad\qquad\quad\annotate{\public}{\Response{[\endd{\response(p,q,w,r,t')}]}}))}{K}
         \end{prog}\\ 
     \mbox{// Effect: $[\check{\public}{n_p}]$}\\ 
     \Match{\mathit{plain}}{r\ty\mathit{Res}(w)}
            {\mathit{rest}\ty(t'\ty\Un,\annotate{\public}{\Response{[\endd{\response(p,q,w,r,t')}]}})}\\ 
     \mbox{// Effect: $[\check{\public}{n_p}]$}\\ 
     \Match{\mathit{rest}}{t}{n_p'\ty\annotate{\public}{\Response{[\endd{\response(p,q,w,r,t)}]}}}\\ 
     \mbox{// Effect: $[\check{\public}{n_p}]$}\\ 
     \CheckNonce{n_p}{n_p'}\\ 
     \mbox{// Effect: $[\endd{\response(p,q,w,r,t)}]$}\\ 
     \End{\response(p,q,w,r,t)}\\ 
     \mbox{// Effect: $[\,]$}\\ 
     \Case{r} \IsTag{\ell}{x}{\Snd{k}{x}}\\ 
     \mbox{// Effect: $[\,]$}
 \end{prog}$
\vspace{1ex}
For the new implementation of web service $w$, rather than giving the
full type derivation, we outline the derivation of effects:
\vspace{1ex}

$\begin{prog}
     \Repl\Rcv{w}{c\ty\Un,\bdy \ty \Un,k_1 \ty \Un}\\
     \mbox{// Effect: $[\,]$}\\
         \Case{\bdy} \IsTag{\request}{\getnonce()}{}\\
         \mbox{// Effect: $[\,]$}\\
         \pDecrypt{c}{p\ty\Un,\mathit{ekp}\ty\annotate{\Enc}{\Key{\AuthEncMsg{p}}}}{\VK{CA}}\\
         \mbox{// Effect: $[\,]$}\\
         \Res{n_q \ty \annotate{\public}{\Challenge{[\,]}}}\\
         \mbox{// Effect: $[\check{\public}{n_q}]$}\\
         \Res{n_K \ty \annotate{\private}{\Challenge{[\,]}}}\\
         \mbox{// Effect: $[\check{\public}{n_q},\check{\private}{n_K}]$}\\
         \Snd{k_1}{(\CertEK{q},\asencr{\msgI(q,n_K)}{\mathit{ekp}},\response(\getnonce(n_q)))};\\
         \mbox{// Effect: $[\check{\public}{n_q},\check{\private}{n_K}]$}\\
\end{prog}$%

$\begin{prog}
         \Rcv{w}{\mathit{cipher}_1\ty\Un,n_p \ty \Un, \mathit{cipher}_2\ty\Un,k_2 \ty \Un}\\
         \mbox{// Effect: $[\check{\public}{n_q},\check{\private}{n_K}]$}\\
         \begin{prog}
           \pDecrypt{\mathit{cipher}_1\\\qquad}{\msgII(\mathit{plain}_1\ty(w\ty\Un,p'\ty\Un,K\ty\Top,\\
              \qquad\qquad\quad\annotate{\private}{\Response{[\trust{K\ty\SKey{p',q,w}}]}}))}{\DK{q}}
         \end{prog}\\
         \mbox{// Effect: $[\check{\public}{n_q},\check{\private}{n_K}]$}\\
         \Match{\mathit{plain}_1}{w}{\mathit{rest}\ty
           (\begin{prog}
             p'\ty\Un,K\ty\Top,\\
             \annotate{\private}{\Response{[\trust{K\ty\SKey{p',q,w}}]}})}
            \end{prog}\\
         \mbox{// Effect: $[\check{\public}{n_q},\check{\private}{n_K}]$}\\
         \Match{\mathit{rest}}{p}{\mathit{rest'}\ty(K\ty\Top,
                   n'_K\ty\annotate{\private}{\Response{[\trust{K\ty\SKey{p,q,w}}]}})}\\
         \mbox{// Effect: $[\check{\public}{n_q},\check{\private}{n_K}]$}\\
         \Split{\mathit{rest'}}{K\ty\Top}{n'_K\ty\annotate{\private}{\Response{[\trust{K\ty\SKey{p,q,w}}]}}}\\
         \mbox{// Effect: $[\check{\public}{n_q},\check{\private}{n_K}]$}\\
         \CheckNonce{n_K}{n_K'}\\
         \mbox{// Effect: $[\check{\public}{n_q},\trust{K\ty\SKey{p,q,w}}]$}\\
         \Trust{K}{K'\ty\SKey{p,q,w}}\\
         \mbox{// Effect: $[\check{\public}{n_q}]$}\\
         \begin{prog}
            \Decrypt{\mathit{cipher}_2}{\request(\mathit{plain}_2\ty(a\ty\mathit{Req}(w),t\ty\Un,\\
            \qquad\qquad\quad\annotate{\public}{\Response{[\endd{\request(p,q,w,a,t)}]}}))}{K'}
         \end{prog}\\
         \mbox{// Effect: $[\check{\public}{n_q}]$}\\
         \Split{\mathit{plain}_2}{a\ty\mathit{Req}(w)}{t\ty\Un,n_q'\ty\annotate{\public}{\Response{[\endd{\request(p,q,w,a,t)}]}}}\\
         \mbox{// Effect: $[\check{\public}{n_q}]$}\\
         \CheckNonce{n_q}{n_q'}\\
         \mbox{// Effect: $[\endd{\request(p,q,w,a,t)}]$}\\
         \End{\request(p,q,w,a,t)}\\
         \mbox{// Effect: $[\,]$}\\
         \LetCall{r \ty \mathit{Res}(w)}{w}{p,a}\\
         \mbox{// Effect: $[\,]$}\\
         \Begin{\response(p,q,w,r,t)} \\
         \mbox{// Effect: $[\endd{\response(p,q,w,r,t)}]$}\\
         \Cast{n_p}{n'_p \ty
            \annotate{\public}{\Response{[\endd{\response(p,q,w,r,t)}]}}}\\
         \mbox{// Effect: $[\,]$}\\
         \Snd{k_2}{\sencr{\response(r,t,n'_p)}{K'}}\\
         \mbox{// Effect: $[\,]$}
     \end{prog}$

\ifJournal \vspace{1ex} \mbox{} \hfill \fi
\end{proof}

 \else

\section{Formal translation of the Object Calculus}
\label{app:translation}

In this appendix, we give the complete translation of our object
calculus into the spi-calculus. The translation acts on both types and
expressions. The translation presented in
Section~\ref{sec:spi-calculus-semantics} was incomplete, in that it
did not address types.

\subsection{Types Translation}
The translations for types is straightforward. 

\begin{display}{Type Translation:}
\clause{\Princ \triangleq \Un}\\
\clause{\Def{\TransV{\Id}}{\Princ}}\\
\clause{\Def{\TransV{c}}
  {\mbox{$\Union(\Null(\Un), c(\Un))$}}}
\end{display}

A consequence of this translation is that $\qq{A}$ is isomorphic to
$\Un$ for all types $A$. Formally,
\begin{lemma}\label{lemma:iso-un}
$\Judge{\qq{A}\tyq\Un}$ for all types $A$.
\end{lemma}
In practice, this means that we can replace $\qq{A}$ by $\Un$ in type
derivations, and vice versa. 

\begin{display}{Environment Translation:}
\clause{\TransV{x_1\ty A_1, \ldots, x_n\ty A_n} \triangleq
  x_1\ty \TransV{A_1}, \ldots, x_n\ty \TransV{A_n}}
\end{display}

If $\As=A_1,\ldots,A_n$ and $\xs=x_1,\ldots,x_n$
we sometimes write $B(\As\:\xs)$ as shorthand
for the signature $B(A_1\:x_1,\ldots,A_n\:x_n)$.

\begin{display}{Request and Response Types:}
\clause{\qq{A_1,\ldots,A_m} \triangleq \qq{A_1},\ldots,\qq{A_m}}\\
\clause{
  \begin{prog}
  \mathit{Req}(w), \mathit{Res}(w) \triangleq
  (\begin{prog}
   \Union(\ell_i(\qq{\As_i}) \For{i \in 1..n}),
   \Union(\ell_i(\qq{B_i}) \For{i \in 1..n}))
  \end{prog} \\ \quad
  \mbox{where $\class(w)=c$ and $\methods(c) =
    \ell_i \mapsto (B_i(\As_i \xs_i),b_i) \For{i \in 1..n}$}
  \end{prog}
}
\end{display}

\subsection{Translation of Expressions}

The translation of expressions really acts on the type derivation of an
expression, not just the expression itself. This means that during the
translation of an expression, we have access to the types of the
subexpressions appearing in the expression. To reduce clutter, we write the
translation as though it is acting on the expression itself, except that when
we need access to the type of a subexpression, we annotate the appropriate
subexpression with its type. For example, the translation of
$\Let{x}{a}\In{b}$ depends on the type of $a$, which is available through the
type derivation of $\Judge{E}{\Let{x}{a}\In{b}:B}$. We write
$\Let{x}{a_A}\In{b}$ to indicate that the type of $a$ is $A$, according to the
type derivation.

\begin{display}{Translation of a Value $v$ to a Message $\qq{v}$:}
\clause{\Def{\TransV{x}}{x}}\\
\clause{\Def{\TransV{\Null}}{\Null()}}\\
\clause{\Def{\TransV{\New\:c(v_1,\ldots,v_n)}}{c(\TransV{v_1},\ldots,\TransV{v_n})}}\\
\clause{\Def{\TransV{p}}{p}}
\end{display}

\begin{display}{Translation of a Method Body $b$ to a Process
$\TransB{b}{p}{k}$:}
\clause{\Def{\TransB{v}{p}{k}}
            {\Snd{k}{\TransV{v}}}}\\
\clause{\Def{\TransB{\Let{x}{a_A} \In{b}}{p}{k}}
            {\Res{k'\ty \Un}
              (\TransB{a}{p}{k'} \parpop \Rcv{k'}{x\ty\Un}\TransB{b}{p}{k})}}\\
\clause{\Def{\TransB{\If u=v \Then a \Else b}{p}{k}}
            {\SpiIf \TransV{u}=\TransV{v}
             \SpiThen \TransB{a}{p}{k} \SpiElse \TransB{b}{p}{k} }}\\
\clause{\Def{\TransB{v_c.f_j}{p}{k}}
            {\Case{\TransV{v}}
             \begin{prog}
               \IsTag{\Null}{y\ty\Un}{\Stop} \\
               \IsTag{c}{y\ty\Un}{\SplitPoly{y}{(x_1\ty\qq{A_1},\ldots,x_n\ty\qq{A_n})}\Snd{k}{x_j}} \\ \qquad
               \mbox{where $\fields(c)=f_i \mapsto A_i \For{i \in 1..n}$, and $j \in 1..n$}
        \end{prog}}}\\
\clause{\Def{\TransB{v_c.\ell(u_1,\ldots,u_n)}{p}{k}}
            {\Case{\TransV{v}}
             \begin{prog}
               \IsTag{\Null}{y\ty\Un}{\Stop} \\
               \IsTag{c}{y\ty\Un}{\Snd{c\_\ell}{(p,\TransV{v},\TransV{u_1},\ldots,\TransV{u_n},k)}}
        \end{prog}}}
\end{display}

\begin{display}{Translation of Method $\ell$ of Class $c$:}
\clause{
I_{\mathit{class}}(c,\ell) \triangleq
     \begin{prog}
     \Repl\Rcv{c\_\ell}{z\ty\Un}\\
     \SplitPoly{z}{(p\ty\Princ,\this\ty\Un,x_1\ty\qq{A_1},\ldots,x_n\ty\qq{A_n},k\ty\Un)}
        \TransB{b}{p}{k}\\ \quad
     \mbox{where $\methods(c)(\ell) = (B(A_1\:x_1,\ldots,A_n\:x_n),b)$}
     \end{prog}}
\end{display}

\begin{display}{Type of Key Shared Between Client $p$ and Server $q$:}
\clause{
  \begin{prog}
    \CSKey(p,q) \triangleq \\ \quad
    \begin{prog}
    \SharedKey {\Union (\begin{prog}
                  \request (\begin{prog}
                            w \ty \Un,
                            a \ty \Un,
                            t \ty \Un,\\
                            n_q \ty \annotate{\public}{\Response{[\endd{\request(p,q,w,a,t)}]}}),
             \end{prog}\\
                  \response (\begin{prog}
                             w \ty \Un,
                             r \ty \Un,
                             t \ty \Un,\\
                             n_p \ty \annotate{\public}{\Response{[\endd{\response(p,q,w,r,t)}]}}))}
              \end{prog}
        \end{prog}
    \end{prog}
   \end{prog}}
\end{display}

\begin{display}{Semantics of Web Method Call:}
\clause{\begin{prog}
          \TransB{w\mo{:}\ell(u_1,\ldots,u_n)}{p}{k} \triangleq {} \\ \quad
             \Res{k_1 \ty \Un, k_2 \ty \Un, t\ty\Un,
                  n_p \ty \annotate{\public}{\Challenge{[\,]}}} \\ \quad
             \Begin{\request(p,q,w,\ell(\qq{u_1},\ldots,\qq{u_n}),t)} \\ \quad
              \Snd{w}{(\request(\getnonce()),k_1)}; \\ \quad
              \Rcv{k_1}{\response(\getnonce(n_q \ty \Un))} \\ \quad
              \Cast{n_q}{n'_q \ty
                \annotate{\public}{\Response{[\endd{\request(p,q,w,\ell(\qq{u_1},\ldots,\qq{u_n}),t)}]}}} \\ \quad
              \Snd{w}{(p,\sencr{\request(w,\ell(\qq{u_1},\ldots,\qq{u_n}),t,n'_q)}{K_{pq}},n_p,k_2)}; \\ \quad
              \Rcv{k_2}{q'\ty\Un,\bdy \ty \Un}\\ \quad
              \Decrypt{\bdy}{\response(\mathit{plain})}{K_{pq}}\\ \quad
              \Match{\mathit{plain}}{w}{\mathit{rest} \ty
                (r \ty \mathit{Res}(w), t'\ty\Un,
                 \annotate{\public}{\Response{[\endd{\response(p,q,w,r,t')}]}})}\\ \quad
              \Split{\mathit{rest}}{r\ty\mathit{Res}(w)}{rest'\ty(t'\ty\Un,n_p'\ty\annotate{\public}{\Response{[\endd{\response(p,q,w,r,t')}]}})}\\ \quad
         \Match{\mathit{rest'}}{t}{n_p'\ty\annotate{\public}{\Response{[\endd{\response(p,q,w,r,t)}]}}}\\ \quad
         \CheckNonce{n_p}{n_p'}\\ \quad
              \End{\response(p,q,w,r,t)}\\ \quad
              \Case{r} \IsTag{\ell}{x}{\Snd{k}{x}}\\
             \mbox{where $q=\owner(w)$}
       \end{prog}}
\end{display}

\clearpage
\begin{display}{Server-Side Invocation of Web Method:}
\clause{
\begin{prog}
\LetCall{x}{w}{p, \mathit{args}}P \triangleq {} \\
   \begin{prog}
   \Res{k}\\
   \quad \Case{\mathit{args}} \\
   \quad (\begin{prog}
          \IsTag{\ell_i}{\xs_i}{} \\ \quad
          \Res{k'}
          (\Snd{c\_\ell_i}{(q,c(p),\xs_i,k')}
          \parpop \Rcv{k'}{r}\Snd{k}{\ell_i(r)})\\
          ) \For{i \in 1..n}
          \end{prog} \\
    \quad \parpop \Rcv{k}{x}P
   \end{prog}\\
    \quad \begin{prog}
          \mbox{where $c=\class(w)$, $q=\owner(w)$,}\\
          \mbox{ and $\methods(c)=\ell_i \mapsto
           (B_i(\As_i,\xs_i),b_i) \For{i \in 1..n}$}
          \end{prog}
\end{prog}}
\end{display}

\begin{display}{Web Service Translation:}
\clause{I_{\mathit{ws}}(w) \triangleq 
     \begin{prog}
     \Repl\Rcv{w}{\bdy \ty \Un,k_1 \ty \Un}\\ \quad
       \begin{prog}
         \Case{\bdy} \IsTag{\request}{\getnonce()}{}\\
         \Res{n_q \ty \annotate{\public}{\Challenge{[\,]}}}\\
         \Snd{k_1}{(\response(\getnonce(n_q)))};\\
         \Rcv{w}{p'\ty\Un,\mathit{cipher}\ty\Un,n_p \ty \Un, k_2 \ty \Un}\\
         \textstyle\prod_{p \in \Prin}\SpiIf p=p' \SpiThen \\
         \Decrypt{\mathit{cipher}}{\request(\mathit{plain})}{K_{pq}}\\
         \Match{\mathit{plain}}{w}{\mathit{rest} \ty \\ \quad
                (a \ty \mathit{Req}(w),t\ty\Un,
                  \annotate{\public}{\Response{[\endd{\request(p,q,w,a,t)}]}})}\\
    \Split{\mathit{rest}}{a\ty\mathit{Req}(w)}{t\ty\Un,n_q'\ty \\ \quad
           \annotate{\public}{\Response{[\endd{\request(p,q,w,a,t)}]}}}\\
         \CheckNonce{n_q}{n_q'}\\
    \End{\request(p,q,w,a,t)}\\
         \LetCall{r \ty \mathit{Res}(w)}{w}{p,a}\\
         \Begin{\response(p,q,w,r,t)} \\
         \Cast{n_p}{n'_p \ty
            \annotate{\public}{\Response{[\endd{\response(p,q,w,r,t)}]}}}\\
         \Snd{k_2}{(q,\sencr{\response(w,r,t,n'_p)}{K_{pq}})}
       \end{prog} \\
     \mbox{where $q=\owner(w)$}
     \end{prog}}
\end{display}

\begin{display}{Implementation of Classes and Web Services:}
\clause{\ClassMethods \triangleq
  \{(c,\ell) ~:~ c\in\Class, \ell\in\dom(\methods(c))\}}\\
\clause{
I_{\mathit{class}} \triangleq
     \IndexedPar{(c,\ell) \in \ClassMethods} I_{\mathit{class}}(c,\ell)}\\
\clause{
  I_{\mathit{ws}} \triangleq
     \IndexedPar{w\in\WebService} I_{\mathit{ws}}(w)}
\end{display}

\begin{display}{Top-Level Environments:}
\clause{
E_{\mathit{class}} \triangleq
  (c\_\ell \ty \Un)\For{(c,\ell) \in \ClassMethods}}\\
\clause{E_{\mathit{keys}}\triangleq(K_{pq}\ty\CSKey(p,q))\For{p,q\in\Prin}}\\
\clause{E_{\mathit{ws}} \triangleq (w\ty\Un) \For{w \in \WebService}}\\
\clause{E_{\mathit{prin}} \triangleq
         p_1 \ty  \Princ, \ldots, p_n \ty  \Princ}
  {where $\Prin=\{p_1, \ldots, p_n\}$}\\
 \clause{E_0 \triangleq E_{\mathit{ws}}, E_{\mathit{prin}},
                      E_{\mathit{class}}, E_{\mathit{keys}}}
\end{display}

Some general remarks on typing are in order. A consequence of
Lemma~\ref{lemma:iso-un}, as well as our general use of types, reveals
that we rely on typing exclusively to show security properties, not to
establish standard safety results. For instance, we do not use types
to ensure that the type of the arguments supplied at method invocation
match the type of the parameters to the method. Indeed, the only
channel type in our translation has itself type $\Un$.

In order to prove Theorem~\ref{t:robust-safety-translation}, we first
establish some lemmas.

\begin{lemma}~\label{lemma:type-preservation-2}
\begin{enumerate}
\item
If $E \vdash v : A$ then $E_{\mathit{prin}},\qq{E} \vdash \qq{v} : \qq{A}$.
\item
If $E \vdash a : A$ and $\Judge{E_0,\qq{E}}{p:\Princ}$ and $k \notin \dom(E_0,\qq{E})$
then: \[E_0, \qq{E}, k \ty \Un \vdash \qq{a}^p_k : [\,]\]
\item
If $c\in\Class$ and $\ell\in\dom(\methods(c))$
then $E_0 \vdash I_{\mathit{class}}(c,\ell) : [\,]$.
\item
If $w\in\WebService$ then $E_0 \vdash I_{\mathit{ws}}(w) : [\,]$.
\end{enumerate}
\end{lemma}
\begin{proof}
\begin{enumerate}

\item We prove this by induction on the height of the type derivation
for $\Judge{E}{v:A}$:
\begin{itemize}

\citem{$v=x$} Since $\Judge{E}{x:A}$, we must have $x\ty A\in E$. By
definition of the translation for environment, $x\ty\qq{A}\in \qq{E}$,
hence $\Judge{E_{\mathit{prin}},\qq{E}}{x:\qq{A}}$,
as required.

\citem{$v=\Null$} We have $\Judge{E}{\Null:c}$. Since
$\qq{c}=\Union(\Null(),c(\Un))$ and $\qq{\Null}=\Null()$, we have
$\Judge{E_{\mathit{prin}},\qq{E}}{\Null():\Union(\Null(\Un),$ $c(\Un))}$,
as required.

\citem{$v=\New\:c(v_1,\ldots,v_n)$} Since $\Judge{E}{v:A}$, where
$A=c$, we have $\fields(c)=f_i\mapsto A_i\For{i\in 1..n}$, and
$\Judge{E}{v_i:A_i}$ for all $i\in 1..n$. Let
$E'=E_{\mathit{prin}},\qq{E}$.  By induction
hypothesis, $\Judge{E'}{\qq{v_i}:\qq{A_i}}$ for all $i\in 1..n$. We
can now derive: 
\[
\infer{\Judge{E'}{c(\qq{v_1},\ldots,\qq{v_n}):\Union(\Null(\Un),c(\Un))}}
      {\infer{\Judge{E'}{\qq{v_1},\ldots,\qq{v_n}:\Un}}
             {\infer{\Judge{E'}{\qq{v_1},\ldots,\qq{v_n}:(\Un,\ldots,\Un)}}
                    {\infer{\Judge{E'}{(\qq{v_1},\ldots,\qq{v_n}):(\qq{A_1},\ldots,\qq{A_n})}}
                           {\Judge{E'}{\qq{v_i}:\qq{A_i}} \quad
                            \forall i\in 1..n}}}}
\]
 
\citem{$v=p$} Since $\Judge{E}{p:A}$ (with $A=\Id$), we have
$p\in\Prin$, hence $p\ty\Princ\in E_{\mathit{prin}}$. Since
$\qq{\Id}=\Princ$, we have
$\Judge{E_{\mathit{prin}},\qq{E}}{p:\Princ}$, as
required. 

\end{itemize}

\item Again, we proceed by induction on the height of the type
derivation for $\Judge{E}{a:A}$. 
\begin{itemize}

\citem{$a=v$} We can apply the result of part (1). Since
$\Judge{E}{v:A}$, then
$\Judge{E_{\mathit{prin}},\qq{E}}{\qq{v}:\qq{A}}$.
We can derive: 
\[
\infer{\Judge{E_0,\qq{E},k\ty\Un}{\Snd{k}{\qq{v}}:[\,]}}
      {\Judge{E_0,\qq{E},k\ty\Un}{k:\Un} &
       \infer{\Judge{E_0,\qq{E}}{\qq{v}:\Un}}
             {\Judge{E_0,\qq{E}}{\qq{v}:\qq{A}}}}
\]

\citem{$a=\Let{x}{a_0}\In{b}$} We have $\Judge{E}{a_0:B}$ for some
$B$, and $\Judge{E,x\ty B}{b:A}$. Applying the induction hypothesis,
we derive $\Judge{E_0,\qq{E},k'\ty\Un}{\TransB{a_0}{p}{k'}:[\,]}$ and
$\Judge{E_0,\qq{E},x\ty\qq{B},k\ty\Un}{\TransB{b}{p}{k}:[\,]}$. Let
$E'=E_0,\qq{E},k\ty\Un$. We can now derive:
\[
\infer{\Judge{E'}
             {\Res{k'\ty\Un}(\TransB{a}{p}{k'}\parpop\Rcv{k'}{x\ty\Un}\TransB{b}{p}{k}):[\,]}}
      {\infer{\Judge{E',k'\ty\Un}{\TransB{a}{p}{k'}\parpop\Rcv{k'}{x\ty\Un}\TransB{b}{p}{k}:[\,]}}
             {\Judge{E',k'\ty\Un}{\TransB{a}{p}{k'}:[\,]} &
              \infer{\Judge{E',k'\ty\Un}{\Rcv{k'}{x\ty\Un}\TransB{b}{p}{k}:[\,]}}
                    {\Judge{E',k'\ty\Un}{k':\Un} & 
                     \infer{\Judge{E',k'\ty\Un,x\ty\Un}{\TransB{b}{p}{k}:[\,]}}
                           {\Judge{E',k'\ty\Un,x\ty\qq{B}}{\TransB{b}{p}{k}:[\,]}}}}}
\]

\citem{$a=\If u=v\Then a_0 \Else a_1$} We have $\Judge{E}{u:B}$,
$\Judge{E}{v:B}$, $\Judge{E}{a_0:A}$, and $\Judge{E}{a_1:A}$. Applying
the induction hypothesis, we derive
$\Judge{E_0,\qq{E},k\ty\Un}{\TransB{a_0}{p}{k}:[\,]}$ and
$\Judge{E_0,\qq{E},k\ty\Un}{\TransB{a_0}{p}{k}:[\,]}$. By (1), we also
have $\Judge{E_0,\qq{E}}{\qq{u}:\qq{B}}$ and
$\Judge{E_0,\qq{E}}{\qq{v}:\qq{B}}$. This gives us
$\Judge{E_0,\qq{E},k\ty\Un}{\SpiIf \qq{u}=\qq{v}\SpiThen
\TransB{a_0}{p}{k} \SpiElse \TransB{a_1}{p}{k}:[\,]}$, as required. 

\citem{$a=v.f_j$} We have $\Judge{E}{v.f_j:A_j}$, where
$\Judge{E}{v:c}$ and $\fields(c)=f_i\mapsto A_i\For{i\in 1..n}$. By
(1), $\Judge{E_0,\qq{E}}{\qq{v}:\qq{c}}$. Let
$E'=E_0,\qq{E},k\ty\Un$. First, let us derive that 
$\Judge{E',y:\Un}{\SplitPoly{y}{(x_1\ty\qq{A_1},\ldots,x_n\ty\qq{A_n})}\Snd{k}{x_j}:[\,]}$.
Let $E'' = x_1\ty\qq{A_1},\ldots,x_n\ty\qq{A_n}$.
(We trim the environments where possible to reduce clutter.)
\[
\infer{\Judge{E',y:\Un}{\SplitPoly{y}{(x_1\ty\qq{A_1},\ldots,x_n\ty\qq{A_n})}\Snd{k}{x_j}:[\,]}}
      {\Judge{E',y:\Un}{y:\Un} & 
       \infer{\Judge{E',y:\Un,E''}{\Snd{k}{x_j}:[\,]}}
             {\Judge{E'}{k:\Un} & 
              \infer{\Judge{E',y:\Un,E''}{x_j:\Un}}
                    {\Judge{E',y:\Un,E''}{x_j:\qq{A_j}}}}}
\]

We can now derive:
\[
\infer{\Judge{E'}{\Case{\qq{v}}
         \begin{prog}
         \IsTag{\Null}{y\ty\Un}{\Stop}\\
         \IsTag{c}{y}{\SplitPoly{y}{(x_1\ty\qq{A_1},\ldots,x_n\ty\qq{A_n})}
                      \Snd{k}{x_j}}:[\,]
    \end{prog}}}
      {\begin{prog}
         \Judge{E'}{\qq{v}:\Union(\Null(\Un),c(\Un))} \\
         \Judge{E',y:\Un}{\Stop:[\,]}\\
         \Judge{E',y:\Un}{\SplitPoly{y}{(x_1\ty\qq{A_1},\ldots,x_n\ty\qq{A_n})}\Snd{k}{x_j}:[\,]}
       \end{prog}}
\]

\citem{$a=v.\ell_j(u_1,\ldots,u_m)$} We have
$\Judge{E}{v.\ell_j(u_1,\ldots,u_m):B}$, where $\Judge{E}{v:c}$,
$\methods(c)=\ell_i\mapsto(\sig_i,b_i)\For{i\in 1..n}$,
$\sig_j=B(A_1\:x_1,$ $\ldots,$ $A_m\:x_m)$, and $\Judge{E}{u_k:A_k}$ for all
$k\in 1..m$. By (1), $\Judge{E_0,\qq{E}}{\qq{u_k}:\qq{A_k}}$ for all
$k\in 1..m$. Let $E'=E_0,\qq{E},k\ty\Un$. First, let us derive that 
$\Judge{E',y\ty\Un}{\Snd{c\_\ell}{(p,\TransV{v},\TransV{u_1},\ldots,\TransV{u_n},k)}:[\,]}$.
\[
\infer{\Judge{E',y\ty\Un}{\Snd{c\_\ell}{(p,\TransV{v},\TransV{u_1},\ldots,\TransV{u_n},k)}:[\,]}}
      {\Judge{E',y\ty\Un}{c\_\ell:\Un} & 
       \Judge{E',y\ty\Un}{(p,\TransV{v},\TransV{u_1},\ldots,\TransV{u_n},k):\Un}}
\]

We can derive:
\[
\infer{\Judge{E'}{\Case{\TransV{v}}
                  \begin{prog}
                  \IsTag{\Null}{y\ty\Un}{\Stop} \\
                  \IsTag{c}{y}{\Snd{c\_\ell}{(p,\TransV{v},\TransV{u_1},\ldots,\TransV{u_n},k)}}:[\,]
        \end{prog}}}
      {\begin{prog}
         \Judge{E'}{\qq{v}:\Union(\Null(\Un),c(\Un))} \\
         \Judge{E',y\ty\Un}{\Stop:[\,]}\\
         \Judge{E',y\ty\Un}{\Snd{c\_\ell}{(p,\TransV{v},\TransV{u_1},\ldots,\TransV{u_n},k)}:[\,]}
       \end{prog}}
\]

\citem{$a=w\mo{:}\ell_j(u_1,\ldots,u_m)$} We have
$\Judge{E}{w\mo{:}\ell_j(u_1,\ldots,u_m):B}$, where $\class(w)=c$,
$\owner(w)=q$, 
$\methods(c)=\ell_i\mapsto(\sig_i,b_i)\For{i\in 1..n}$, 
$\sig_j=B(A_1\:x_1,\ldots,A_m\:x_m)$, and $\Judge{E}{u_k:A_k}$ for all
$k\in 1..m$. By (1), $\Judge{E_0,\qq{E}}{\qq{u_k}:\qq{A_k}}$ for all
$k\in 1..m$. Rather than giving the full type derivation for the
translation of a web service call, we outline the derivation of
effects:

$\begin{prog}\quad
             \Res{k_1 \ty \Un, k_2 \ty \Un, t\ty\Un,
                  n_p \ty \annotate{\public}{\Challenge{[\,]}}} \\ \quad
             \mbox{// Effect: $[\check{\public}{n_p}]$}\\ \quad
             \Begin{\request(p,q,w,\ell(\qq{u_1},\ldots,\qq{u_n}),t)} \\ \quad
             \mbox{// Effect: $[\check{\public}{n_p},\endd{\request(p,q,w,\ell(\qq{u_1},\ldots,\qq{u_n}),t)}]$}\\ \quad
              \Snd{w}{(\request(\getnonce()),k_1)}; \\ \quad
             \mbox{// Effect: $[\check{\public}{n_p},\endd{\request(p,q,w,\ell(\qq{u_1},\ldots,\qq{u_n}),t)}]$}\\ \quad
              \Rcv{k_1}{\response(\getnonce(n_q \ty \Un))} \\ \quad
              \mbox{// Effect: $[\check{\public}{n_p},\endd{\request(p,q,w,\ell(\qq{u_1},\ldots,\qq{u_n}),t)}]$}\\ \quad
              \Cast{n_q}{n'_q \ty
                \annotate{\public}{\Response{[\endd{\request(p,q,w,\ell(\qq{u_1},\ldots,\qq{u_n}),t)}]}}} \\ \quad
              \mbox{// Effect: $[\check{\public}{n_p}]$}\\ \quad
              \Snd{w}{(p,\sencr{\request(w,\ell(\qq{u_1},\ldots,\qq{u_n}),t,n'_q)}{K_{pq}},n_p,k_2)}; \\ \quad
              \mbox{// Effect: $[\check{\public}{n_p}]$}\\ \quad
              \Rcv{k_2}{q'\ty\Un,\bdy \ty \Un}\\ \quad
              \mbox{// Effect: $[\check{\public}{n_p}]$}\\ \quad
              \Decrypt{\bdy}{\response(\mathit{plain})}{K_{pq}}\\ \quad
              \mbox{// Effect: $[\check{\public}{n_p}]$}\\ \quad
              \Match{\mathit{plain}}{w}{\mathit{rest} \ty \\ \qquad
                (r \ty \mathit{Res}(w), t'\ty\Un,
                 \annotate{\public}{\Response{[\endd{\response(p,q,w,r,t')}]}})}\\ \quad
              \mbox{// Effect: $[\check{\public}{n_p}]$}\\ \quad
              \Split{\mathit{rest}}{r\ty\mathit{Res}(w)}{rest'\ty \\ \qquad
                (t'\ty\Un,n_p'\ty\annotate{\public}{\Response{[\endd{\response(p,q,w,r,t')}]}})}\\ \quad
              \mbox{// Effect: $[\check{\public}{n_p}]$}\\ \quad
         \Match{\mathit{rest'}}{t}{n_p'\ty\annotate{\public}{\Response{[\endd{\response(p,q,w,r,t)}]}}}\\ \quad
              \mbox{// Effect: $[\check{\public}{n_p}]$}\\ \quad
         \CheckNonce{n_p}{n_p'}\\ \quad
              \mbox{// Effect: $[\endd{\response(p,q,w,r,t)}]$}\\ \quad
              \End{\response(p,q,w,r,t)}\\ \quad
              \mbox{// Effect: $[\,]$}\\ \quad
              \Case{r} \IsTag{\ell}{x}{\Snd{k}{x}}\\ \quad
              \mbox{// Effect: $[\,]$}
  \end{prog}$

\end{itemize}

\item Recall that we assume that method bodies are well-typed, that
is, we assume for $c,\ell_j$ with
$\methods(c)=\ell_i\mapsto(\sig_i,b_i)$ and
$\sig_j=B(A_1\:a_1,\ldots,A_m\:x_m)$, that $\Judge{\this\ty c,x_1\ty
A_1,\ldots,x_m\ty A_m}{b_j:B}$. By clause (2) above, this means that
$\Judge{E_0,\this\ty\qq{c},x_1\ty \qq{A_1},\ldots,x_m\ty
\qq{A_m},k\ty\Un}{\TransB{b_j}{p}{k}:[\,]}$. Applying
Lemma~\ref{lemma:iso-un}, we derive $\Judge{E_0,\this\ty\Un,x_1\ty
\Un,\ldots,x_m\ty\Un,k\ty\Un}{\TransB{b_j}{p}{k}:[\,]}$. We can now easily derive the following:
\[
\infer{\Judge{E_0}{I_{\mathit{class}}(c,\ell) : [\,]}}
      {\infer{\Judge{E_0}{\Repl{\Rcv{c\_\ell}{z}
                          \SplitPoly{z}{(p\ty\Princ,\this\ty\Un,x_1\ty\Un,\ldots,x_n\ty\Un,k\ty\Un)}
                          \TransB{b_j}{p}{k}:[\,]}}}
             {\Judge{E_0}{c\_\ell:\Un} & 
              \infer{\Judge{E_0,z\ty\Un}{\SplitPoly{z}{(p\ty\Princ,\this\ty\Un,x_1\ty\Un,\ldots,x_n\ty\Un,k\ty\Un)}
                                         \TransB{b_j}{p}{k}:[\,]}}
                    {\begin{prog}
                     \Judge{E_0,z\ty\Un}{z:(\Princ,\Un,\ldots,\Un)} \\
                     \Judge{E_0,z\ty\Un,p\ty\Princ,\this\ty\Un,x_1\ty\Un,\ldots,x_n\ty\Un,k\ty\Un}
                           {\TransB{b_j}{p}{k}:[\,]}
           \end{prog}}}}
\]

\item Let $w\in\WebService$, with $\owner(w)=q$. First, note that the following
derivation is admissible:
\[\infer{\Judge{E_0,E}{\LetCall{r \ty \mathit{Res}(w)}{w}{p,a}P:\es}}
       {\Judge{E_0,E}{p:\Prin} &
        \Judge{E_0,E}{a:\mathit{Req}(w)} &
        \Judge{E_0,E,r\ty\mathit{Res}(w)}{P:\es}}\]
(The proof is a straightforward, if longish, type derivation.)
Rather than giving the full type derivation for the
implementation of web service $w$, we outline the derivation of
effects:

$\begin{prog}\quad
     \Repl\Rcv{w}{\bdy \ty \Un,k_1 \ty \Un}\\ \quad
     \mbox{// Effect: $[\,]$}\\ \quad

         \Case{\bdy} \IsTag{\request}{\getnonce()}{}\\ \quad
         \mbox{// Effect: $[\,]$}\\  \quad
         \Res{n_q \ty \annotate{\public}{\Challenge{[\,]}}}\\ \quad
         \mbox{// Effect: $[\check{\public}{n_q}]$}\\  \quad
         \Snd{k_1}{(\response(\getnonce(n_q)))};\\ \quad
         \mbox{// Effect: $[\check{\public}{n_q}]$}\\  \quad
         \Rcv{w}{p'\ty\Un,\mathit{cipher}\ty\Un,n_p \ty \Un, k_2 \ty \Un}\\ \quad
         \mbox{// Effect: $[\check{\public}{n_q}]$}\\  \quad
         \textstyle\prod_{p \in \Prin}\SpiIf p=p' \SpiThen \\ \quad
         \mbox{// Effect: $[\check{\public}{n_q}]$}\\  \quad
         \Decrypt{\mathit{cipher}}{\request(\mathit{plain})}{K_{pq}}\\ \quad
         \mbox{// Effect: $[\check{\public}{n_q}]$}\\  \quad
         \Match{\mathit{plain}}{w}{\mathit{rest} \ty
                (\begin{prog}
                 a \ty \mathit{Req}(w),t\ty\Un,\\
                  \annotate{\public}{\Response{[\endd{\request(p,q,w,a,t)}]}})}
                 \end{prog}\\ \quad
         \mbox{// Effect: $[\check{\public}{n_q}]$}
  \end{prog}$

$\begin{prog}\quad
    \Split{\mathit{rest}}{a\ty\mathit{Req}(w)}{t\ty\Un,n_q'\ty\annotate{\public}{\Response{[\endd{\request(p,q,w,a,t)}]}}}\\ \quad
         \mbox{// Effect: $[\check{\public}{n_q}]$}\\  \quad
         \CheckNonce{n_q}{n_q'}\\ \quad
         \mbox{// Effect: $[\endd{\request(p,q,w,a,t)}]$}\\ \quad
    \End{\request(p,q,w,a,t)}\\ \quad
         \mbox{// Effect: $[\,]$}\\ \quad
         \LetCall{r \ty \mathit{Res}(w)}{w}{p,a}\\ \quad
         \mbox{// Effect: $[\,]$}\\ \quad
         \Begin{\response(p,q,w,r,t)} \\ \quad
         \mbox{// Effect: $[\endd{\response(p,q,w,r,t)}]$}\\ \quad
         \Cast{n_p}{n'_p \ty
            \annotate{\public}{\Response{[\endd{\response(p,q,w,r,t)}]}}}\\ \quad
         \mbox{// Effect: $[\,]$}\\ \quad
         \Snd{k_2}{(q,\sencr{\response(w,r,t,n'_p)}{K_{pq}})}\\ \quad
         \mbox{// Effect: $[\,]$}
  \end{prog}$

\end{enumerate}\hfill
\end{proof}

\begin{lemma}\label{lemma:type-preservation-3}
If $\emptyset \vdash a : A$ and $p \in \Prin$ and $k \notin \dom(E_0)$
then: \[E_{\mathit{ws}}, E_{\mathit{prin}} \vdash
  \Res{E_{\mathit{class}}, E_{\mathit{keys}}}
    (I_{\mathit{class}} \parpop
     I_{\mathit{ws}} \parpop
     \Res{k \ty \Un} \qq{a}^p_k) : [\,]\]
\end{lemma}
\begin{proof}
This is a corollary of Lemma~\ref{lemma:type-preservation-2}.
Specifically, we can derive:  
\[
\infer{\Judge{E_{\mathit{ws}}, E_{\mathit{prin}}}
             {\Res{E_{\mathit{class}}, E_{\mathit{keys}}}
              (I_{\mathit{class}} \parpop
               I_{\mathit{ws}} \parpop
               \Res{k \ty \Un} \qq{a}^p_k) : [\,]}}
      {\infer{\Judge{E_0}{I_{\mathit{class}} \parpop
                          I_{\mathit{ws}} \parpop
                          \Res{k \ty \Un} \qq{a}^p_k : [\,]}}
             {\infer{\Judge{E_0}{I_{\mathit{class}}:[\,]}}
                    {\Judge{E_0}{I_{\mathit{class}}(c,\ell):[\,]}\For{(c,\ell)\in\ClassMethods}} &
              \infer{\Judge{E_0}{I_{\mathit{ws}}:[\,]}}
                    {\Judge{E_0}{I_{\mathit{ws}}(w):[\,]\For{w\in\WebService}}} &
              \infer{\Judge{E_0}{\Res{k \ty \Un} \qq{a}^p_k : [\,]}}
                    {\Judge{E_0,k\ty\Un}{\qq{a}^p_k:[\,]}}}}
\]\hfill
\end{proof}

We can now rephrase Theorem~\ref{t:robust-safety-translation}
formally, and prove it.
\begin{theorem}\label{t:rephrased-robust-safety-translation}
If $\Judge{\emptyset}{a : A}$ and $p \in \Prin$ and $k \notin \dom(E_0)$ then the system
\[\Res{E_{\mathit{class}}, E_{\mathit{keys}}}
    (I_{\mathit{class}} \parpop
     I_{\mathit{ws}} \parpop
     \Res{k \ty \Un} \qq{a}^p_k)\]
is robustly safe.
\end{theorem}
\begin{proof}
By Lemma~\ref{lemma:type-preservation-3},
    \[\Judge{E_{\mathit{ws}},E_{\mathit{prin}}}
    {\Res{E_{\mathit{class}}, E_{\mathit{keys}}}  
                 (I_{\mathit{class}} \parpop
                 I_{\mathit{ws}} \parpop
                 \Res{k \ty \Un} \qq{a}^p_k):[\,]}.\]
 Robust safety of the system follows by Theorem~\ref{Thm:robust-safety}.   \hfill
\end{proof}
\fi%
\fi%

\ifLong
 \ifJournal\else
\section{Implementation Using Asymmetric Cryptography}
\label{app:asymmetric}
The security abstraction we describe in Section~\ref{sec:sec-abs}
relies on shared keys between principals. This is hardly a reasonable
setup in modern systems. In this appendix, we show that our approach
can easily accommodate public-key infrastructures.

\subsection{Authenticated Web Methods} We start by describing the
protocol and implementation for authenticated web methods. Hence, for
now, we assume that all the exported methods of a web service are
annotated with \texttt{Auth}.

Consider a simple public-key infrastructure for digital
signatures. Each principal $p$ has a signing key $\SK{p}$ and a
verification key $\VK{p}$. The signing key is kept private, while the
verification key is public. To bind the name of a principal with their
verification key, we assume a \emph{certification authority} $\CA$
(itself with a signing key $\SK{CA}$ and verification key $\VK{CA}$)
that can sign certificates $\CertVK{p}$ of the form
$\asencr{p,\VK{p}}{\SK{CA}}$. (The notation $\asencr{\cdot}{K}$
is used to represent both asymmetric encryption and signature, 
differentiating it from symmetric encryption. In the case where
$\asencr{M}{K}$ represent a signature, this is simply notation for $M$
along with a token representing the signature of $M$ with asymmetric
key $K$.)

Here is a protocol that uses digital signatures to authenticate
messages, for $p$ making a web service call
$w\mo{:}\ell(u_1,\ldots,u_n)$ to service $w$ owned by $q$, including
the names of continuation channels used at the spi level.  Again, we
assume that in addition to the methods of $\class(w)$, each web
service also supports a method $\getnonce$, which we implement
specially. 
\[\begin{array}{l}
p \rightarrow q ~\mbox{on}~ w: \CertVK{p}, n_p, \request(\getnonce()),k_1\\
q \rightarrow p ~\mbox{on}~ k_1: \CertVK{q}, \response(\getnonce(n_q))\\
p \rightarrow q ~\mbox{on}~ w: p, \asencr{\request(w,\ell(u_1,\ldots,u_n),t,q,n_q)}{\SK{p}},k_2\\
q \rightarrow p ~\mbox{on}~ k_2: q, \asencr{\response(w,\ell(r),t,p,n_p)}{\SK{q}}
\end{array}\]

\begin{display}{Type of Signing Keys:}
\clause{
  \begin{prog}
    \AuthMsg{p} \triangleq \\ \quad
    \begin{prog}
    \Union (\begin{prog}
                  \request (\begin{prog}
                            w \ty \Un,
                            a \ty \Un,
                            t \ty \Un,
                            q \ty \Un,
                            n_q \ty \annotate{\public}{\Response{[\endd{\request(p,q,w,a,t)}]}}),
             \end{prog}\\
                  \response (\begin{prog}
                             w \ty \Un,
                             r \ty \Un,
                             t \ty \Un,
                             q \ty \Un,
                             n_q \ty \annotate{\public}{\Response{[\endd{\response(q,p,w,r,t)}]}}))   
              \end{prog}
             \end{prog}
    \end{prog}
  \end{prog}}    \\
\clause{\AuthKeys{p} \triangleq \KeyPair{\AuthMsg{p}}}\\
\clause{\AuthCert \triangleq (p:\Un, \annotate{\Dec}{\Key{\AuthMsg{p}}})}\\
\clause{\AuthCertKeys \triangleq \KeyPair{\AuthCert}}
\end{display}

We will represent the key pair of a signing key and verification key
for principal $p$ by a pair $\DS{p}$, of type $\AuthKeys{p}$. The key
pair for the certification authority will be represented by a pair
$\DS{CA}$. We use the following abbreviations: 
\begin{display}{Key and Certificates Abbreviations:}
\clause{\SK{p} \triangleq \extract{\Enc}{\DS{p}}}{$p$'s signing key}\\
\clause{\VK{p} \triangleq \extract{\Dec}{\DS{p}}}{$p$'s verification key}\\
\clause{\CertVK{p} \triangleq \asencr{p,\VK{p}}{\SK{CA}}}{$p$'s certificate}
\end{display}

With that in mind, we can amend the translation of
Section~\ref{sec:spi-calculus-semantics} to accommodate the new
protocol. First, we give a new translation for a web method call
$w\mo{:}\ell(u_1,\ldots,u_n)$: 
\begin{display}{New Semantics of Web Method Call:}
\clause{\begin{prog}
          \TransB{w\mo{:}\ell(u_1,\ldots,u_n)}{p}{k} \triangleq {} \\ \quad
             \Res{k_1 \ty \Un, k_2 \ty \Un, t\ty\Un,
                  n_p \ty \annotate{\public}{\Challenge{[\,]}}} \\ \quad
             \Begin{\request(p,q,w,\ell(\qq{u_1},\ldots,\qq{u_n}),t)} \\ \quad
              \Snd{w}{(\CertVK{p},n_p,\request(\getnonce()),k_1)}; \\ \quad
              \Rcv{k_1}{c\ty\Un,\response(\getnonce(n_q \ty \Un))} \\ \quad
              \pDecrypt{c}{\mathit{cert}\ty(q'\ty\Un,\annotate{\Dec}{\Key{\AuthMsg{q'}}})}{\VK{CA}}\\ \quad
              \Match{\mathit{cert}}{q}{\mathit{vkq}\ty\annotate{\Dec}{\Key{\AuthMsg{q}}}}\\ \quad
              \Cast{n_q}{n'_q \ty
                \annotate{\public}{\Response{[\endd{\request(p,q,w,\ell(\qq{u_1},\ldots,\qq{u_n}),t)}]}}} \\ \quad
              \Snd{w}{(p,\asencr{\request(w,\ell(\qq{u_1},\ldots,\qq{u_n}),t,q,n'_q)}{\SK{p}},k_2)}; \\ \quad
              \Rcv{k_2}{q''\ty\Un,\bdy \ty \Un}\\ \quad
              \begin{prog}
                \pDecrypt{\bdy}{\response(\mathit{plain}\ty(w'\ty\Un,r\ty\Un,t'\ty\Un,p'\ty\Un,\\
                        \qquad\qquad\quad\annotate{\public}{\Response{[\endd{\response(p',q,w',r,t')}]}}))}{\mathit{vkq}}
         \end{prog}\\ \quad
              \Match{\mathit{plain}}{w}{\mathit{rest} \ty
                (\begin{prog}
                 r \ty \mathit{Res}(w), t'\ty\Un, p'\ty\Un,\\
                 \annotate{\public}{\Response{[\endd{\response(p',q,w,r,t')}]}})}
                 \end{prog}\\ \quad
         \Split{\mathit{rest}}{r\ty\mathit{Res}(w)}{\mathit{rest'}\ty
                (\begin{prog}
                 t'\ty\Un,p'\ty\Un,\\
                 \annotate{\public}{\Response{[\endd{\response(p',q,w,r,t')}]}})}
       \end{prog}\\ \quad
              \Match{\mathit{rest'}}{t}{\mathit{rest''}\ty(p'\ty\Un,\annotate{\public}{\Response{[\endd{\response(p',q,w,r,t)}]}})}\\ \quad
              \Match{\mathit{rest''}}{p}{n_p'\ty\annotate{\public}{\Response{[\endd{\response(p,q,w,r,t)}]}}}\\ \quad
         \CheckNonce{n_p}{n_p'}\\ \quad
              \End{\response(p,q,w,r,t)}\\ \quad
              \Case{r} \IsTag{\ell}{x}{\Snd{k}{x}}\\
             \mbox{where $q=\owner(w)$}
       \end{prog}}
\end{display}

We also need to give a new implementation for web services, again to
take into account the different messages being exchanged:
\begin{display}{New Web Service Translation:}
\clause{\begin{prog}
I_{\mathit{ws}}(w) \triangleq \\
   \quad  \begin{prog}
     \Repl\Rcv{w}{c\ty\Un, n_p\ty\Un,\bdy \ty \Un,k_1 \ty \Un}\\ %

         \Case{\bdy} \IsTag{\request}{\getnonce()}{}\\
         \pDecrypt{c}{p\ty\Un,\mathit{vkp}\ty\annotate{\Dec}{\Key{\AuthMsg{p}}}}{\VK{CA}}\\ 
         \Res{n_q \ty \annotate{\public}{\Challenge{[\,]}}}\\
         \Snd{k_1}{(\CertVK{q},\response(\getnonce(n_q)))};\\
         \Rcv{w}{p'\ty\Un,\mathit{cipher}\ty\Un,k_2 \ty \Un}\\
         \SpiIf p=p' \SpiThen \\
         \begin{prog}
             \pDecrypt{\mathit{cipher}}{\request(\mathit{plain}\ty(w\ty\Un,a\ty\Un,t\ty\Un,q'\ty\Un,\\
                   \qquad\qquad\quad\annotate{\public}{\Response{[\endd{\request(p,q',w,a,t)}]}}))}{\mathit{vkp}}
         \end{prog}\\
         \Match{\mathit{plain}}{w}{\mathit{rest} \ty
                (\begin{prog}
                 a \ty \mathit{Req}(w),t\ty\Un,q'\ty\Un,\\
                  \annotate{\public}{\Response{[\endd{\request(p,q',w,a,t)}]}})}
                 \end{prog}\\
    \Split{\mathit{rest}}{\begin{prog}a\ty\mathit{Req}(w)}{\\t\ty\Un,\mathit{rest'}\ty(q'\ty\Un,\annotate{\public}{\Response{[\endd{\request(p,q',w,a,t)}]}})}\end{prog}\\
         \Match{\mathit{rest'}}{q}{n_q'\ty\annotate{\public}{\Response{[\endd{\request(p,q,w,a,t)}]}}}\\
         \CheckNonce{n_q}{n_q'}\\
    \End{\request(p,q,w,a,t)}
\end{prog}\end{prog}$

\\

$\begin{prog}\quad\begin{prog}
         \LetCall{r \ty \mathit{Res}(w)}{w}{p,a}\\
         \Begin{\response(p,q,w,r,t)} \\
         \Cast{n_p}{n'_p \ty
            \annotate{\public}{\Response{[\endd{\response(p,q,w,r,t)}]}}}\\
         \Snd{k_2}{(q,\asencr{\response(w,r,t,p,n'_p)}{\SK{q}})}

     \end{prog}\\
     \mbox{where $q=\owner(w)$}
   \end{prog}}
\end{display}

Finally, we need to change the top-level environment to account for
the new keys, and to add a channel through which we will publish the public keys. 
\begin{display}{Top-Level Environments:}
\clause{
E_{\mathit{class}} \triangleq
  (c\_\ell \ty \Un)\For{(c,\ell) \in \ClassMethods}}\\
\clause{E_{\mathit{keys}}\triangleq \DS{CA}\ty\AuthCertKeys,(\DS{p}\ty\AuthKeys{p})\For{p\in\Prin}}\\
\clause{E_{\mathit{ws}} \triangleq (w\ty\Un) \For{w \in \WebService}}\\
\clause{E_{\mathit{prin}} \triangleq
         p_1 \ty  \Princ, \ldots, p_n \ty  \Princ}
  {where $\Prin=\{p_1, \ldots, p_n\}$}\\
\clause{E_{\mathit{net}} \triangleq \mathit{net}\ty\Un}\\
\clause{E_0 \triangleq E_{\mathit{ws}}, E_{\mathit{prin}},E_{\mathit{net}},
                       E_{\mathit{class}}, E_{\mathit{keys}}}
\end{display}

Publishing can be achieved by simply sending the public keys on a
public channel, here $\mathit{net}$:
\begin{display}{Public Keys Publishing:}
\clause{
I_{\mathit{net}} \triangleq \Snd{\mathit{net}}{(\VK{CA},(\VK{p})\For{p\in\Prin})}}
\end{display}

We can now establish that the resulting system is robustly safe: 
\begin{theorem}
If $\Judge{\emptyset}{a : A}$ and $p \in \Prin$ and $k \notin \dom(E_0)$ then the system
\[\Res{E_{\mathit{class}}, E_{\mathit{keys}}}
    (I_{\mathit{net}} \parpop I_{\mathit{class}} \parpop
     I_{\mathit{ws}} \parpop
     \Res{k \ty \Un} \qq{a}^p_k)\]
is robustly safe.
\end{theorem}
\begin{proof}
Rather than giving a full proof, we point out the parts of the proof
of Theorem~\ref{t:rephrased-robust-safety-translation} that need to be
updated. Essentially, we need to show that the new semantics for web
method invocations is effect-free, and similarly for the new
implementation of web services. These occur in the proof of
Lemma~\ref{lemma:type-preservation-2}, part (2) and (4). 

As we did in Lemma~\ref{lemma:type-preservation-2}, rather than giving
the full type derivation for the translation of a web service call, we
outline the derivation of effects:

\vspace{1ex}
$\begin{prog}
             \Res{k_1 \ty \Un, k_2 \ty \Un, t\ty\Un,
                  n_p \ty \annotate{\public}{\Challenge{[\,]}}} \\ 
             \mbox{// Effect: $[\check{\public}{n_p}]$}\\ 
             \Begin{\request(p,q,w,\ell(\qq{u_1},\ldots,\qq{u_n}),t)} \\ 
             \mbox{// Effect: $[\check{\public}{n_p},\endd{\request(p,q,w,\ell(\qq{u_1},\ldots,\qq{u_n}),t)}]$}\\ 
              \Snd{w}{(\CertVK{p},n_p,\request(\getnonce()),k_1)}; \\ 
             \mbox{// Effect: $[\check{\public}{n_p},\endd{\request(p,q,w,\ell(\qq{u_1},\ldots,\qq{u_n}),t)}]$}\\ 
              \Rcv{k_1}{c\ty\Un,\response(\getnonce(n_q \ty \Un))} \\ 
             \mbox{// Effect: $[\check{\public}{n_p},\endd{\request(p,q,w,\ell(\qq{u_1},\ldots,\qq{u_n}),t)}]$}\\ 

\end{prog}$

$\begin{prog}
              \pDecrypt{c}{\mathit{cert}\ty(q'\ty\Un,\annotate{\Dec}{\Key{\AuthMsg{q'}}})}{\VK{CA}}\\ 
             \mbox{// Effect: $[\check{\public}{n_p},\endd{\request(p,q,w,\ell(\qq{u_1},\ldots,\qq{u_n}),t)}]$}\\ 
              \Match{\mathit{cert}}{q}{\mathit{vkq}\ty\annotate{\Dec}{\Key{\AuthMsg{q}}}}\\ 
             \mbox{// Effect: $[\check{\public}{n_p},\endd{\request(p,q,w,\ell(\qq{u_1},\ldots,\qq{u_n}),t)}]$}\\ 
              \Cast{n_q}{n'_q \ty
                \annotate{\public}{\Response{[\endd{\request(p,q,w,\ell(\qq{u_1},\ldots,\qq{u_n}),t)}]}}} \\ 
             \mbox{// Effect: $[\check{\public}{n_p}]$}\\ 
              \Snd{w}{(p,\asencr{\request(w,\ell(\qq{u_1},\ldots,\qq{u_n}),t,q,n'_q)}{\SK{p}},k_2)}; \\ 
             \mbox{// Effect: $[\check{\public}{n_p}]$}\\ 
              \Rcv{k_2}{q''\ty\Un,\bdy \ty \Un}\\ 
             \mbox{// Effect: $[\check{\public}{n_p}]$}\\ 
              \begin{prog}
                \pDecrypt{\bdy}{\response(\mathit{plain}\ty(w'\ty\Un,r\ty\Un,t'\ty\Un,p'\ty\Un,\\
                        \qquad\qquad\quad\annotate{\public}{\Response{[\endd{\response(p',q,w',r,t')}]}}))}{\mathit{vkq}}
         \end{prog}\\ 
             \mbox{// Effect: $[\check{\public}{n_p}]$}\\ 
              \Match{\mathit{plain}}{w}{\mathit{rest} \ty
                (\begin{prog}
                 r \ty \mathit{Res}(w), t'\ty\Un, p'\ty\Un,\\
                 \annotate{\public}{\Response{[\endd{\response(p',q,w,r,t')}]}})}
                 \end{prog}\\ 
             \mbox{// Effect: $[\check{\public}{n_p}]$}\\ 
         \Split{\mathit{rest}}{r\ty\mathit{Res}(w)}{\mathit{rest'}\ty
                (\begin{prog}
                 t'\ty\Un,p'\ty\Un,\\
                 \annotate{\public}{\Response{[\endd{\response(p',q,w,r,t')}]}})}
       \end{prog}\\ 
             \mbox{// Effect: $[\check{\public}{n_p}]$}\\ 
              \Match{\mathit{rest'}}{t}{\mathit{rest''}\ty(p'\ty\Un,\annotate{\public}{\Response{[\endd{\response(p',q,w,r,t)}]}})}\\ 
             \mbox{// Effect: $[\check{\public}{n_p}]$}\\ 
              \Match{\mathit{rest''}}{p}{n_p'\ty\annotate{\public}{\Response{[\endd{\response(p,q,w,r,t)}]}}}\\ 
             \mbox{// Effect: $[\check{\public}{n_p}]$}\\ 
         \CheckNonce{n_p}{n_p'}\\ 
             \mbox{// Effect: $[\endd{\response(p,q,w,r,t)}]$}\\ 
              \End{\response(p,q,w,r,t)}\\ 
             \mbox{// Effect: $[\,]$}\\ 
              \Case{r} \IsTag{\ell}{x}{\Snd{k}{x}} \\ 
             \mbox{// Effect: $[\,]$}
       \end{prog}$

\vspace{1ex}
For the new implementation of web service $w$, rather than giving the
full type derivation, we outline the derivation of effects:
\vspace{1ex}

$\begin{prog}
     \Repl\Rcv{w}{c\ty\Un, n_p\ty\Un,\bdy \ty \Un,k_1 \ty \Un}\\
     \mbox{// Effect: $[\,]$}\\
         \Case{\bdy} \IsTag{\request}{\getnonce()}{}\\
         \mbox{// Effect: $[\,]$}\\
         \pDecrypt{c}{p\ty\Un,\mathit{vkp}\ty\annotate{\Dec}{\Key{\AuthMsg{p}}}}{\VK{CA}}\\ 
         \mbox{// Effect: $[\,]$}\\
         \Res{n_q \ty \annotate{\public}{\Challenge{[\,]}}}\\
         \mbox{// Effect: $[\check{\public}{n_q}]$}\\
         \Snd{k_1}{(\CertVK{q},\response(\getnonce(n_q)))};\\
         \mbox{// Effect: $[\check{\public}{n_q}]$}\\
         \Rcv{w}{p'\ty\Un,\mathit{cipher}\ty\Un,k_2 \ty \Un}\\
         \mbox{// Effect: $[\check{\public}{n_q}]$}\\
         \SpiIf p=p' \SpiThen \\
         \mbox{// Effect: $[\check{\public}{n_q}]$}\\
 \end{prog}$

$\begin{prog}
         \begin{prog}
             \pDecrypt{\mathit{cipher}}{\request(\mathit{plain}\ty(w\ty\Un,a\ty\Un,t\ty\Un,q'\ty\Un,\\
                   \qquad\qquad\quad\annotate{\public}{\Response{[\endd{\request(p,q',w,a,t)}]}}))}{\mathit{vkp}}
    \end{prog}\\
         \mbox{// Effect: $[\check{\public}{n_q}]$}\\
         \Match{\mathit{plain}}{w}{\mathit{rest} \ty
                (\begin{prog}
                 a \ty \mathit{Req}(w),t\ty\Un,q'\ty\Un,\\
                  \annotate{\public}{\Response{[\endd{\request(p,q',w,a,t)}]}})}
                 \end{prog}\\
         \mbox{// Effect: $[\check{\public}{n_q}]$}\\
    \Split{\mathit{rest}}{\begin{prog}a\ty\mathit{Req}(w)}{\\t\ty\Un,\mathit{rest'}\ty(q'\ty\Un,\annotate{\public}{\Response{[\endd{\request(p,q',w,a,t)}]}})}\end{prog}\\
         \mbox{// Effect: $[\check{\public}{n_q}]$}\\
         \Match{\mathit{rest'}}{q}{n_q'\ty\annotate{\public}{\Response{[\endd{\request(p,q,w,a,t)}]}}}\\
         \mbox{// Effect: $[\check{\public}{n_q}]$}\\
         \CheckNonce{n_q}{n_q'}\\
         \mbox{// Effect: $[\endd{\request(p,q,w,a,t)}]$}\\
    \End{\request(p,q,w,a,t)}\\
         \mbox{// Effect: $[\,]$}\\ 
         \LetCall{r \ty \mathit{Res}(w)}{w}{p,a}\\
         \mbox{// Effect: $[\,]$}\\ 
         \Begin{\response(p,q,w,r,t)} \\
         \mbox{// Effect: $[\endd{\response(p,q,w,r,t)}]$}\\ 
         \Cast{n_p}{n'_p \ty
            \annotate{\public}{\Response{[\endd{\response(p,q,w,r,t)}]}}}\\
         \mbox{// Effect: $[\,]$}\\ 
         \Snd{k_2}{(q,\asencr{\response(w,r,t,p,n'_p)}{\SK{q}})}\\
         \mbox{// Effect: $[\,]$}
     \end{prog}$
\hfill
\end{proof}

The protocol we give above to provide authentication has some
undesirable properties. Specifically, it requires the server to
remember the certificate $\CertVK{p}$ and nonce $n_p$ at the time
when a nonce is requested. Since anyone can request a nonce, and no
authentication is performed at that stage of the protocol, this makes
the server severely vulnerable to denial-of-service attacks.  The
following variation on the protocol achieves the same guarantees, but 
pushes the exchange of certificates and nonces to later messages,
basically just when they are needed.
\[\begin{array}{l}
p \rightarrow q ~\mbox{on}~ w: \request(\getnonce()),k_1\\
q \rightarrow p ~\mbox{on}~ k_1: \response(\getnonce(n_q))\\
p \rightarrow q ~\mbox{on}~ w: p, \CertVK{p}, n_p, \asencr{\request(w,\ell(u_1,\ldots,u_n),t,q,n_q)}{\SK{p}},k_2\\
q \rightarrow p ~\mbox{on}~ k_2: q, \CertVK{q}, \asencr{\response(w,\ell(r),t,p,n_p)}{\SK{q}}
\end{array}\]

\subsection{Authenticated and Encrypted Web Methods} We now describe a protocol and 
implementation for authenticated and encrypted web methods. Hence, for
now, we assume that all the exported methods of a web service are
annotated with \texttt{AuthEnc}.

The public-key infrastructure we consider for this case is similar
to the one for authenticated web methods, except that now we have
encryption and decryption keys, as opposed to signing and verification
keys. Each principal $p$ has an encryption key $\EK{p}$ and a
decryption key $\DK{p}$. The decryption key is kept private, while the
encryption key is public. To bind the name of a principal with their
encryption key, we again assume a \emph{certification authority} $\CA$
(with a signing key $\SK{CA}$ and verification key $\VK{CA}$)
that can sign certificates $\CertEK{p}$ of the form
$\asencr{p,\EK{p}}{\SK{CA}}$.

Here is a protocol for $p$ making a web service call
$w\mo{:}\ell(u_1,\ldots,u_n)$ to service $w$ owned by $q$, including
the names of continuation channels used at the spi level.  Again, we
assume that in addition to the methods of $\class(w)$, each web
service also supports a method $\getnonce$, which we implement
specially. The protocol uses public key encryption to exchange a
session-specific shared key $K$ used to encrypt the actual method
call. 
\[\begin{array}{l}
p \rightarrow q ~\mbox{on}~ w: \CertEK{p}, \request(\getnonce()),k_1\\
q \rightarrow p ~\mbox{on}~ k_1: \CertEK{q}, \asencr{\msgI(q,n_K)}{\EK{p}}, \response(\getnonce(n_q))\\
p \rightarrow q ~\mbox{on}~ w: \asencr{\msgII(w,p,K,n_K)}{\EK{q}},n_p, \sencr{\request(\ell(u_1,\ldots,u_n),t,n_q)}{K},k_2\\
q \rightarrow p ~\mbox{on}~ k_2: \sencr{\response(\ell(r),t,n_p)}{K}
\end{array}\]

\begin{display}{Type of Keys:}
\clause{
  \begin{prog}
    \SKey{p,q,w} \triangleq \\ \quad
      \SharedKey{\Union(\begin{prog}
        \request(a \ty \Un, 
                 t \ty \Un,
                 n_q \ty \annotate{\public}{\Response{[\endd{\request(p,q,w,a,t)}]}}),\\
        \response(r \ty \Un, 
                  t \ty \Un,
                  n_p \ty \annotate{\public}{\Response{[\endd{\response(p,q,w,r,t)}]}}))}
         \end{prog}
  \end{prog}}\\
\clause{
  \begin{prog}
    \AuthEncMsg{p} \triangleq \\ \quad
    \begin{prog}
    \Union (\begin{prog}
                  \msgI (\begin{prog}
                         q \ty \Un,
                         n_K \ty \annotate{\private}{\Challenge{[\,]}}),
                    \end{prog}\\
                  \msgII (\begin{prog}
                             w \ty \Un,
                             q \ty \Un,
                             K \ty \Top,\\
                             n_K \ty \annotate{\private}{\Response{[\trust{K\ty\SKey{p,q,w}}]}}))
              \end{prog}
             \end{prog}
    \end{prog}
  \end{prog}}    \\
\clause{\AuthEncKeys{p} \triangleq \KeyPair{\AuthEncMsg{p}}}\\
\clause{\AuthEncCert \triangleq (p\ty\Un, \annotate{\Enc}{\Key{\AuthEncMsg{p}}})}\\
\clause{\AuthEncCertKeys \triangleq \KeyPair{\AuthEncCert}}
\end{display}

We will represent the key pair of an encryption key and decryption key
for principal $p$ by a pair $\PK{p}$, of type $\AuthEncKeys{p}$. The
signing key pair for the certification authority will be represented
by a pair $\DS{CA}$. We use the following abbreviations:
\begin{display}{Key and Certificates Abbreviations:}
\clause{\EK{p} \triangleq \extract{\Enc}{\PK{p}}}{$p$'s encryption key}\\
\clause{\DK{p} \triangleq \extract{\Dec}{\PK{p}}}{$p$'s decryption key}\\
\clause{\CertEK{p} \triangleq \asencr{p,\EK{p}}{\SK{CA}}}{$p$'s certificate}
\end{display}

Again, we can amend the translation of
Section~\ref{sec:spi-calculus-semantics} to accommodate the new
protocol. First, we give a new translation for a web method call
$w\mo{:}\ell(u_1,\ldots,u_n)$: 
\begin{display}{New Semantics of Web Method Call:}
\clause{\begin{prog}
          \TransB{w\mo{:}\ell(u_1,\ldots,u_n)}{p}{k} \triangleq {} \\ \quad
             \Res{k_1 \ty \Un, k_2 \ty \Un, t\ty\Un,
                  n_p \ty \annotate{\public}{\Challenge{[\,]}}} \\ \quad
             \Begin{\request(p,q,w,\ell(\qq{u_1},\ldots,\qq{u_n}),t)} \\ \quad
              \Snd{w}{(\CertEK{p},\request(\getnonce()),k_1)}; \\ \quad
              \Rcv{k_1}{c\ty\Un,\mathit{cipher}\ty\Un,\response(\getnonce(n_q \ty \Un))}

         \end{prog}}\\
\clause{\begin{prog}\quad

              \pDecrypt{c}{\mathit{cert}\ty(q'\ty\Un,\annotate{\Enc}{\Key{\AuthEncMsg{q'}}})}{\VK{CA}}\\ \quad
              \Match{\mathit{cert}}{q}{\mathit{ekq}\ty\annotate{\Enc}{\Key{\AuthEncMsg{q}}}}\\ \quad
              \pDecrypt{\mathit{cipher}}{\msgI(q'\ty\Un,n_K\ty\Un)}{\DK{p}}\\ \quad
              \SpiIf q=q' \SpiThen\\ \quad
              \Cast{n_q}{n'_q \ty
                \annotate{\public}{\Response{[\endd{\request(p,q,w,\ell(\qq{u_1},\ldots,\qq{u_n}),t)}]}}} \\ \quad
              \Res{K \ty \SKey{p,q,w}}\\ \quad
              \Witness{K\ty\SKey{p,q,w}}\\ \quad
              \Cast{n_K}{n'_K \ty
                \annotate{\private}{\Response{[\trust{K\ty\SKey{p,q,w}}]}}}\\ \quad
              \Snd{w}{(\asencr{\msgII(w,p,t,K,n_K')}{\mathit{ekq}},n_p,
                       \sencr{\request(w,\ell(\qq{u_1},\ldots,\qq{u_n}),t,n'_q)}{K},k_2)}; \\ \quad
              \Rcv{k_2}{\bdy \ty \Un}\\ \quad
              \begin{prog}
                 \Decrypt{\bdy}{\response(\mathit{plain}\ty(r\ty\mathit{Res}(w),t'\ty\Un,\\
                   \qquad\qquad\quad\annotate{\public}{\Response{[\endd{\response(p,q,w,r,t')}]}}))}{K}
         \end{prog}\\ \quad
              \Match{\mathit{plain}}{r\ty\mathit{Res}(w)}
                    {\mathit{rest}\ty(t'\ty\Un,\annotate{\public}{\Response{[\endd{\response(p,q,w,r,t')}]}})}\\ \quad
              \Match{\mathit{rest}}{t}{n_p'\ty\annotate{\public}{\Response{[\endd{\response(p,q,w,r,t)}]}}}\\ \quad
         \CheckNonce{n_p}{n_p'}\\ \quad
              \End{\response(p,q,w,r,t)}\\ \quad
              \Case{r} \IsTag{\ell}{x}{\Snd{k}{x}}\\
             \mbox{where $q=\owner(w)$}
       \end{prog}}
\end{display}

We also need to give a new implementation for web services, again to
take into account the different messages being exchanged:
\begin{display}{New Web Service Translation:}
\clause{\begin{prog}
I_{\mathit{ws}}(w) \triangleq \\ \quad
     \begin{prog}
     \Repl\Rcv{w}{c\ty\Un,\bdy \ty \Un,k_1 \ty \Un}\\
         \Case{\bdy} \IsTag{\request}{\getnonce()}{}\\
         \pDecrypt{c}{p\ty\Un,\mathit{ekp}\ty\annotate{\Enc}{\Key{\AuthEncMsg{p}}}}{\VK{CA}}\\
         \Res{n_q \ty \annotate{\public}{\Challenge{[\,]}}}\\
         \Res{n_K \ty \annotate{\private}{\Challenge{[\,]}}}\\
         \Snd{k_1}{(\CertEK{q},\asencr{\msgI(q,n_K)}{\mathit{ekp}},\response(\getnonce(n_q)))};\\
         \Rcv{w}{\mathit{cipher}_1\ty\Un,n_p \ty \Un, \mathit{cipher}_2\ty\Un,k_2 \ty \Un}\\
         \begin{prog}
           \pDecrypt{\mathit{cipher}_1\\\qquad}{\msgII(\mathit{plain}_1\ty(w\ty\Un,p'\ty\Un,K\ty\Top,\\
              \qquad\qquad\quad\annotate{\private}{\Response{[\trust{K\ty\SKey{p',q,w}}]}}))}{\DK{q}}
         \end{prog}\\
         \Match{\mathit{plain}_1}{w}{\mathit{rest}\ty
           (\begin{prog}
             p'\ty\Un,K\ty\Top,\\
             \annotate{\private}{\Response{[\trust{K\ty\SKey{p',q,w}}]}})}
            \end{prog}\\
         \Match{\mathit{rest}}{p}{\mathit{rest'}\ty(K\ty\Top,
                   \annotate{\private}{\Response{[\trust{K\ty\SKey{p,q,w}}]}})}\\
         \Split{\mathit{rest'}}{K\ty\Top}{n'_K\ty\annotate{\private}{\Response{[\trust{K\ty\SKey{p,q,w}}]}}}\\
         \CheckNonce{n_K}{n_K'}\\
         \Trust{K}{K'\ty\SKey{p,q,w}}\\
         \begin{prog}
            \Decrypt{\mathit{cipher}_2}{\request(\mathit{plain}_2\ty(a\ty\mathit{Req}(w),t\ty\Un,\\
            \qquad\qquad\quad\annotate{\public}{\Response{[\endd{\request(p,q,w,a,t)}]}}))}{K'}
         \end{prog}\\
    \Split{\mathit{plain}_2}{a\ty\mathit{Req}(w)}{t\ty\Un,n_q'\ty\annotate{\public}{\Response{[\endd{\request(p,q,w,a,t)}]}}}\\
         \CheckNonce{n_q}{n_q'}\\
    \End{\request(p,q,w,a,t)}\\

\end{prog}\end{prog}}\\
\clause{\begin{prog}\quad\begin{prog}

         \LetCall{r \ty \mathit{Res}(w)}{w}{p,a}\\
         \Begin{\response(p,q,w,r,t)} \\
         \Cast{n_p}{n'_p \ty
            \annotate{\public}{\Response{[\endd{\response(p,q,w,r,t)}]}}}\\
         \Snd{k_2}{\sencr{\response(r,t,n'_p)}{K'}}
     \end{prog}\\
     \mbox{where $q=\owner(w)$}
     \end{prog}}
\end{display}

Finally, we need to change the top-level environment to account for
the new keys, and to add a channel through which we will publish the public keys. 
\begin{display}{Top-Level Environments:}
\clause{
E_{\mathit{class}} \triangleq
  (c\_\ell \ty \Un)\For{(c,\ell) \in \ClassMethods}}\\
\clause{E_{\mathit{keys}}\triangleq \DS{CA}\ty\AuthEncCertKeys,(\PK{p}\ty\AuthEncKeys{p})\For{p\in\Prin}}\\
\clause{E_{\mathit{ws}} \triangleq (w\ty\Un) \For{w \in \WebService}}\\
\clause{E_{\mathit{prin}} \triangleq
         p_1 \ty  \Princ, \ldots, p_n \ty  \Princ}
  {where $\Prin=\{p_1, \ldots, p_n\}$}\\
\clause{E_{\mathit{net}} \triangleq \mathit{net}\ty\Un}\\
\clause{E_0 \triangleq E_{\mathit{ws}}, E_{\mathit{prin}},E_{\mathit{net}},
                       E_{\mathit{class}}, E_{\mathit{keys}}}
\end{display}

Publishing can be achieved by simply sending the public keys on a
public channel, here $\mathit{net}$:
\begin{display}{Public Keys Publishing:}
\clause{
I_{\mathit{net}} \triangleq \Snd{\mathit{net}}{(\VK{CA},(\EK{p})\For{p\in\Prin})}}
\end{display}

We can now establish that the resulting system is robustly safe: 
\begin{theorem}
If $\Judge{\emptyset}{a : A}$ and $p \in \Prin$ and $k \notin \dom(E_0)$ then the system
\[\Res{E_{\mathit{class}}, E_{\mathit{keys}}}
    (I_{\mathit{net}} \parpop I_{\mathit{class}} \parpop
     I_{\mathit{ws}} \parpop
     \Res{k \ty \Un} \qq{a}^p_k)\]
is robustly safe.
\end{theorem}
\begin{proof}
Rather than giving a full proof, we point out the parts of the proof
of Theorem~\ref{t:rephrased-robust-safety-translation} that need to be
updated. Essentially, we need to show that the new semantics for web
method invocations is effect-free, and similarly for the new
implementation of web services. These occur in the proof of
Lemma~\ref{lemma:type-preservation-2}, part (2) and (4). 

As we did in Lemma~\ref{lemma:type-preservation-2}, rather than giving
the full type derivation for the translation of a web service call, we
outline the derivation of effects:
\vspace{1ex}

$\begin{prog}
     \Res{k_1 \ty \Un, k_2 \ty \Un, t\ty\Un,
          n_p \ty \annotate{\public}{\Challenge{[\,]}}} \\
     \mbox{// Effect: $[\check{\public}{n_p}]$}\\
     \Begin{\request(p,q,w,\ell(\qq{u_1},\ldots,\qq{u_n}),t)} \\ 
     \mbox{// Effect: $[\check{\public}{n_p},\endd{\request(p,q,w,\ell(\qq{u_1},\ldots,\qq{u_n}),t)}]$}\\
     \Snd{w}{(\CertEK{p},\request(\getnonce()),k_1)}; \\
     \mbox{// Effect: $[\check{\public}{n_p},\endd{\request(p,q,w,\ell(\qq{u_1},\ldots,\qq{u_n}),t)}]$}
\end{prog}$

$\begin{prog}
     \Rcv{k_1}{c\ty\Un,\mathit{cipher}\ty\Un,\response(\getnonce(n_q \ty \Un))} \\ 
     \mbox{// Effect: $[\check{\public}{n_p},\endd{\request(p,q,w,\ell(\qq{u_1},\ldots,\qq{u_n}),t)}]$}\\ 
     \pDecrypt{c}{\mathit{cert}\ty(q'\ty\Un,\annotate{\Enc}{\Key{\AuthEncMsg{q'}}})}{\VK{CA}}\\ 
     \mbox{// Effect: $[\check{\public}{n_p},\endd{\request(p,q,w,\ell(\qq{u_1},\ldots,\qq{u_n}),t)}]$}\\ 
     \Match{\mathit{cert}}{q}{\mathit{ekq}\ty\annotate{\Enc}{\Key{\AuthEncMsg{q}}}}\\ 
     \mbox{// Effect: $[\check{\public}{n_p},\endd{\request(p,q,w,\ell(\qq{u_1},\ldots,\qq{u_n}),t)}]$}\\ 
     \pDecrypt{\mathit{cipher}}{\msgI(q'\ty\Un,n_K\ty\Un)}{\DK{p}}\\ 
     \mbox{// Effect: $[\check{\public}{n_p},\endd{\request(p,q,w,\ell(\qq{u_1},\ldots,\qq{u_n}),t)}]$}\\ 
     \SpiIf q=q' \SpiThen\\ 
     \mbox{// Effect: $[\check{\public}{n_p},\endd{\request(p,q,w,\ell(\qq{u_1},\ldots,\qq{u_n}),t)}]$}\\ 
     \Cast{n_q}{n'_q \ty
         \annotate{\public}{\Response{[\endd{\request(p,q,w,\ell(\qq{u_1},\ldots,\qq{u_n}),t)}]}}} \\ 
     \mbox{// Effect: $[\check{\public}{n_p}]$}\\ 
     \Res{K \ty \SKey{p,q,w}}\\ 
     \mbox{// Effect: $[\check{\public}{n_p}]$}\\ 
     \Witness{K\ty\SKey{p,q,w}}\\ 
     \mbox{// Effect: $[\check{\public}{n_p},\trust{K\ty\SKey{p,q,w}}]$}\\ 
     \Cast{n_K}{n'_K \ty
           \annotate{\private}{\Response{[\trust{K\ty\SKey{p,q,w}}]}}}\\ 
     \mbox{// Effect: $[\check{\public}{n_p}]$}\\ 
     \Snd{w}{(\asencr{\msgII(w,p,t,K,n_K')}{\mathit{ekq}},n_p,
              \sencr{\request(w,\ell(\qq{u_1},\ldots,\qq{u_n}),t,n'_q)}{K},k_2)}; \\ 
     \mbox{// Effect: $[\check{\public}{n_p}]$}\\ 
     \Rcv{k_2}{\bdy \ty \Un}\\ 
     \mbox{// Effect: $[\check{\public}{n_p}]$}\\ 
              \begin{prog}
                 \Decrypt{\bdy}{\response(\mathit{plain}\ty(r\ty\mathit{Res}(w),t'\ty\Un,\\
                   \qquad\qquad\quad\annotate{\public}{\Response{[\endd{\response(p,q,w,r,t')}]}}))}{K}
         \end{prog}\\ 
     \mbox{// Effect: $[\check{\public}{n_p}]$}\\ 
     \Match{\mathit{plain}}{r\ty\mathit{Res}(w)}
            {\mathit{rest}\ty(t'\ty\Un,\annotate{\public}{\Response{[\endd{\response(p,q,w,r,t')}]}})}\\ 
     \mbox{// Effect: $[\check{\public}{n_p}]$}\\ 
     \Match{\mathit{rest}}{t}{n_p'\ty\annotate{\public}{\Response{[\endd{\response(p,q,w,r,t)}]}}}\\ 
     \mbox{// Effect: $[\check{\public}{n_p}]$}\\ 
     \CheckNonce{n_p}{n_p'}\\ 
     \mbox{// Effect: $[\endd{\response(p,q,w,r,t)}]$}\\ 
     \End{\response(p,q,w,r,t)}\\ 
     \mbox{// Effect: $[\,]$}\\ 
     \Case{r} \IsTag{\ell}{x}{\Snd{k}{x}}\\ 
     \mbox{// Effect: $[\,]$}
 \end{prog}$
\vspace{1ex}

For the new implementation of web service $w$, rather than giving the
full type derivation, we outline the derivation of effects:
\vspace{1ex}

$\begin{prog}
     \Repl\Rcv{w}{c\ty\Un,\bdy \ty \Un,k_1 \ty \Un}\\
     \mbox{// Effect: $[\,]$}\\
         \Case{\bdy} \IsTag{\request}{\getnonce()}{}\\
         \mbox{// Effect: $[\,]$}\\
         \pDecrypt{c}{p\ty\Un,\mathit{ekp}\ty\annotate{\Enc}{\Key{\AuthEncMsg{p}}}}{\VK{CA}}\\
         \mbox{// Effect: $[\,]$}\\
         \Res{n_q \ty \annotate{\public}{\Challenge{[\,]}}}\\
         \mbox{// Effect: $[\check{\public}{n_q}]$}\\
 \end{prog}$

$\begin{prog}
         \Res{n_K \ty \annotate{\private}{\Challenge{[\,]}}}\\
         \mbox{// Effect: $[\check{\public}{n_q},\check{\private}{n_K}]$}\\
         \Snd{k_1}{(\CertEK{q},\asencr{\msgI(q,n_K)}{\mathit{ekp}},\response(\getnonce(n_q)))};\\
         \mbox{// Effect: $[\check{\public}{n_q},\check{\private}{n_K}]$}\\
         \Rcv{w}{\mathit{cipher}_1\ty\Un,n_p \ty \Un, \mathit{cipher}_2\ty\Un,k_2 \ty \Un}\\
         \mbox{// Effect: $[\check{\public}{n_q},\check{\private}{n_K}]$}\\
         \begin{prog}
           \pDecrypt{\mathit{cipher}_1\\\qquad}{\msgII(\mathit{plain}_1\ty(w\ty\Un,p'\ty\Un,K\ty\Top,\\
              \qquad\qquad\quad\annotate{\private}{\Response{[\trust{K\ty\SKey{p',q,w}}]}}))}{\DK{q}}
         \end{prog}\\
         \mbox{// Effect: $[\check{\public}{n_q},\check{\private}{n_K}]$}\\
         \Match{\mathit{plain}_1}{w}{\mathit{rest}\ty
           (\begin{prog}
             p'\ty\Un,K\ty\Top,\\
             \annotate{\private}{\Response{[\trust{K\ty\SKey{p',q,w}}]}})}
            \end{prog}\\
         \mbox{// Effect: $[\check{\public}{n_q},\check{\private}{n_K}]$}\\
         \Match{\mathit{rest}}{p}{\mathit{rest'}\ty(K\ty\Top,
                   n'_K\ty\annotate{\private}{\Response{[\trust{K\ty\SKey{p,q,w}}]}})}\\
         \mbox{// Effect: $[\check{\public}{n_q},\check{\private}{n_K}]$}\\
         \Split{\mathit{rest'}}{K\ty\Top}{n'_K\ty\annotate{\private}{\Response{[\trust{K\ty\SKey{p,q,w}}]}}}\\
         \mbox{// Effect: $[\check{\public}{n_q},\check{\private}{n_K}]$}\\
         \CheckNonce{n_K}{n_K'}\\
         \mbox{// Effect: $[\check{\public}{n_q},\trust{K\ty\SKey{p,q,w}}]$}\\
         \Trust{K}{K'\ty\SKey{p,q,w}}\\
         \mbox{// Effect: $[\check{\public}{n_q}]$}\\
         \begin{prog}
            \Decrypt{\mathit{cipher}_2}{\request(\mathit{plain}_2\ty(a\ty\mathit{Req}(w),t\ty\Un,\\
            \qquad\qquad\quad\annotate{\public}{\Response{[\endd{\request(p,q,w,a,t)}]}}))}{K'}
         \end{prog}\\
         \mbox{// Effect: $[\check{\public}{n_q}]$}\\
         \Split{\mathit{plain}_2}{a\ty\mathit{Req}(w)}{t\ty\Un,n_q'\ty\annotate{\public}{\Response{[\endd{\request(p,q,w,a,t)}]}}}\\
         \mbox{// Effect: $[\check{\public}{n_q}]$}\\
         \CheckNonce{n_q}{n_q'}\\
         \mbox{// Effect: $[\endd{\request(p,q,w,a,t)}]$}\\
         \End{\request(p,q,w,a,t)}\\
         \mbox{// Effect: $[\,]$}\\
         \LetCall{r \ty \mathit{Res}(w)}{w}{p,a}\\
         \mbox{// Effect: $[\,]$}\\
         \Begin{\response(p,q,w,r,t)} \\
         \mbox{// Effect: $[\endd{\response(p,q,w,r,t)}]$}\\
         \Cast{n_p}{n'_p \ty
            \annotate{\public}{\Response{[\endd{\response(p,q,w,r,t)}]}}}\\
         \mbox{// Effect: $[\,]$}\\
         \Snd{k_2}{\sencr{\response(r,t,n'_p)}{K'}}\\
         \mbox{// Effect: $[\,]$}
     \end{prog}$
\hfill
\end{proof}

We can note some further possibilities, with respect to the protocols
implemented in this section:
\begin{itemize}
\item
The protocol implementing authenticated and encrypted invocation uses
certificates to essentially negotiate a symmetric key with which to
actually perform the encryption. It is straightforward to apply the
same idea to the authenticated-only case, negotiating a symmetric key
with which to hash the content of the method call (instead of relying
on public-key signatures).
\item
In the above protocol, a new symmetric key is negotiated at every
method invocation. A more efficient variation would be to re-use a
negotiated symmetric key over multiple web method calls. Once a
symmetric key has been negotiated, it can effectively act as a shared
key between the two principals, which is the case we investigated in
the body of this paper. We can therefore use the above protocol for
the first web method call between a principal and a particular
service, and the shared-key protocol for subsequent web method calls. 
\end{itemize}
 \fi%
\fi%

\ifLong
\section{First-Class Web Services}\label{app:first-class-services}

The model of web services captured by our calculus in
Section~\ref{sec:formal-model} does not consider web services to be
values. This reflects the fact that current WSDL does not allow for
web services to be passed as requests or results. On the other hand, a 
web service has a simple representation as a string, namely the URL
used to access the web service, and this string \emph{can} be passed
as a request or a result. Hence, it is possible, in a sense, to pass
web services as values given the current web services
infrastructure. In this section, we explore an extension of our object 
calculus that allows web services as first-class values. The main
point here is to show that there is no real difficulty in modelling
this aspect of the web services infrastructure.
Our main result is type safety.
We expect it would be straightforward to translate this extended calculus into
the spi-calculus, but we do not describe this in detail.

For the sake of keeping this section essentially self-contained, we
give the full syntax and semantics of the extended object calculus.

\subsection{Syntax}

We assume finite sets $\Prin$, $\WebService$, $\Class$, $\Field$,
$\Meth$ of principal, web service, class, field, and method
names, respectively.
\ifJournal\clearpage\fi
\begin{renewcommand}{\ratio}{.5}
\begin{display}{Classes, Fields, Methods, Principals, Web Services:}
\clause{c \in \Class}{class name}\\
\clause{f \in \Field}{field name}\\
\clause{\ell \in \Meth}{method name}\\
\clause{p \in \Prin}{principal name}\\
\clause{w \in \WebService}{web service name}
\end{display}
\end{renewcommand}

There are now three kinds of data type: $\Id$ is the type of principal 
identifiers, $c \in \Class$ is the type of instances of class $c$, and 
$\WS(c)$ is the type of web services with implementation class
$c\in\Class$. A method signature specifies the types of its arguments
and result. 
\begin{display}{Types and Method Signatures:}
\Category{A,B \in \Type}{type}\\
\entry{\Id}{principal identifier}\\
\entry{c}{object}\\
\entry{\WS(c)}{web service}\\
\clause{\sig \in \Sig ::= B(A_1\:x_1,\ldots,A_n\:x_n)}
  {method signature ($x_i$ distinct)}
\end{display}

As in Section~\ref{sec:formal-model}, an execution environment defines 
the services and code available in the distributed system.
\begin{renewcommand}{\ratio}{.49}
\begin{display}{Execution Environment: $(\fields,\methods,\owner,\class)$}
\clause{\fields \in \Class \to (\Field \finmap \Type)}
  {fields of a class}\\
\clause{\methods \in \Class \to (\Meth \finmap \Sig \times \Body)}
  {methods of a class}\\
\clause{\owner \in \WebService \to \Prin}
  {service owner}\\
\clause{\class \in \WebService \to \Class}
  {service implementation}
\end{display}
\end{renewcommand}

The owner and implementation class of a web service need not be globally
known. We can assume that the representation of a web service $w$ carries
representations of its owner and its implementation class, which $\class$ and
$\owner$ simply read off. Since we assume web services are given, and we do
not provide for ways to actually create new web services, there is no loss of
generality in taking this particular approach.

The syntax of method bodies and values is that of the original object
calculus, with the differences that web services are values, and that
we do not assume that web service invocations require a fixed web
service.
\begin{renewcommand}{\ratio}{.34}
\begin{display}{Values and Method Bodies:}
\clause{x,y,z}{name: variable, argument}\\
\Category{u,v \in \Value}{value}\\
\entry{x}{variable}\\
\entry{\Null}{null}\\
\entry{\New\:c(v_1,\ldots,v_n)}{object}\\
\entry{p}{principal identifier}\\
\entry{w}{web service}\\
\Category{a,b \in \Body}{method body}\\
\entry{v}{value}\\
\entry{\Let{x}{a} \In{b}}{let-expression}\\
\entry{\If u=v \Then a \Else b}{conditional}\\
\entry{v.f}{field lookup}\\
\entry{v.\ell(u_1,\ldots,u_n)}{method call}\\
\entry{v\mo{:}\ell(u_1,\ldots,u_n)}{service call}\\
\entry{p[a]}{body $a$ running as $p$}
\end{display}
\end{renewcommand}

We again require a method body of the form $p[a]$, meaning $p$ running 
body $a$, to keep track of which principal is running a method body in 
the upcoming operational semantics. 

\subsection{Operational Semantics} 

The operational semantics is defined by a transition relation, written
$a \to^p a'$, where $a$ and $a'$ are method bodies, and $p$ is the
principal evaluating the body $a$.
\begin{display}{Transitions:}
\staterule
  {(Red Let 1)}
  {a \to^p a'}
  {\Let{x}{a} \In{b} \to^p \Let{x}{a'} \In{b}}
\staterule
  {(Red Let 2)}
  {}
  {\Let{x}{v} \In{b} \to^p b\SUB{x \GETS v}}
\ifJournal\else \\[\GAP]\fi
\staterule
  {(Red If)}
  {}
  {\If u=v \Then a_{\mathit{true}} \Else a_{\mathit{false}} \to^p a_{u=v}}
  \\[\GAP]
\staterule
  {(Red Field)}
  {\fields(c) = f_i \mapsto A_i \For{i \in 1..n} \quad j \in 1..n}
  {(\New\:c(v_1,\ldots,v_n)).f_j \to^p v_j}
  \\[\GAP]
\staterule
  [(where $v=\New\:c(v_1,\ldots,v_n)$)]
  {(Red Invoke)}
  {\methods(c) = \ell_i \mapsto (\sig_i,b_i) \For{i \in 1..n} \quad
   j \in 1..n \quad \sig_j=B(A_1\:x_1,\ldots,A_m\:x_m)}
  {v.\ell_j(u_1,\ldots,u_m) \to^p
     b_j\SUB{\mathit{this} \GETS v,x_k \GETS u_k \For{k \in 1..m}}}
  \\[\GAP]
\staterule
  {(Red Remote)}
  {\owner(w)=q \quad \class(w)=c}
  {w\mo{:}\ell(u_1,\ldots,u_n) \to^p q[\New\:c(p).\ell(u_1,\ldots,u_n)]}
\ifJournal\else \\[\GAP]\fi
\staterule
  {(Red Prin 1)}
  {a \to^q a'}
  {q[a] \to^p q[a']}
\staterule
  {(Red Prin 2)}
  {}
  {q[v] \to^p v}
\end{display}

\subsection{Type System} 

The judgments of our type system all depend on an \emph{environment}
$E$, that defines the types of all variables in scope. An environment
takes the form $x_1\ty A_1,\ldots,x_n\ty A_n$ and defines the type
$A_i$ for each variable $x_i$. The domain $\dom(E)$ of an environment
$E$ is the set of variables whose types it defines. 

\begin{display}{Environments:}
\Category{D,E}{environment}\\
\entry{\emptyset}{empty}\\
\entry{E,x\ty A}{entry}\\
\clause{\dom(x_1\ty A_1,\ldots,x_n\ty A_n) \triangleq \{x_1,\ldots,x_n\}}{domain of an environment}
\end{display}

The following are the two judgments of our type system. They are
inductively defined by rules presented in the following tables. 

\begin{display}{Judgments $\Judge{E}{\mathcal{J}}$:}
\clause{\JudgeOK{E}}{good environment}\\
\clause{\Judge{E}{a:A}}{good expression $a$ of type $A$}
\end{display}%
We write $\Judge{E}{\mathcal{J}}$ when we want to talk about both
kinds of judgments, where $\mathcal{J}$ stands for either $\diamond$
or $a:A$.

The following rules define an environment
$x_1\ty A_1,\ldots,x_n\ty A_n$ to be well-formed if each of the
names $x_1,\ldots,x_n$ are distinct. 

\begin{display}{Rules for Environments:}
\typerule
  {(Env $\emptyset$)}
  {}
  {\JudgeOK{\emptyset}}
\typerule
  [(where $x\not\in\dom(E)$)]
  {(Env $x$)}
  {\JudgeOK{E}}
  {\JudgeOK{E,x\ty A}}
\end{display}

We present the rules for deriving the judgment $\Judge{E}{a:A}$ that
assigns a type $A$ to a value or method body $a$. These rules are
split into two tables, one for values, and one for method bodies. 

\begin{display}{Rules for Typing Values:}
\typerule
  {(Val $x$)}
  {E=E_1,x\ty A,E_2 \quad \JudgeOK{E}}
  {\Judge{E}{x : A}}
\typerule
  {(Val $\Null$)}
  {\JudgeOK{E}}
  {\Judge{E}{\Null:c}}
\typerule
  {(Val WS)}
  {\JudgeOK{E} \quad \class(w)=c}
  {\Judge{E}{w : \WS(c)}}
  \\[\GAP]
\typerule
  {(Val Object)}
  {\fields(c)=f_i \mapsto A_i \For{i \in 1..n} \quad
  \Judge{E}{v_i:A_i} \quad \forall i\in 1..n}
  {\Judge{E}{\New\:c(v_1,\ldots,v_n):c}}
\typerule
  {(Val Princ)}
  {\JudgeOK{E}}
  {\Judge{E}{p:\Id}}
\end{display}

\begin{display}{Rules for Typing Method Bodies:}
\typerule
  {(Body Let)}
  {\Judge{E}{a:A} \quad \Judge{E,x\ty A}{b:B}}
  {\Judge{E}{\Let{x}{a} \In{b}:B}}
\ifJournal\else \\[\GAP]\fi
\typerule
  {(Body If)}
  {\Judge{E}{u:A} \quad \Judge{E}{v:A} \quad \Judge{E}{a:B} \quad \Judge{E}{b:B}}
  {\Judge{E}{\If u=v \Then a \Else b:B}}
  \\[\GAP]
\typerule
  {(Body Field)}
  {\Judge{E}{v:c} \quad \fields(c)=f_i \mapsto A_i \For{i \in 1..n} \quad j \in 1..n}
  {\Judge{E}{v.f_j : A_j}}
  \\[\GAP]
\typerule
  {(Body Invoke)}
  {\Judge{E}{v:c} \quad \methods(c)=\ell_i \mapsto (\sig_i,b_i) \For{i \in 1..n} \quad j \in 1..n \\
   \sig_j = B (A_1\:x_1,\ldots,A_m\:x_m) \quad
   \Judge{E}{u_k:A_k} \quad \forall k \in 1..m}
  {\Judge{E}{v.\ell_j(u_1,\ldots,u_m) : B}}
  \\[\GAP]
\typerule
  {(Body Remote)}
  {\Judge{E}{v : \WS(c)} \\
  \methods(c)=\ell_i \mapsto (\sig_i,b_i) \For{i \in 1..n}
  \quad j \in 1..n \\
   \sig_j = B (A_1\:x_1,\ldots,A_m\:x_m) \quad
   \Judge{E}{u_i:A_i} \quad \forall i \in 1..m}
  {\Judge{E}{v\mo{:}\ell_j(u_1,\ldots,u_m) : B}}
\typerule
   {(Body Princ)}
   {\Judge{E}{a : A}}
   {\Judge{E}{p[a] : A}}
\end{display}

We make the following assumption on the execution environment.
\begin{display}{Assumptions on the Execution Environment:}
(1) \mbox{For each $w \in \WebService$, $\fields(\class(w)) =
    \mathit{CallerId}:\Id$.}\\
(2) \begin{prog}
    \mbox{No tagged expression $p[a]$ occurs within the body of any
method;}\\
    \mbox{such expressions occur only at runtime, to track the call
    stack of principals.}
    \end{prog}\\
(3) \begin{prog}
    \mbox{for each $c \in \Class$ and each $\ell \in
       \dom(\methods(c))$,}\\
    \mbox{if $\methods(c)(\ell) = (B
       (A_1\:x_1,\ldots,A_n\:x_n),b)$,}\\
    \mbox{then $\Judge{\this\ty c,x_1\ty A_1,\ldots,x_n\ty A_n}{b:B}$.}
    \end{prog}
\end{display}

We can establish the soundness of the type system of this extended
object calculus by essentially the same way we established the
soudness of the type system of the original object calculus. Recall
that a method body is null-blocked if it is of the form $\Null.f_j$,
$\Null.\ell(u_1,\ldots,u_n)$, $\Let{x}{a}\In{b}$ (where $a$ is
null-blocked), or $q[a]$ (where $a$ is null-blocked). A method body is 
stuck if $a$ is not a value, $a$ is not null-blocked, and there is no $a'$
and $p$ such that $a\to^p a'$. We write $a\to^{*}a'$ to mean that
there exists a sequence $a_1,\ldots,a_n$ and principals
$p_1,\ldots,p_{n+1}$ such that
$a\to^{p_1}a_1\to^{p_2}\cdots\to^{p_{n}}a_n\to^{p_{n+1}}a'$.
\begin{theorem}[Soundness]
If $\Judge{\emptyset}{a:A}$, and $a\to^{*}a'$, then $a'$ is not
stuck. 
\end{theorem}
\begin{proof}
A straightforward adaptation of the proof of
Theorem~\ref{thm:soundness}, via corresponding Preservation and
Progress theorems.
\hfill
\end{proof}

To illustrate the usefulness of first-class web services, consider the 
following simple example, where the fact that web services can be
passed as arguments to methods is quite natural. Suppose, as we did in 
Section~\ref{sec:formal-model}, that there are two principals
$\Alice,\Bob\in\Prin$, and a web service
$\mathit{cal}=\textit{http://mycalendar.com/CalendarService}$, where
we have $\class(\mathit{cal})=\mathit{CalendarServiceClass}$. The web
service $\mathit{cal}$ maintains an appointment calendar for
principals. It offers web methods to query a principal's calendar for a 
free time slot, and to reserve time slots. More precisely, the service 
has the following interface:
\[\begin{prog}
\mathit{class}\ \mathit{CalendarServiceClass}\\ \quad
  \begin{prog}
  \Id\:\mathit{CallerId}\\
  \mathit{Bool}\:\mathit{Available}(\Id\:\mathit{account}, 
                                    \mathit{Time}\:\mathit{from},
                                    \mathit{Time}\:\mathit{to})\\ \quad
    \langle\mbox{\textit{check if selected time slot if free for account}}\rangle\\
  \mathit{Void}\:\mathit{Reserve}(\Id\:\mathit{account},
                                  \mathit{Time}\:\mathit{from},
                                  \mathit{Time}\:\mathit{to})\\ \quad
    \langle\mbox{\textit{reserve time slot for account}}\rangle\\
  \end{prog}
\end{prog}\]
(We assume that the classes $\mathit{Bool}$, $\mathit{Time}$, and
$\mathit{Void}$ are provided in the execution environment. The details 
of their implementation are irrelevant to our discussion.)

Suppose that Alice has an account on $\mathit{cal}$, and that she
wants to make an appointment with a calendar-enabled banking
service---that is, a banking service that offers a web method for
scheduling appointments with a bank advisor via a calendar
service. Consider a calendar-enabled version of the banking service of
Section~\ref{sec:formal-model}. Let
$w=\textit{http://bob.com/BankingService}$, where we have
$\owner(w)=\Bob$ and $\class(w)=\mathit{BankingServiceClass}$.  We add
a web method $\mathit{MakeAppt}$ to $\mathit{BankingServiceClass}$
that takes as argument a time period during which the appointment is
sought, and a calendar service that the banking service can query to
confirm that a common free time slot is available between the client
and the bank advisor. The interface of the augmented banking service
is as follows:
\[\begin{prog}
\mathit{class}\ \mathit{BankingServiceClass}\\ \quad
  \begin{prog}
  \Id\:\mathit{CallerId}\\
  \mathit{Num}\:\mathit{Balance}(\mathit{Num}\:\mathit{account})\\ \quad
    \begin{prog}
    \If \mathit{account}=12345 \Then \\ \quad
     \If \mathit{this}.\mathit{CallerId}=\Alice \Then 100 \Else \Null \\
    \Else \ldots
    \end{prog} \\
  \mathit{Time}\:\mathit{MakeAppt}(\mathit{Time}\:\mathit{from},
                                   \mathit{Time}\:\mathit{to},
                                   \WS(\mathit{CalendarService})\:\mathit{cs})\\ \quad
    \ldots \mathit{cs}.\mathit{Available}(\mathit{CallerId},\ldots)\ldots
  \end{prog}
\end{prog}\]
Hence, if Alice wants to make an appointment sometime within the next
week, 
\ifJournal
  she could issue the web method call
\else
  she could call
\fi
$w\mo{:}\mathit{MakeAppt}(\textit{18/11/02:08:00},\textit{23/11/02:17:00},\mathit{cal})$.
(We assume appropriate syntax for constants of type $\mathit{Time}$.)
During the evaluation of this web method invocation, the
implementation of $\mathit{MakeAppt}$ will make calls to
$\mathit{cal}\mo{:}\mathit{Available}$ to find a time slot suitable to
Alice, and finally a call to $\mathit{cal}\mo{:}\mathit{Reserve}$ to
reserve a time slot. A principal with an account on a different
calendar service $c$ would call $w\mo{:}\mathit{MakeAppt}$ passing in
$c$ as the calendar service.

\fi%

\ifLong
 \ifJournal\else
  \clearpage
 \fi
\fi

\bibliographystyle{plain}
\bibliography{srpc}
\end{document}